\documentclass[onecolumn,sort&compress,numbers]{els-mrw} 

\usepackage{amsmath,amssymb,amsfonts,amsthm,makeidx,graphicx}
\usepackage{txfonts}
\usepackage{helvet}
\usepackage{graphics,psfrag,empheq,float}
\usepackage{mnsymbol} 
\usepackage[export]{adjustbox} 
\usepackage{axodraw2} 
\usepackage[colorlinks=true,linkcolor=black]{hyperref}

\newcommand{\bg}{\begin{align}}
\newcommand{\eeg}{\end{align}}
\newcommand{\be}{\begin{equation}}
\newcommand{\ee}{\end{equation}}
\newcommand{\ba}{\begin{eqnarray}}
\newcommand{\ea}{\end{eqnarray}}

\newcommand{\nn}{\nonumber}

\newcommand{\ra}{\rangle}

\newcommand{\hvp}{{\hat{\boldsymbol{p}}}}

\newcommand{\vp}{{\boldsymbol{p}}}
\newcommand{\vq}{{\boldsymbol{q}}}

\newcommand{\Ima}{{\rm Im}}
\newcommand{\Rea}{{\rm Re}}

\newcommand{\cD}{{\cal D}}
\newcommand{\cN}{{\cal N}}
\newcommand{\cdd}{M_{\rm CDD}}
\newcommand{\cS}{{\cal S}}

\newcommand{\ep}{\epsilon}

\newcommand{\vep}{\varepsilon}

\newcommand{\ve}{\varepsilon}
\newcommand{\vh}{\varphi}

\begin{document}




\chapter{Coupled-channel formalism}\label{chap1}


\author[1]{Jos\'e Antonio Oller}%


\address[1]{\orgname{Universidad de Murcia}, \orgdiv{Departmento de F\'{\i}sica}, \orgaddress{Campus de Espinardo, E-30071 Murcia, Spain}}

\articletag{Account of coupled-channel scattering formalism that shares pedagogical and reviewal purposes.
An account of coupled-channel scattering formalism  that serves pedagogical or review purposes.}

\maketitle

\begin{abstract}[Abstract]
This article provides a pedagogical exploration of  coupled-channel scattering in hadronic physics, focusing on theoretical  methodologies, and practical applications. It covers key concepts such as the $N/D$ method, Castillejo-Dalitz-Dyson poles, and the generalization of these techniques to coupled channels. The article also presents a parameterization for partial wave amplitudes, ensuring unitarity and analyticity, and addresses scattering near the threshold of dominant channels, with use also of the Lippmann-Schwinger equation. Additionally, it explores final-state interactions, Omnès and Muskhelishvily solutions, the Khuri-Treiman approach, and the two-potential model for resonant event distributions, highlighting the role of Riemann sheets in resonance behavior. These methods are also applicable to related fields like nuclear and atomic physics.
\end{abstract}

\begin{keywords}
  Scattering \sep
  Unitarity\sep
Analyticity\sep
Resonances\sep
Final-state interactions
\end{keywords}

\section*{Objectives}

\begin{itemize}
\item Section~\ref{sec.181104.1} introduces the uncoupled $N/D$ method, Castillejo-Dalitz-Dyson poles, and associated linear integral equations. It also examines simplifications when the crossed-channel cut is absent, and extends these concepts to coupled channels.
\item Section~\ref{sec.241102.1} presents a general parameterization for partial wave amplitudes in coupled channels, ensuring unitarity and compatibility with analyticity. The discussion includes the process of Riemann sheet transitions, estimates of the subtraction constant in the unitarity loop functions, and the role of Castillejo-Dalitz-Dyson poles in identifying resonance nature.
\item Section~\ref{sec.241031.1} focuses on scattering near the threshold of dominant channels, emphasizing non-relativistic dynamics. It applies previous methods to this context, alongside the effective-range expansion and Flatté parameterization, and explores the Lippmann-Schwinger equation for both contact  and finite-range potentials involving the exchange of bare states.
\item Section~\ref{sec.241103.1} addresses final-state interactions in reactions initiated by external probes, often weaker than the strong interactions. It covers constraints from analyticity and unitarity, the uncoupled Omnès solution, the coupled-channel Muskhelishvily-Omnès extension, and the Khuri-Treiman approach for $\eta \to 3\pi$ decays. Additionally, it discusses the use of the Lippmann-Schwinger equation in the non-relativistic case.
\item Section~\ref{sec.190224.1} examines the two-potential model for studying resonant event distributions. It highlights the importance of the Riemann sheet containing the resonance pole and its impact on physical signals, such as event distributions, and branching ratios.
\end{itemize}

\section{Introduction}

Coupled-channel scattering is an indispensable framework for understanding the complex and multifaceted nature of hadronic interactions. In these systems, the interactions between particles are typically not confined to a single channel but involve multiple channels, including elastic and inelastic scattering processes, bound states and resonance formation in unphysical Riemann sheets. These interactions are often highly interdependent, with each channel influencing the others in subtle ways. This complexity necessitates rather sophisticated approaches generalizing  single-channel models, making the coupled-channel formalism crucial for the accurate description of many scattering phenomena in  the rich realm of hadronic physics.

This article aims to provide a rather comprehensive exploration of coupled-channel scattering in hadronic-physics scattering, with a focus on its theoretical foundations, key methodologies, and phenomenological applications. Needless to say that these methods could also be of use in other related areas like nuclear or atomic/molecular physics.

This chapter is pretended to be  rather self-contained,  requiring only a basic understanding of the topic. Classical references on the subjects here treated are \cite{martin.200705.1,Barton:1965dr,Weinberg:1995mt,Schweber:1961zz,Weinberg:2013qm,olive.181102.1}, which have significantly influenced the content of these lectures. Connecting ones are  author's previous book \cite{Oller:2019rej}, reviews \cite{Oller:2020guq,Oller:2019opk}, and lectures \cite{Oller:2024lrk}.

\section{$N/D$ method and CDD poles}
\def\theequation{\arabic{section}.\arabic{equation}}
\setcounter{equation}{0}   
\label{sec.181104.1}

We begin with the elastic case and then generalize for coupled partial-wave amplitudes (PWAs). Let $T(s)$ denote a PWA with total angular momentum $J$ and other quantum numbers. Above threshold $s_{\text{th}}$, unitarity implies:
\begin{align}
\label{181105.2a}
\Ima T(s)&=|T(s)|^2\rho(s)\longrightarrow \Ima T^{-1}(s)=-\rho(s)\equiv-\frac{|\hvp|}{8\pi\sqrt{s}},~s\geq s_{\text{th}}~.
\end{align}
Along the LHC for $s<s_{\rm L}$, Schwarz reflection leads to:
\begin{equation}
\label{lhc}
T(s+i\epsilon)-T(s-i\epsilon)=2 i \Ima T(s),~s<s_{\rm L}\,.
\end{equation}
Crossed cuts in the complex $s$ plane appear in two forms. For processes with equal masses, like $\pi\pi$ scattering, there is only a LHC for $s<s_{\rm L}$, where $s_{\rm L}=0$ for $\pi\pi$. However, for processes like $a+b\rightarrow a+b$ with $m_1=m_3=m_a$ and $m_2=m_4=m_b$, in addition to a LHC, there is also a circular cut at $|s|=m_2^2-m_1^2$ \cite{martin.200705.1}, with $m_2>m_1$. For simplicity, we treat the LHC as encompassing all unphysical cuts, which suffices for our purposes.

The $N/D$ method \cite{Chew:1960iv} provides a solution for $T(s)$ by expressing it as a quotient:
\begin{equation}
\label{n/d}
T(s)=\frac{N_L(s)}{D_L(s)}~,
\end{equation}
where $D(s)$ has the unitarity cut for $s>s_{\text{th}}$, and $N(s)$ has the LHC for $s<s_{\rm L}$. From Eqs.~(\ref{181105.2a}) and \eqref{n/d}, these functions satisfy:
\begin{align}
\label{eqs1}
\Ima  D(s)&=\Ima T^{-1}(s) N(s)=-\rho(s) N(s), &s>s_{\text{th}} \\
\Ima D(s)&=0,   &s<s_{\text{th}}\nn\\
\label{eqs2}
\Ima  N(s)&=\Ima T(s) \; D(s), &s<s_{\rm L}  \\
\Ima N(s)&=0,  &s>s_{\rm L}\,\nn
\end{align}
with
\begin{align}
  \label{241205.1}
  \rho(s)&=\frac{1}{16\pi} \sqrt{\frac{\lambda(s,m_1^2,m_2^2)}{s^2}}~,\\
  \lambda(s,m_1^2,m_2^2)&=(s - (m_{1} + m_{2})^2)(s - (m_{1} - m_{2})^2)\,.\nn
  \end{align}
Next, we derive an $n$-time subtracted dispersion relation (DR) for $D(s)$, assuming $D(s)/s^n\to 0$ for $s\to\infty$.  Along the unitarity cut, from Eq.~\eqref{eqs1} it follows that $\lim_{s\to\infty}N(s)/s^n=0$ (since $\rho(s)\to \text{constant}$ as $s\to\infty$). This enables an $n$-times subtracted DR for $N(s)$ as well:
\begin{align}
  \label{240814.7}
  \oint dz\frac{D(z)}{(z-s)(z-s_0)^n}&=
  2\pi i\frac{D(s)}{(s-s_0)^n}
  +2\pi i\frac{1}{(n-1)!}
  \left.\frac{d^{n-1}}{dz^{n-1}}\frac{D(z)}{z-s}\right|_{z=s_0}\,.
\end{align}
Dropping the contribution from the circle at infinity, we have:
\begin{align}
\label{240814.8}
\oint dz\frac{D(z)}{(z-s)(z-s_0)^n}&=\int_{s_{\text{th}}}^{+\infty} ds'\frac{D(s'+i\ep)-D(s'-i\ep)}{(s'-s)(s'-s_0)^n}=2i\int_{s_{\text{th}}}^{+\infty} ds'\frac{\Ima D(s')}{(s'-s)(s'-s_0)^n}\,.
\end{align}
Equating Eqs.~\eqref{240814.7} and \eqref{240814.8} and using Eq.~\eqref{eqs1}, we find the dispersive expression for $D(s)$:
\begin{align}
\label{240814.9}
D(s)&=-\frac{(s-s_0)^n}{\pi}\int_{s_{\text{th}}}^{+\infty} ds'\frac{\rho(s')N(s')}{(s'-s)(s'-s_0)^n}-\frac{(s-s_0)^n}{(n-1)!}\left.\frac{d^{n-1}}{dz^{n-1}}\frac{D(z)}{z-s}\right|_{z=s_0}\,.
\end{align}
The last term is a polynomial of degree $n-1$, and Eq.~\eqref{240814.9} becomes:
\begin{equation}
\label{d'}
D(s)=-\frac{(s-s_0)^n}{\pi}\int^{+\infty}_{s_{\text{th}}} ds' 
\frac{\rho(s') {N}(s')}{(s'-s)(s'-s_0)^n}+\sum_{m=0}^{n-1}  a _m s^m\,.
\end{equation}
Similarly, for $N(s)$, we close the integration contour at infinity with a circle that engulfs the LHC for $s<s_{\rm L}$: 
\begin{equation}
\label{n2}
N(s)=\frac{(s-s_0)^{n}}{\pi}\int_{-\infty}^{s_{\rm L}} ds' 
\frac{{\Ima  T}(s') {D}(s')}{ (s'-s)(s'-s_0)^{n}}+
\sum_{m=0}^{n-1}  b_m s^m~.
\end{equation}
Equations~(\ref{d'}) and (\ref{n2}) form a system of coupled integral equations (IEs) for $N(s)$ and $D(s)$, with input ${\Ima T}(s)$ along the LHC, and the coefficients $a_m$ and $b_m$. However, these solutions are not the most general. The inverse of a PWA along the unitarity cut is undefined when it crosses a zero. Thus, Eq.~\eqref{eqs1} is ambiguous at such points. Other zeros in the complex plane can be represented by zeros in $N(s)$, except for those along the unitarity cut, which must be handled separately. These zeros are the Castillejo-Dalitz-Dyson (CDD) poles~\cite{Castillejo:1955ed}. Let $\{s_i\}$ denote the zeros along the unitarity cut, where ${T}(s_i)=0$, then one has \cite{Oller:1998zr} the more general DRs, 
\begin{align}
\label{d'2}
D(s) &= -\frac{(s-s_0)^n}{\pi}\int_{s_{\text{th}}}^{+\infty} \frac{\rho(s') {N}(s')}{(s'-s)(s'-s_0)^n}ds' + \sum_{m=0}^{n-1}{a}_m s^m + \sum_i \frac{\gamma_i}{s-s_i}~, \\
N(s) &= \frac{(s-s_0)^{n}}{\pi}\int_{-\infty}^{s_{\rm L}} ds' \frac{{\Ima T}(s') {D}(s')}{(s'-s)(s'-s_0)^{n}} + \sum_{m=0}^{n-1}  b_m s^m~.\nn
\end{align}
Every CDD pole term has two free parameters: its position $s_i$ and residue $\gamma_i$, which is real as implied by unitarity. For an ordinary potential  the Lippmann-Schwinger (LS) equation can be solved without any free parameters. This corresponds to take an unsubtracted DR for $N(s)$ and a once subtracted DR for $D(s)$, fixing its normalization such that  $D(0) = 1$:
\begin{align}
\label{240814.11}
D(s) &= 1 - \frac{s}{\pi}\int_{s_{\text{th}}}^{+\infty} \frac{\rho(s')N(s')}{s'(s'-s)}\, , \\
N(s) &= \frac{1}{\pi}\int_{-\infty}^{s_{\rm L}}ds'\frac{\Ima T(s')D(s')}{s'-s}\,.\nn
\end{align}
Any additional free parameters in Eq.~\eqref{d'2} compared to Eq.~\eqref{240814.11} reflect microscopic dynamics in the scattering process, such as underlying degrees of freedom. In hadron physics, these may correspond to QCD dynamics, as discussed in Refs.~\cite{Entem:2016ipb,Oller:2018zts,Sanchez:2024xzl}. Reference~\cite{Castillejo:1955ed} originally linked CDD poles to the internal structure of the scatterer, analogous to the Wigner-Eisenbud formula, which expresses all information about a scattering process without knowing the scatterer internal structure. 
A CDD pole can be added to $D(s)$ to reproduce a resonance or bound state by adjusting its parameters, which are related to the coupling constant and mass of an `elementary' particle, not generated by the rescattering of interacting particles due to exchange forces.

The procedure to solve the $N/D$ equations involves substituting the DR for $N(s)$ into that for $D(s)$, resulting in a linear IE for $D(s)$ for $s<s_{\rm L}$. We illustrate this with Eq.~\eqref{240814.11}:
\begin{align}
  \label{240905.1}
  D(s) &= 1 - \frac{s}{\pi^2} \int_{s_{\text{th}}}^{+\infty} ds' \frac{\rho(s')}{s'(s'-s)} \int_{-\infty}^{s_{\rm L}} ds'' \frac{\Ima T(s'') D(s'')}{s'' - s'}.
\end{align}
Exchanging the order of integration gives
\begin{align}
  \label{240905.2}
  D(s) &= 1 + \frac{s}{\pi^2} \int_{-\infty}^{s_{\rm L}} ds'' \Ima T(s'') D(s'') \int_{s_{\text{th}}}^{+\infty} ds' \frac{\rho(s')}{s'(s'-s)(s'-s'')}.
\end{align}
Simplifying the denominator,
\begin{align}
  \label{240907.1}
  -\frac{s}{\pi} \int_{s_{\text{th}}}^{+\infty} ds' \frac{\rho(s')}{s'(s'-s)(s'-s'')} &= -\frac{s}{\pi(s-s'')} \int_{s_{\text{th}}}^{+\infty} ds' \rho(s') \left( \frac{1}{s'(s'-s)} - \frac{1}{s'(s'-s'')} \right).
\end{align}
Each of these integrals is convergent, as discussed in the next section (cf. Eq.~\eqref{240815.3}). This leads to a linear IE for $D(s)$ for $s < s_{\rm L}$. After solving it, numerically if necessary, one can use the same DR and Eq.~\eqref{240814.11} to find $D(s)$ and $N(s)$ for any $s \in \mathbb{C}$. Then, $T(s) = N(s)/D(s)$ is also determined.

Neglecting the LHC, assuming it is sufficiently smooth or weak (as shown for meson-meson scattering in Ref.~\cite{Oller:1998zr}), we can take ${\Ima T}(s) = 0$ in Eq.~(\ref{n2}),
and $N(s)$ becomes a polynomial of degree $n-1$, $N(s) = \mathcal{C} \prod_{j=1}^{n-1} (s - \tilde{s}_j)$\,. Thus, apart from the constant $\mathcal{C}$, $N(s)$ introduces $n-1$ zeros in $T(s)$. One can divide both $N(s)$ and $D(s)$ by the former, so that we can then set $N(s) = 1$ and treat the zeros of $T(s)$ as poles of $D(s)$, in addition to the CDD poles in Eq.~\eqref{d'2}. These poles contribute extra terms to the DR for $D(s)$, which must be added to Eq.~\eqref{240814.7}. After simplification, they contribute similarly to the CDD poles in Eq.~\eqref{d'2}, and we obtain:
\begin{align}
  \label{fin/d}
  T(s) &= \frac{1}{D(s)}, \\
  N(s) &= 1, \nn\\
  D(s) &= -\frac{(s - s_0)^n}{\pi} \int_{s_{\text{th}}}^{+\infty} ds' \frac{\rho(s')}{(s' - s)(s' - s_0)^n} + \sum_{m=0}^{n-1} a_m s^m + \sum_i \frac{\gamma_i}{s - s_i}.\nn
\end{align}
Since $\rho(s')/s'$ vanishes as $s' \to \infty$, the integral simplifies by canceling $n-1$ factors of $s' - s_0$ in the denominator using
\begin{align}
  \label{240815.2}
  \frac{(s - s_0)^n}{(s' - s)(s' - s_0)^n} =  \frac{(s - s_0)^{n-1}}{(s' - s)(s' - s_0)^{n-1}} - \frac{(s - s_0)^{n-1}}{(s' - s_0)^n}.
\end{align}
The last term is a polynomial of degree $n-1$, reabsorbed in the sum $\sum_m a_m s^m$ of Eq.~\eqref{fin/d}. Repeating this process $n-1$ times, we get:
\begin{align}
  \label{fin/dd}
  T(s) &= \frac{1}{{D}(s)}, \\
  N(s) &= 1, \nn\\
  D(s) &= -\frac{s - s_0}{\pi} \int_{s_{\text{th}}}^{+\infty} ds' \frac{\rho(s')}{(s' - s_0)(s' - s)} + \sum_{m=0}^{n-1} a_m s^m + \sum_i \frac{\gamma_i}{s - s_i}.\nn
\end{align}
For a complex pole $s_i$ with residue $\gamma_i$, its complex conjugate $s_i^*$ also appears as a pole with residue $\gamma_i^*$. A pair of complex conjugate poles thus adds four parameters instead of two, as with a real CDD pole. In Eq.~\eqref{fin/dd}, $N(s)$ has been normalized to 1, unlike in Eq.~\eqref{240814.11}, where $D(0) = 1$. Equation~\eqref{fin/dd}, first derived in Ref.~\cite{Oller:1998zr}, represents the general form of an elastic PWA, with arbitrary $J$, neglecting the LHC. The parameters can be fitted to experimental data or derived from underlying theory. In Ref.~\cite{Oller:1998zr}, the dynamics are based on QCD, but Eq.~\eqref{fin/dd} applies to other interactions, such as those in Refs.~\cite{Blas:2020dyg, Blas:2020och} for graviton-graviton interactions.
\begin{figure}
  \begin{center}
  \includegraphics[width=0.25\textwidth]{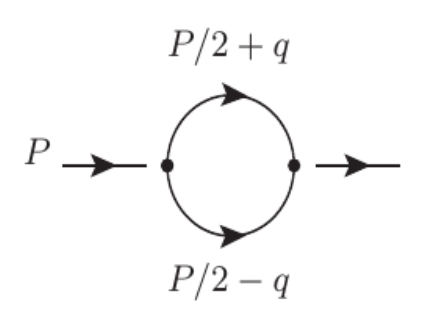}
  \caption{{\small The loop function $g(s)$ in Eq.~\eqref{240815.4}, with momentum $P$.}\label{fig.240815.1}}
  \end{center}
\end{figure}
The integral along the unitarity cut in Eq.~\eqref{fin/dd} is denoted as $g(s)$:
\begin{align}
  \label{240815.3}
  g(s) &= g(s_{\text{th}}) - \frac{s - s_{\text{th}}}{\pi} \int_{s_{\text{th}}}^{+\infty} ds' \frac{\rho(s')}{(s' - s_{\text{th}})(s' - s)}.
\end{align}
This integral corresponds to the loop integral shown in Fig.~\ref{fig.240815.1}:
\begin{align}
  \label{240815.4}
  g(s) &= i \int \frac{d^4q}{(2\pi)^4} \frac{1}{[(P/2 - q)^2 - m_1^2 + i\ep][(P/2 + q)^2 - m_2^2 + i\ep]} = \frac{1}{(4\pi)^2} \left\{ a(\mu) + \ln \frac{m_1^2}{\mu^2} - x_+ \ln \frac{x_+ - 1}{x_+} - x_- \ln \frac{x_- - 1}{x_-} \right\}, 
\end{align}
where $x_\pm = \frac{E_2 \pm |\vp|}{\sqrt{s}}$. The algebraic expression for $g(s)$ is obtained by calculating in dimensional regularization  the loop in Eq.~\eqref{240815.3} and reabsorbing the divergence $1/(d-4)$ in the subtraction constant $a(\mu)$. From Eq.~\eqref{240815.4}, we have:
\begin{align}
  \label{240815.5}
  g(s_{\text{th}}) &= \frac{a(\mu)}{(4\pi)^2} + \frac{m_1 \ln(m_1/\mu) + m_2 \ln(m_2/\mu)}{8\pi^2(m_1 + m_2)}.
\end{align}
In terms of $g(s)$, the DR for $D(s)$ in Eq.~\eqref{fin/dd} simplifies (with the polynomial term reduced to a constant, which is pertinent  for a meaningful DR with $N(s)=1$):
\begin{align}
  \label{240815.6}
  D(s) &= \alpha + \sum_i \frac{\gamma_i}{s-s_i} + g(s)\,, \\
  \label{240815.6b}
  T(s) &= \frac{1}{\alpha + \sum_i \frac{\gamma_i}{s-s_i} + g(s)}= \frac{\left(\alpha + \sum_i \frac{\gamma_i}{s-s_i}\right)^{-1}}{1 + \left(\alpha + \sum_i \frac{\gamma_i}{s-s_i}\right)^{-1} g(s)}\,.
\end{align}
The constant $\alpha$ can also be seen as a CDD pole at infinity, with the ratio $-\gamma_k/s_k \to \alpha$.

To extend the $N/D$ method to coupled channels, we use a matrix formalism \cite{Bjorken:1960zz,Sanchez:2024xzl}.  We neglect unphysical cuts and consider ${T}$ to have only a unitarity cut. The matrix relation for ${T}$ becomes:
\begin{equation}
\label{A.4}
T(s) = D(s)^{-1} N(s)\,,
\end{equation}
where ${N}(s)$ contains only LHC (currently neglected), and ${D}(s)$ has only the unitarity cut. ${N}(s)$ can be chosen free of poles, with the corresponding zeroes absorbed into ${D}(s)$ to account for bound states. Thus, ${N}(s)$ is a polynomial matrix of degree $n-1$:
\begin{equation}
\label{A.5}
N(s) = \hat{Q}_{n-1}(s)\,.
\end{equation}
Unitarity in coupled channels implies in matrix language that for $s>s_{\text{th};1}$, the lightest threshold,
\begin{align}
  \label{240813.3e}
\Ima {T}(s)^{-1}&=-{\rho}(s)\,,
\end{align}
where $\rho(s)$ now is a diagonal matrix with the phase space for each open channel. From Eqs.~(\ref{240813.3e}) and (\ref{A.4}), we obtain:
\begin{equation}
\label{A.6}
\Ima D(s) = -N(s) {\rho}(s)\,.
\end{equation}
Since ${N}(s)$ is polynomial, it can be reabsorbed into ${D}$, so that $D^{-1} N = (N^{-1} D)^{-1}$. Neglecting the LHC, we take ${N}(s) = I$ and ${D}(s)$ satisfying Eq.~\eqref{A.6}. Additionally, we have:
\begin{align}
  \label{240815.7}
N^{-1} &= \frac{1}{\det{N}} \text{adj}{N}\,,
\end{align}
where $\text{adj}{N}$ is the adjoint matrix and $\det{N}$ is the determinant. Thus, $\displaystyle{N^{-1} D = \frac{\text{adj}{N}\, D}{\det{N}}}$, which has poles at the zeroes of $\det{N}(s)$. We also define the diagonal matrix ${G}(s)$, where each diagonal element corresponds to $g(s)$ in Eq.~\eqref{240815.4} for every channel, using the relevant masses and subtraction constants $a_i(\mu)$. Next, analogous to the single-channel case in Eq.~\eqref{fin/dd}, we write for the coupled case:
\begin{align}
\label{A.8}
T(s) &= D^{-1}\,, \\
N(s) &= I\,, \nonumber \\
D(s) &= \underbrace{G(\hat{s}_{\rm th}) - \frac{s-\hat{s}_{\rm th}}{\pi} \int_{\hat{s}_{\rm th}}^{+\infty} ds' \frac{{\rho}(s')}{(s'-\hat{s}_{\rm th})(s'-s)}}_{G(s)} + \sum_{m=0}^n \hat{a}_m s^m + \sum_i \frac{\hat{\gamma}_i}{s-s_i}\,,
\end{align}
where the subtractive coefficients $\hat{a}_m$ and CDD-pole residues $\hat{\gamma}_i$ are symmetric matrices, and  the matrix $\hat{s}_{\rm th}$ contains the threshold values for each channel. 
Finally, ${T}(s)$ can be expressed as:
\begin{align}
  \label{240815.8}
  T(s) &= \left(I + \left( \sum_{m=0}^n \hat{a}_m s^m + \sum_i \frac{\hat{\gamma}_i}{s-s_i} \right)^{-1} G(s) \right)^{-1} \left( \sum_{m=0}^n \hat{a}_m s^m + \sum_i \frac{\hat{\gamma}_i}{s-s_i} \right)^{-1}\,.
\end{align}

The interested reader can consider the explicit inclusion of the LHC for the coupled $N/D$ method as done, for instance, in Refs.~\cite{Sanchez:2024xzl,Albaladejo:2012sa,Oller:2018zts}. Other more advanced topics not covered in this brief summary of the $N/D$ method are the first-iterated $N/D$ method \cite{martin.200705.1,Gulmez:2016scm,Du:2018gyn} and the {\it exact} $N/D$ method \cite{Entem:2016ipb,Oller:2018zts}. 

The $S$- and $T$-matrices in PWAs are related by 
\begin{align}
  \label{241201.1}
  S(s)&=I+2i\rho(s)^\frac{1}{2}T(s)\rho(s)^\frac{1}{2}\,,
\end{align}
and  $S$ is unitary and symmetric, $SS^\dagger=S^\dagger S=I$. 
In the two-coupled channel case, the phase shifts $\delta_i(s)$ and inelasticity parameter $\eta(s)$  parameterize the unitary $S$ matrix  as
\begin{align}
\label{181120.3}
S(s) = \left(
\begin{array}{ll}
\eta e^{2i\delta_1} & i\sqrt{1-\eta^2}e^{i(\delta_1+\delta_2)} \\
i\sqrt{1-\eta^2}e^{i(\delta_1+\delta_2)} & \eta e^{2i\delta_2}
\end{array}
\right)~.
\end{align}

\section{General Parameterization for PWAs and Access to Unphysical Riemann Sheets}
\def\theequation{\arabic{section}.\arabic{equation}}
\setcounter{equation}{0}
\label{sec.241102.1}

\begin{figure}
  \begin{center}
    \includegraphics[width=0.35\textwidth]{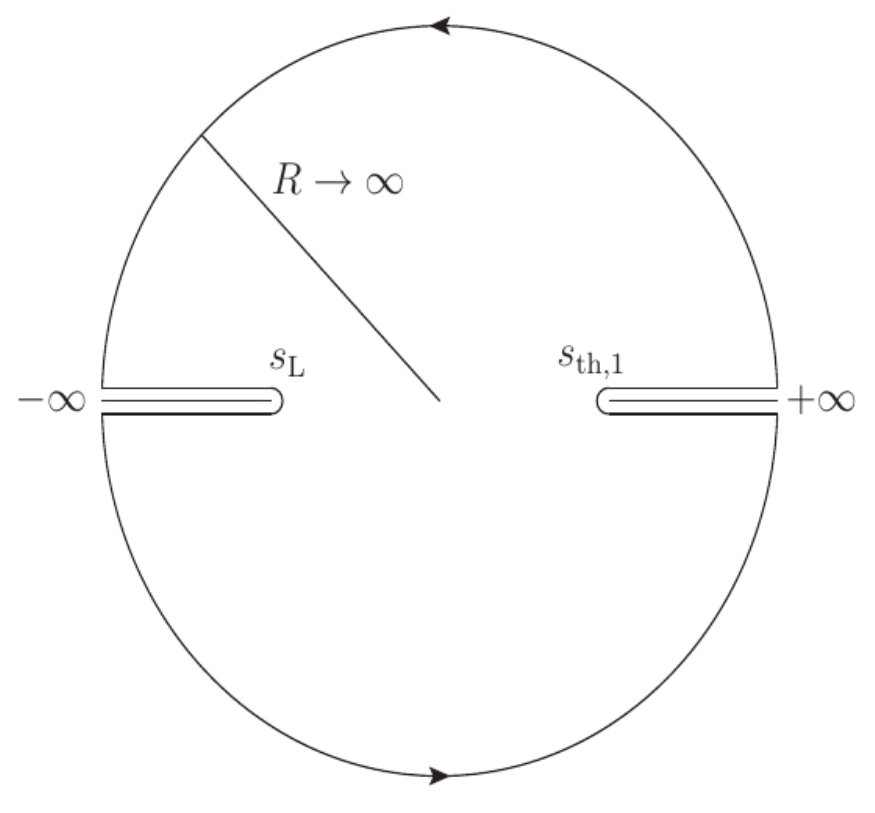}
    \caption{{\small Contour ${\cal C}$ used for the parameterization in Eq.~\eqref{181109.6} of $T(s)$.}\label{fig.240815.3}}
  \end{center}
\end{figure}

We begin by deriving a generic parameterization for the $T$ matrix of PWAs, isolating the unitarity cut and respecting the related analytic properties. To this end, we use a DR for the inverse of $T(s)$ taking into account unitarity in the form  given in Eq.~\eqref{240813.3e}. This leads to the diagonal matrix ${G}(s)$ containing the functions $g_i(s)$,  cf. Eq.~\eqref{240815.4}. Additionally, there are crossed cuts, represented by a LHC, as seen in Fig.~\ref{fig.240815.3}. The contribution from these cuts generates the matrix ${\cal N}(s)^{-1}$. We then arrive to the result:
\begin{align}
  \label{181109.6}
  {T}(s) &= \left({\cal N}(s)^{-1} + {G}(s)\right)^{-1}\,, \\
  \label{181109.6b}
  {T}(s) &= \left[I + {\cal N}(s){G}(s)\right]^{-1}{\cal N}(s)\,.
\end{align}
If  the crossed cuts are neglected, Eq.~\eqref{240815.8} becomes applicable, leading to:
\begin{align}
  \label{240815.9}
  {\cal N}(s)^{-1} &= \sum_{m=0}^n \hat{a}_m s^m + \sum_i \frac{\hat{\gamma}_i}{s - s_i}\,.
\end{align}

\begin{figure}
  \begin{center}
    \includegraphics[width=0.8\textwidth]{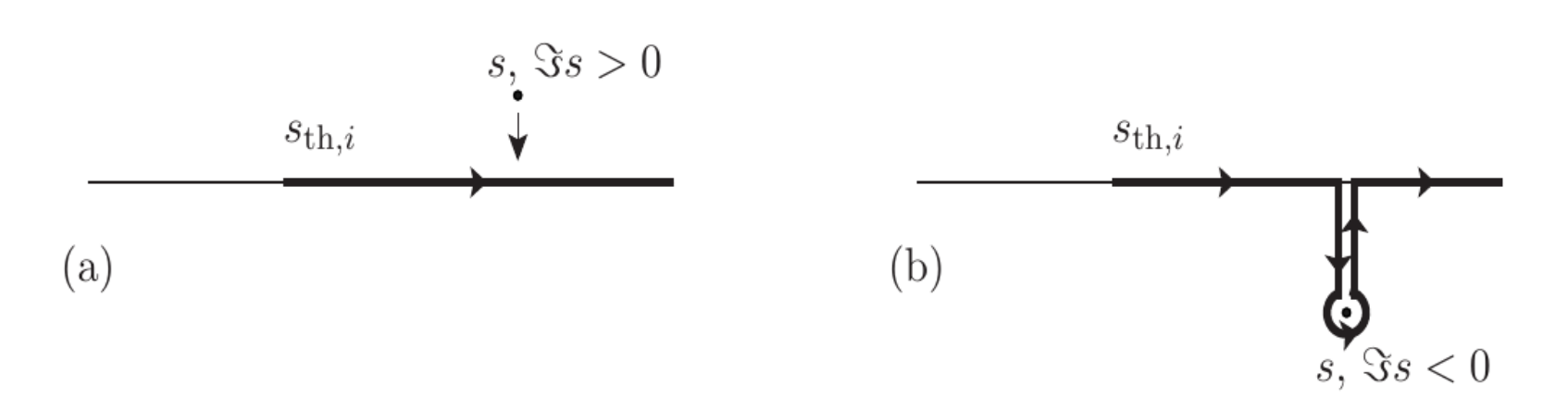}
    \caption{{\small Contour deformation (solid line) for passing to the second RS of $g_i(s)$, crossing the RHC from top (a) to bottom (b). The deformation avoids the singularity at $s' = s$.}}\label{fig.240815.2}
  \end{center}
\end{figure}

Next, we turn our attention to accessing unphysical Riemann sheets (RSs), which is essential for identifying poles corresponding to resonances and antibound states. These states reside in unphysical RSs, as bound states appear in the first or physical RS. To access the second RS of $g_i(s)$, we perform an analytic continuation across the RHC, also referred as unitarity cut. This requires deforming the integration contour in the representation of $g_i(s)$, as shown in Fig.~\ref{fig.240815.2}, ensuring the singularity at $s' = s$ is bypassed. We denote the function $g_i(s)$ in the second RS by $g_{II,i}(s)$. They are related by:
\begin{align}
  \label{181110.1}
  g_{II,i}(s) &= g_i(s) - 2i \rho_{II,i}(s) = g_i(s) + 2i \rho_{I,i}(s)\,,
\end{align}
where $\rho_{I,i}(s)$ represents the function in the first RS, Eq.~\eqref{241205.1},  
with $\sqrt{z}$ taken in the first RS ($\arg z \in [0, 2\pi)$). The change of sign in front of $\rho_i(s)$ in the last two terms  in Eq.~\eqref{181110.1} reflects that $\rho_{II,i}(s) = -\rho_{I,i}(s)$.  This two-sheet structure implies that after a complete circle around the threshold and a subsequent crossing of the RHC, we return to the first RS. This process indicates that the singularity is of square-root type, similar to the momentum in the center-of-mass (CM) system: 
\begin{align}
  \label{181110.3}
  |\vp(s)| &= \pm \sqrt{\frac{\lambda(s, m_1^2, m_2^2)}{4s}}\,,
\end{align}
and the RSs are labeled based on the sign in front of the square root, e.g., $(+, +, \dots)$ for the first RS, $(-, +, \dots)$ for the second, and so on.

\begin{figure}[H]
  \begin{center}
  \begin{tabular}{lr}
    \hspace{-1cm} \includegraphics[width=0.4\textwidth]{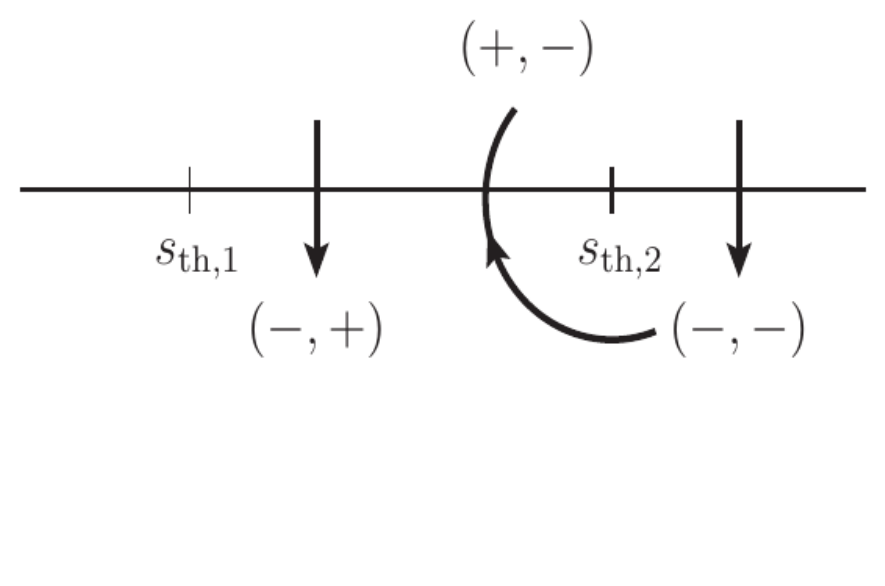} &
    \hspace{1cm} \includegraphics[width=0.5\textwidth]{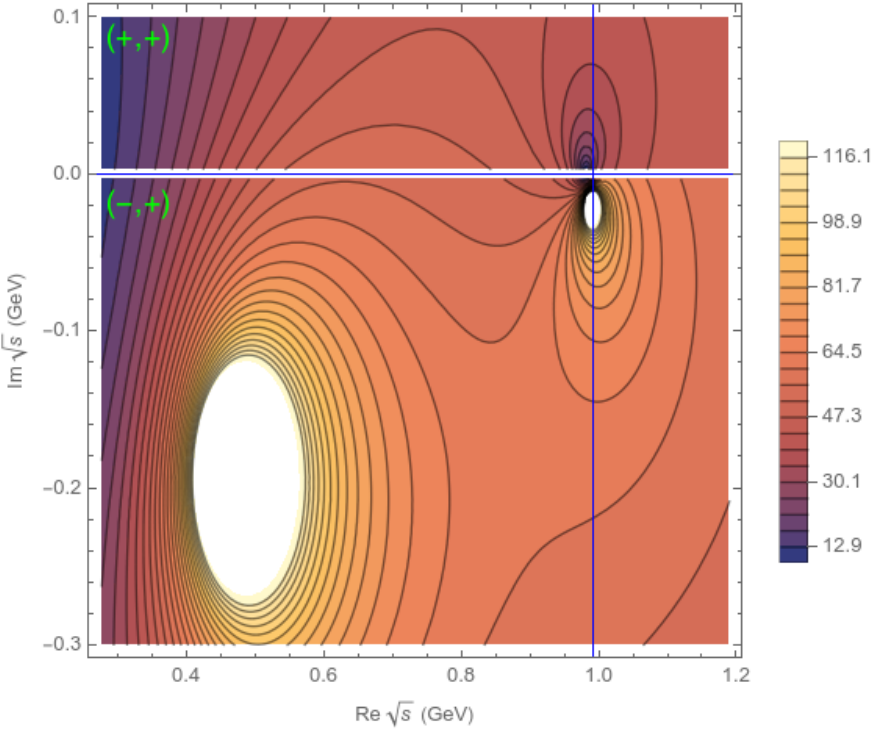}
  \end{tabular}
  \caption{{\small Left: Two-channel case with four RSs. Right: Resonances in the second RS ($\Ima s < 0$) affecting the first RS ($\Ima s > 0$). The $\sigma/f_0(500)$ and $f_0(980)$ poles are depicted.}}\label{fig.240816.1}
\end{center}
\end{figure}
The left panel of Fig.~\ref{fig.240816.1} illustrates the case of two channels and four RSs. Notice that the third RS is not contiguous to the  first RS in any segment of the physical $s$ axis. In the right panel, a resonance in the second RS affects the physical RS above the real axis for $s_{\text{th};2}>s>s_{\text{th};1}$. 
A second pole near the threshold of the heavier channel, also in the second RS, directly affects the first RS only below the threshold of the heavier channel, but not above it. As a result, the resonance signal is not symmetric around $s_{\text{th};2}$. This analysis demonstrates the complex structure of multiple RSs and their interrelations, especially in coupled-channel systems like $S$-wave $\pi\pi$ and $K\bar{K}$ scattering, the example shown in Fig.~\ref{fig.240816.1}.

\subsection{Example from $\pi\pi$ scattering}
\label{sec.241120.2}

The lightest QCD resonance is the $\sigma$ or $f_0(500)$, while the dominant one below 1~GeV is the $\rho(770)$ \cite{ParticleDataGroup:2024cfk}. The low-energy effective field theory of QCD for $s\ll 1$~GeV$^2$ is Chiral Perturbation Theory (ChPT) \cite{Gasser:1984gg}, with an expected expansion scale of $\sim 1$~GeV$^2$. The $\sigma$ resonance has a pole at
\begin{align}
\label{240816.10}
\sqrt{s}_\sigma\approx 450-i\,250~\text{MeV},
\end{align}
indicating a convergence radius for ChPT much smaller than expected, $|s_\sigma|\approx 0.25\ll 1$~GeV$^2$. Historically, this  challenged the acceptance of the $\sigma$ resonance pole. This pole can be understood from chiral symmetry and unitarity, with $m_\pi$ and $f_\pi\approx 92$~MeV, the pion mass and weak-decay constant, respectively,  as inputs \cite{Oller:1997ti,Oller:2000wa}. The $\pi^0\pi^0\to\pi^+\pi^-$ scattering amplitude at leading order in ChPT is
\begin{align}
\label{240816.1}
A(s,t,u) = \frac{s-m_\pi^2}{f_\pi^2}\left(1 + \mathcal{O} \left(\frac{s}{\Lambda^2}\right)\right)\,.
\end{align}
A pedagogical derivation of the $\pi\pi$ scattering amplitudes with definite isospin  by employing crossing and isospin symmetry from $A(s,t,u)$ is given in  Sec.~3 of Ref.~\cite{Oller:2024lrk}.  Then,  the isoscalar $\pi\pi$ amplitude is 
\begin{align}
\label{240816.2}
A_0(s,t,u) = 2 \frac{s-m_\pi^2/2}{f_\pi^2}\,.
\end{align}
Its partial-wave projection in  $S$-wave is trivial,  $A_{00}(s)=A_0(s,t,u)$. This PWA has no LHC and we then use Eq.~\eqref{240815.6b} for its unitarization. The zero at $s=m_\pi^2/2$ implies the need of a CDD pole with residue $\gamma_1=f_\pi^2$. From Eq.~\eqref{240815.6b} we have
\begin{align}
\label{240816.4}
T_{00}(s) = \left(\frac{f_\pi^2}{s-m_\pi^2/2} + g(s)\right)^{-1}\,.
\end{align}
For equal masses, $g(s)$ simplifies as
\begin{align}
\label{240816.5}
g(s) &= \frac{1}{16\pi^2}\left(a(\mu) + \ln\frac{m^2}{\mu^2} - \sigma(s)\ln\frac{\sigma(s)-1}{\sigma(s)+1}\right)\,,\\
\sigma(s) &= \sqrt{1-\frac{4m_\pi^2}{s}}\,.\nn
\end{align}
We now estimate the subtraction constant $a(\mu)$ with a natural-size assumption, relevant for low-energy effective field theory. Using a momentum cutoff $\Lambda$, we express $g_\Lambda(s)$ as
\begin{align}
\label{240816.6}
g_\Lambda(s) = i \int^\Lambda \frac{d^4 q}{(2\pi)^4} \frac{1}{[(P/2-q)^2 - m_\pi^2 + i\epsilon][(P/2+q)^2 - m_\pi^2 + i\epsilon]} = \int_0^\Lambda \frac{p^2 dp}{2\pi^2 w(p)} \frac{1}{s-4w(p)^2+i\epsilon}\,.
\end{align}
Matching $g(s)$ and $g_\Lambda(s)$ at threshold $s_{\text{th}}$, where the momentum is zero, gives
\begin{align}
\label{240816.7}
g_\Lambda(s_{\text{th}}) = -\frac{1}{8\pi^2} \left( \ln\left(1 + \sqrt{1 + \frac{m_\pi^2}{\Lambda^2}}\right) - \ln\frac{m_\pi}{\Lambda} \right)\,.
\end{align}
Equating with $g(s_{\text{th}})$ from Eq.~\eqref{240815.5}, we get for $m_1 = m_2 = m_\pi$
\begin{align}
\label{240816.8a}
a(\mu) = -2 \ln \left(1 + \sqrt{1 + \frac{m_\pi^2}{\Lambda^2}}\right) - \ln \frac{\Lambda^2}{\mu^2}\,.
\end{align}
For $\mu = \Lambda = 1$~GeV, this gives
\begin{align}
\label{240816.8}
a(\Lambda) \approx -2 \ln 2 = -1.40\,.
\end{align}
This is the natural value for the subtraction constant, as discussed in Ref.~\cite{Oller:2000fj}.  Extrapolating Eq.~\eqref{240816.4}  to the second RS, so that $g(s)\to g_{II}(s)$,  the $\sigma$ pole is found at
\begin{align}
  \label{181106.9b}
  s_\sigma &= (0.47 - i\, 0.20)^2~\text{GeV}^2,
\end{align}
in excellent agreement with Eq.~\eqref{240816.10} and compatible with the PDG estimate \cite{ParticleDataGroup:2024cfk}, $s_\sigma = (0.4 - 0.5 - i\,(0.20 - 0.35))^2~\text{GeV}^2\,.$ 
This demonstrates that the $\sigma$ resonance is primarily driven by $\pi\pi$ self-interactions dictated by chiral symmetry, unitarity, and analyticity.

For $P$-wave $\pi\pi$ interactions, which are isovector because of the Bose-Einstein symmetry obeyed by pions, using $A(s,t,u)$ from Eq.~\eqref{240816.1} in the formula for the $I=1$ $\pi\pi$ scattering amplitude \cite{Oller:2024lrk}, and then projected in $P$ wave, we obtain
\begin{align}
  \label{240816.11}
A_{11}(s) &= \frac{s - 4m_\pi^2}{6f_\pi^2}\left(1 + \mathcal{O}\left(\frac{s}{\Lambda^2}\right)\right)\,.
\end{align}
Compared to $A_{00}(s)$ for $I=J=0$, this PWA is suppressed by a factor of 6 for $|s| \gg m_\pi^2$. Not surprisingly, a CDD pole is needed at threshold $4m_\pi^2$ due to the $P$-wave nature of the interaction, with its residue $6f_\pi^2$ fixed by chiral symmetry. Thus, we have
\begin{eqnarray}
  \label{240816.12}
T_{11}(s) = \left(\frac{6f_\pi^2}{s - 4m_\pi^2} + g(s)\right)^{-1} &\to&  
T^{II}_{11}(s) = \left(\frac{6f_\pi^2}{s - 4m_\pi^2} + g_{II}(s)\right)^{-1}\,.
\end{eqnarray}
However, the subtraction constant $a(\Lambda)$ in Eq.~\eqref{240816.8} is too small to yield a good $\rho$ pole. This is because of the suppression by a factor of 6 in $A_{11}(s)$, making its inverse six times bigger. Therefore, a much larger subtraction constant is needed to locate the $\rho$ pole. For $a(\mu) = -14$, $\mu = 1$~GeV, $T^{II}_{11}(s)$ has a pole at
\begin{align}
  \label{181106.10}
s_\rho &= (0.777 - i\, 0.072)^2~\text{GeV}^2\,,
\end{align}
matching the $\rho(770)$ pole at $s_\rho = (0.761 - 0.762 - i\,(0.071 - 0.074))^2~\text{GeV}^2$ from PDG \cite{ParticleDataGroup:2024cfk}. Achieving $a(\mu) = -14$ via Eq.~\eqref{240816.8a} with $\mu = 1$~GeV requires an unreasonably large cutoff $\Lambda = 600$~GeV. Hence, the $\rho(770)$ cannot be explained as a $\pi\pi$-dynamically generated resonance, indicating a different nature compared to the $\sigma$.

The tree-level leading-order ChPT amplitude, Eq.~\eqref{240816.11}, along with the bare $\rho$ exchange in the $s$ channel, can be computed as \cite{Oller:1998zr,Bernard:1991zc}
\begin{equation}
\label{t1n2}
A_{11}(s) = \frac{2}{3} \frac{s - 4m_\pi^2}{f_\pi^2}\left[1 + g_v^2 \frac{s}{M_\rho^2 - s}\right]~.
\end{equation}
The KSFR relation \cite{Kawarabayashi:1966kd,Riazuddin:1966sw} implies $g_v = 1$. We can match this amplitude by introducing two CDD poles in Eq.~\eqref{240815.6b}. The new CDD pole location and its residue in $T_{11}(s)$ are
\begin{align}
\label{181108.1}
s_2 &= \frac{M_\rho^2}{1 - g_v^2}\,, \\
\gamma_2 &= \frac{6f_\pi^2}{1 - g_v^2} \frac{g_v^2 M_\rho^2}{M_\rho^2 - 4(1 - g_v^2) m_\pi^2}\,. \nn
\end{align}
As $g_v^2 \to 1$, $s_2 \to \infty$, leading to
\begin{align}
\label{181108.2}
\lim_{s_2 \to \infty} \frac{\gamma_2}{s - s_2} = - \frac{6f_\pi^2}{M_\rho^2}~.
\end{align}
This is why we can generate the $\rho(770)$ pole by adjusting the subtraction constant $a(\mu)$ in $g(s)$, Eq.~\eqref{240816.12}. Indeed, Eq.~\eqref{181108.2} times $16\pi^2$ gives $-13.6$, which matches the value for $a(1~\text{GeV})\approx -14$ above. This highlights the elementary nature of the $\rho(770)$ resonance in terms of pionic degrees of freedom. Similar discussions apply  to the $\kappa$ and $K^*(890)$ in $K\pi$ scattering \cite{Oller:1998zr} as to the $f_0(500)$ and $\rho(770)$ in $\pi\pi$ scattering, respectively.

\section{Non-relativistic Scattering Equations}
\label{sec.241031.1}

Non-relativistic scattering is relevant when the momenta of dominant channels are much smaller than their energy, so that the velocity $|p_i/E_i|$ is small compared to the speed of light. Various unitarization techniques based on the analytical properties of PWAs, such as the $N/D$ method, effective-range expansion (ERE), and Flatté parameterizations, stress the role of the unitary cut. We also describe methods using the Lippmann-Schwinger equation for exactly soluble potentials with only contact interactions, and for other finite-range  potentials that could include  bare-state exchanges too.

\subsection{CDD Poles, Effective-Range Expansion, and the One-Channel Flatté Parameterizations}
\label{sec.241101.2}

We evaluate non-relativistic scattering near the threshold, assuming the LHC is distant or weak enough. The general results from Sec.~\ref{sec.181104.1}, particularly Eq.~\eqref{fin/d}, are applied under these assumptions, and we specially focus on $S$-wave scattering. For non-relativistic scattering, the kinetic energy \( E \) is $E = {p^2}/{2\mu}$\,, 
with $\mu$ the reduced mass, and \( \sqrt{s} = m_1 + m_2 + E + {\cal O}(p^4) \). The unitary loop function from the RHC integral is explicitly given in Eq.~\eqref{240815.4}, and expanding this function in powers of \( p \) gives the leading terms:
\begin{align}
\label{181204.1}
g(p) &= \frac{a}{16\pi^2} + \frac{1}{8\pi^2(m_1+m_2)}\bigg(m_1\log\frac{m_1}{\mu} + m_2\log\frac{m_2}{\mu}\bigg) 
- i\frac{p}{8\pi(m_1+m_2)} + {\cal O}\left(\frac{p^2}{Q^2}\right)~,
\end{align}
where \( Q \) represents a characteristic mass scale. Nonetheless, for a standard PWA \( t(E) \), the usual normalization is:
\begin{align}
\label{190212.1}
t(E) &= \frac{1}{p\cot\delta - ip}~,
\end{align}
to which we shift along this section. 
The constant term on the right-hand side of Eq.~\eqref{181204.1}, multiplied by \( 8\pi(m_1+m_2) \), is denoted by \( \beta \):
\begin{align}
\label{181204.3}
\beta &= \frac{a(m_1+m_2)}{2\pi} + \frac{1}{\pi}\left(m_1\log\frac{m_1}{\mu} + m_2\log\frac{m_2}{\mu}\right)~.
\end{align}
Considering Eq.~\eqref{fin/d} in $S$ wave, the structures near the threshold, in addition to the branch-point singularity, arise from CDD pole contributions. As in Ref.~\cite{Kang:2016jxw}, we include a single CDD pole:
\begin{align}
\label{181204.4}
t(E) &= \left(\frac{\gamma}{E - M_{\rm CDD}} + \beta - ip(E)\right)^{-1}~.
\end{align}
This equation extends the ERE, which for orbital angular momentum $\ell$ reads:
\begin{align}
\label{181113.4}
p^{2\ell+1}\cot\delta_\ell &= -\frac{1}{a} + \frac{1}{2}r p^2 + \sum_{i=2}^\infty v_i p^{2i}~.
\end{align}
 Any values of \( a \), \( r_2 \), and \( v_2 \) in the ERE can be reproduced by selecting \(\gamma\), \( M_{\rm CDD} \), and \(\beta\):
\begin{align}
\label{181204.5}
\frac{1}{a} &= \frac{\gamma}{M_{\rm CDD}} - \beta~,\\
r_2 &= -\frac{\gamma}{\mu M_{\rm CDD}^2}~,\nn\\
v_2 &= -\frac{\gamma}{4\mu^2 M_{\rm CDD}^3}~.\nn
\end{align}
The ERE is valid near the threshold and up to the nearest singularity of \( t(E)^{-1} \), considering the threshold branch point by removing the term \(-ip=-i\sqrt{p^2}\) in \( t(E)^{-1} \). For \( NN \) scattering, this corresponds to the one-pion exchange threshold. In other systems, the radius of convergence may be determined by the nearest threshold zero in \( t(E) \), often attributed to a CDD pole, which limits the ERE radius of convergence to \( 2\mu |\cdd| \). 
A significant consequence of Eq.~\eqref{181204.5} is that a near-threshold CDD pole, with \( \cdd \to 0 \), leads to large values of \( r \) (as \( \gamma \) can be either positive or negative). This behavior also affects higher-order ERE parameters like \( v_2 \), as shown in Eq.~\eqref{181204.5}. In this limit, the scattering length \( a \) approaches zero.

An alternative commonly used method for near-threshold scattering and an associated resonance is the Flatté parameterization~\cite{Flatte:1976xu}: 
\begin{align}
\label{181206.1}
t_F(E) &= \frac{g^2/2}{E_f - E - i\frac{1}{2}\tilde{\Gamma}(E)}~,\\
\tilde{\Gamma}(E) &= g^2 p(E)~, \hspace{0.4cm} E > 0~, \nn \\
\tilde{\Gamma}(E) &= i g^2 |p(E)|~, \quad E < 0~,\nn
\end{align}
where \( g^2 \geq 0 \) ensures that \( \tilde{\Gamma}(E) \geq 0 \) for \( E > 0 \), and the Flatté mass \( E_f \) is where the real part of \( t_F(E)^{-1} \) vanishes. Notably, \( t_F(E) \) is a particular case of the ERE when considered up to \( \mathcal{O}(p^2) \), with \( a \) and \( r \) related to \( E_f \) and \( g \) as:
\begin{align}
\label{181206.2}
a &= -\frac{g^2}{2E_f}~, \\
r &= -\frac{2}{g^2\mu}~.\nn
\end{align}
A limitation of the Flatté parameterization is that it only accommodates \( r < 0 \).  The pole structure of \( t_F(E) \) depends on the sign of \( E_f \). Solving for the poles gives:
\begin{align}
\label{181206.3b}
p_{i} &= -i\frac{g^2\mu}{2} \left( 1 +(-1)^{i+1} \sqrt{1 - \frac{8E_f}{g^4\mu}} \right)~,
\end{align}

For \( E_f < 0 \), the poles are imaginary, with \( p_1 \) corresponding to a virtual state and \( p_2 \) to a bound state. The magnitude of \( p_1 \) exceeds that of \( p_2 \). At \( E_f = 0 \), the bound state reaches the threshold. For \( 0 < E_f < g^4\mu/8 \), the bound state becomes another virtual state closer to the threshold. When \( E_f \to g^4\mu/8 \), the poles merge into a double virtual-state pole~\cite{Kang:2016jxw}.

For \( E_f > g^4\mu/8 \), the poles correspond to a resonance with equal imaginary parts and symmetric real parts, following the Schwarz reflection principle. A final limitation of the Flatté model is that it cannot produce poles of multiplicity greater than two.
The resonances poles in energy are located at 
\begin{align}
\label{181206.5}
E_{1,2} &= E_f - \frac{g^4 \mu}{4} \mp i\frac{g^4\mu}{4} \sqrt{\frac{8E_f}{g^4\mu} - 1}\equiv M_R\pm i\frac{1}{2}\Gamma_{\text{pole}}~, \quad E_f > \frac{g^4\mu}{8}\,, 
\end{align}
with $M_R$ the resonance mass, and $\Gamma_{\text{pole}}$ its pole width. 
This width is observable and distinct from the bare one. It matches \( g^2\sqrt{2\mu M_R} \) from Eq.~\eqref{181206.1} when \( E_f \gg g^4\mu \), indicating the narrow-resonance limit. The residues of \( t_F(E) \) in the complex momentum plane are:
\begin{align}
\label{181206.7}
\gamma_{k_i}^2 &= -\lim_{p \to p_i} (p - p_i) t_F(p^2 / 2\mu)= (-1)^{i+1}\frac{\mu g^2}{p_1 - p_2} =  \frac{(-1)^{i+1}}{\sqrt{\frac{8E_f}{g^4\mu} - 1}}~,
\end{align}
In the narrow-resonance limit, \( E_f \gg g^4\mu/8 \), \( \gamma_{k_i}^2 \to 0 \), while as \( E_f \to g^4\mu/8 \), \( \gamma_{k_i}^2 \) diverges, signaling a double virtual-state pole \cite{Kang:2016jxw}. For \( \Rea E_{1,2} \geq 0 \), the relation \( 1 \geq |\gamma_{k_i}|^2 \geq 0 \) holds. As shown in Refs.~\cite{Guo:2015daa, Kang:2016ezb}, \( |\gamma_{k_{1,2}}|^2 \) can be interpreted probabilistically, representing the compositeness of the resonance as the contribution from two-body continuum states. Notably, \( |\gamma_{k_1}|^2 = |\gamma_{k_2}|^2 \).

The ERE for a partial wave amplitude \( t(p^2) \) up to \( {\cal O}(p^2) \), denoted as \( t_r(E) \), is a special case of Eq.~\eqref{181113.4} when keeping the \( a \) and \( r \) terms:
\begin{align}
\label{181206.10}
t_r(E) &= \frac{1}{-\frac{1}{a} + \frac{1}{2} r p(E)^2 - i p(E)}~.
\end{align}
The quadratic dependence on \( p \) in the denominator leads to two pole positions $p_i =  \left( i + (-1)^i \sqrt{\frac{2r}{a} - 1} \right)/r$, corresponding to resonance poles when $r/a > 1/2$ and $r < 0$\,.  The residues of \( t_r(E) \) at the poles are:
\begin{align}
\label{181206.13}
\gamma_{k_i}^2 &= \frac{1}{r p_i - i} = (-1)^i \frac{1}{\sqrt{\frac{2r}{a} - 1}}~.
\end{align}
The positions of the poles in energy are given by:
\begin{align}
\label{181206.15}
E_i &= \frac{p_i^2}{2\mu} = \frac{1}{ar\mu}\left( 1 - \frac{a}{r} \right) + i(-1)^i \frac{1}{r^2\mu} \sqrt{\frac{2r}{a} - 1}~.
\end{align}
Thus, \( \Rea E_{1,2} \geq 0 \) is satisfied if $\frac{r}{a} \geq 1$~.
 This condition also allows for a probabilistic interpretation of \( |\gamma_{k_1}|^2 = |\gamma_{k_2}|^2  \in [0, 1] \), as discussed in Refs.~\cite{Guo:2015daa,Kang:2016ezb}, see Eq.~\eqref{181206.13}. 

The resonance width from Eq.~\eqref{181206.15} is $\Gamma_{\text{pole}} = \frac{2}{r^2 \mu} \sqrt{\frac{2r}{a} - 1}$\,. 
The values of \( a \) and \( r \) for a PWA containing one CDD pole are given by Eq.~\eqref{181204.5}. As \( \cdd \to 0 \), the ratio \( |r/a| \) increases, and \( |\gamma_{k_i}|^2 \) decreases, Eq.~\eqref{181206.13}. In this probabilistic interpretation, the \( \cdd \to 0 \) limit corresponds to a purely "elementary" state, where the two-body continuum states contribute nothing to the resonance. This provides another approach to Morgan's pole-counting rule for resonance nature \cite{Morgan:1992ge}. In the \( \cdd \to 0 \) limit, Eq.~\eqref{181206.15} shows that the resonance poles approach:
\begin{align}
\label{181206.17}
E_i &\xrightarrow[\cdd \to 0]{} -\frac{\cdd^3}{\lambda^2} + (-1)^{i+1} \frac{(-\cdd)^{7/2} \sqrt{2\mu}}{\lambda^2}~.
\end{align}
Here, the width diminishes faster than the mass by a factor of \( (-\cdd)^{1/2} \), corresponding to the narrow-resonance limit. This establishes a connection between near-threshold weakly coupled resonances and CDD poles, often associated with the explicit exchange of a bare ``elementary" resonance.  Contrarily, in the limit \( r/a \to 1 \), \( |\gamma_{k_i}|^2 \to 1 \),  the resonance mass approaches the threshold, as shown by Eq.~\eqref{181206.15}, while the resonance width stays put.   This shows that in a such a limit the resonance is composed of two-body continuum states, as discussed in Ref.~\cite{Kang:2016ezb}. We have briefly mentioned results for quantifying the compositeness and elementariness of a pole in the \( S \)-matrix. For a comprehensive approach, see Ref.~\cite{Oller:2017alp}. Additional valuable references include \cite{Weinberg:1962hj, Weinberg:1965zz, Baru:2003qq, Hyodo:2011qc, Aceti:2012dd, Kang:2016ezb, Guo:2015daa, Kang:2016jxw, Sekihara:2016xnq, Hernandez:1984zzb, Kamiya:2015aea, Sekihara:2015gvw, Gao:2018jhk}.

It is important to highlight that Eq.~\eqref{181204.4} is more general than an ERE up to \( \mathcal{O}(p^4) \), which fails to capture scattering behavior beyond a near-threshold zero. Therefore, the generality of Eq.~\eqref{181204.4} extends beyond the Flatté parameterization, which is a specific case of the ERE truncated at \( \mathcal{O}(p^2) \) with $r<0$.

{\bf Coupled channels:} A Flatté parameterization is particularly suitable for modeling a resonance that couples predominantly to a channel close to its threshold. For example, this is the case for the \( f_0(980) \) and \( a_0(980) \) resonances \cite{Flatte:1976xu, Baru:2003qq} regarding the $K\bar{K}$, particularly true for the former. 
The Flatté parameterization  models the dressing of a bare resonance propagator, $1/D(E)$, with self-energy contributions from intermediate channels 1 and 2:
\begin{align}
\label{210703.1}
D(E) &= E - E_f + i \frac{\widetilde{\Gamma}_1}{2} + \frac{i}{2} g^2_2 \sqrt{m_K E}.
\end{align}
Here, \(E\) is the center-of-mass energy relative to the two-kaon threshold (\(E \equiv \sqrt{s} - 2m_K\));  the parameter \(g_a\) denotes the  bare coupling of the resonance to channel \(a\), and the bare width \(\widetilde{\Gamma}_1\) for channel 1 is given by:
\begin{align}
\label{211009.2}
\widetilde{\Gamma}_1 &= \frac{p_1(m_R)g^2_1}{8\pi m_R^2}.
\end{align}
The Flatté parameterization introduces three free parameters (\(E_f\), \(\widetilde{\Gamma}_1\), and \(g_2\)), which can be determined using the resonance  pole position, alongside additional input, such as the experimental branching ratio \(r_{\text{exp}}\) to the lighter channel or the total compositeness \(X\) \cite{Wang:2022vga}.

To locate the resonance pole, we solve \(D(E_R) = 0\):
\begin{align}
\label{210703.4}
E_R &= E_f - \frac{1}{8}m_Kg_2^4 - \frac{i}{2}\widetilde{\Gamma}_1 + \sigma\sqrt{\frac{m_Kg_2^4}{4}}\sqrt{\frac{m_Kg_2^4}{16} - E_f + \frac{i}{2}\widetilde{\Gamma}_1},
\end{align}
where \(\sigma = \pm 1\) denotes the two solutions, with \(\sigma = +1 (-1)\) corresponding to the pole in RS II (RS IV) \cite{Wang:2022vga}. We define  \(Z\) as the modulus of the residue of \(1/D(E)\) at the resonance pole:
\begin{align}
\label{210913.2}
Z &= \left|\lim_{E \to E_R} \frac{E - E_R}{D(E)} \right| = \left| \frac{1}{1 + \frac{i g_2^2}{4} \sqrt{\frac{m_K}{E_R}}} \right| = \frac{\sqrt{8u}}{\left( g_2^4 m_K + 8u + 4 \sigma g_2^2 \sqrt{m_K(u - 2 M_R)} \right)^{1/2}}~,
\end{align}
with $u \equiv \sqrt{4 M_R^2 + \Gamma_R^2}$\,. The squared renormalized coupling \(|\gamma_i^2|\) is related to the bare coupling \(g^2_a\) through the residue of \(g_a^2 / D(E)\):  
\begin{align}
\label{210801.12}
|\gamma_1|^2 &= g^2_1 Z~, \\
|\gamma_2|^2 &= 32\pi m_K^2 g^2_2 Z~.\nn
\end{align}
To our knowledge, the distinction between bare and dressed couplings in a Flatté parameterization has not been adequately addressed until Ref.~\cite{Wang:2022vga}, despite its significant implications. For instance, this explains why the values of \(\widetilde{\Gamma}_{\pi\pi}\) for the \(f_0(980)\) in Table~2 of Ref.~\cite{Baru:2003qq} are often larger than 100 MeV. A similar observation holds for most of the \(\widetilde{\Gamma}_{\pi\eta}\) entries in Table~1 for the \(a_0(980)\). Still another consideration applies to poles in the RS II, where the pole width stems by subtracting the partial-decay widths instead of adding them, as explained in Sec.~\ref{sec.241115.1}, cf. Eq.~\eqref{eq:GIInottot}.

\subsection{Contact Interactions}
\label{sec.190728.1}

When the de Broglie wavelength in the CM frame of two colliding particles is much larger than the effective interaction range, the potential can be approximated using Dirac delta functions and their derivatives \cite{Phillips:1997xu,vanKolck:1998bw}. Here, we review the solution to the Lippmann-Schwinger (LS) equation in this context, following Ref.~\cite{Oller:2017alp}. 

The potential coupling channels \( \alpha \) and \( \beta \) in a partial-wave expansion is expressed as:
\begin{align}
\label{180319.1}
v_{\alpha\beta}(k_\alpha,p_\beta) &= k_\alpha^{\ell_\alpha} p_\beta^{\ell_\beta} \sum_{i,j}^N v_{\alpha\beta;ij} k_\alpha^{2i} p_\beta^{2j}~,
\end{align}
where \( \alpha \) and \( \beta \) represent the channels (ranging from 1 to \( n_c \)), and \( N \geq 0 \) is the maximum expansion order. The factor \( k_\alpha^{\ell_\alpha} p_\beta^{\ell_\beta} \) ensures the correct centrifugal factor. Equation~\eqref{180319.1} can also include couplings between different PWAs of the same channel. The coefficients \( v_{\alpha\beta;ij} \) may depend on energy but not on the off-shell three-momenta. Using matrix notation, Eq.~\eqref{180319.1} becomes:
\begin{align}
\label{180319.4}
v_{\alpha\beta}(k_\alpha,p_\beta) &= [k_\alpha]^T \cdot [v] \cdot [p_\beta]~,
\end{align}
where \( [v] \) is an \( Nn_c \times Nn_c \) matrix:
\begin{align}
\label{180319.2}
[v] &= \begin{pmatrix}
[v_{11}] & [v_{12}] & \cdots & [v_{1n}] \\
[v_{21}] & [v_{22}] & \cdots & [v_{2n}] \\
\vdots & \vdots & \ddots & \vdots \\
[v_{n1}] & [v_{n2}] & \cdots & [v_{nn}]
\end{pmatrix}~,
\end{align}
with each \( [v_{\alpha\beta}] \) being an \( N \times N \) submatrix containing the coefficients \( v_{\alpha\beta;ij} \). The vectors \( [k_\alpha] \) are \( Nn_c \)-dimensional column vectors:
\begin{align}
\label{180319.3}
[k_\alpha]^T &= (\underbrace{0,\ldots,0}_{N(\alpha-1) \text{ entries}}, k^{\ell_\alpha}, k^{\ell_\alpha+2}, \ldots, k^{\ell_\alpha+2N}, 0,\ldots,0)~.
\end{align}

Similarly, the solution for the LS equation $T=V+(H_0-E)^{-1}T$, \( t_{\alpha\beta}(k_\alpha,p_\beta;E) \), can be expressed as:
\begin{align}
\label{180319.5}
t_{\alpha\beta}(k_\alpha,p_\beta;E) &= [k_\alpha]^T \cdot [t(E)] \cdot [p_\beta]~,
\end{align}
where \( [t(E)] \) satisfies the algebraic equation:
\begin{align}
\label{180319.6}
[t(E)] &= [v(E)] - [v(E)] \cdot [G(E)] \cdot [t(E)]~,
\end{align}
with \( [G(E)] \) being the block-diagonal matrix of unitarity one-loop functions:
\begin{align}
\label{180319.6b}
[G(E)] &= \sum_\alpha [G_\alpha(E)]~.
\end{align}
The matrix \( [G_\alpha(E)] \) for channel \( \alpha \) is given by:
\begin{align}
\label{180319.7}
[G_\alpha(E)] &= -\frac{2}{\pi} \int_0^\infty dq \frac{q^2}{q^2 - 2m_\alpha E} [q_\alpha] \cdot [q_\alpha]^T~,
\end{align}
where \( m_\alpha \) is the reduced mass for channel \( \alpha \), and \( [q_\alpha] \) is the momentum vector. It can be calculated upon regularization. From Eq.~\eqref{180319.6}, the solution for \( [t(E)] \) is:
\begin{align}
\label{180319.8}
[t(E)] &= [D(E)]^{-1}~, \\
[D(E)] &= [v(E)]^{-1} + [G(E)]~.\nn
\end{align}

\subsection{Lippmann-Schwinger Equation Involving Bare Resonance States}
\label{sec.190210.1}

A common approach for characterizing near-threshold scattering involves solving the LS equation with a potential that could include  the exchange of a bare resonance \cite{Weinberg:1962hj}. The total potential \( V_T(\vp, \vp', E) \) consists of an energy-independent potential \( V(\vp, \vp') \), which describes direct scattering between two-body continuum states, plus the exchange of a bare resonance \cite{Hanhart:2007yq}:
\begin{align}
\label{181207.1}
V_T(\vp,\vp',E) &= V(\vp,\vp') + \frac{f(\vp) f(\vp')}{E - E_0}~.
\end{align}
Here, \( E_0 \) is the bare mass, and \( f(\vp) \) is a real function denoting the bare coupling to two-body states. The scattering amplitude is found by solving the LS equation in momentum space:
\begin{align}
\label{181207.2}
T(\vp,\vp',E) &= V_T(\vp,\vp',E) + \int \frac{d^3 q}{(2\pi)^3} \frac{V_T(\vp, \vq, E) T(\vq, \vp', E)}{q^2/2\mu - E - i\epsilon}~.
\end{align}
This uses the standard normalization to \( (2\pi)^3 \delta(\vp' - \vp) \) for one-particle states in non-relativistic quantum mechanics.

\begin{figure}
 	\begin{center}
\includegraphics[width=.4\textwidth]{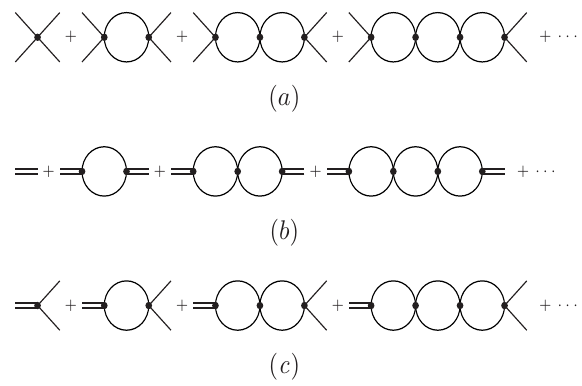}
\caption{{\small A graphical representation of the solution to the Lippmann-Schwinger equation in Eq.~\eqref{181207.2} for \(T(\vp,\vp',E)\). 
Panel $(a)$ shows the solution for direct scattering \(T_V(\vp,\vp',E)\) from Eq.~\eqref{181207.3}.
Panel $(b)$ illustrates the self-energy for the bare propagator, while panel $(c)$ depicts the dressing of the bare coupling due to final-state interactions caused by direct scattering between the two particles.}
 			\label{fig.181207.1}}
 	\end{center}
\end{figure}

The graphical representation of the solution involves diagrams that do not include bare-state propagators, shown in Fig.~\ref{fig.181207.1}$(a)$. These correspond to iterations of the potential \( V(\vp, \vp') \), yielding the direct scattering amplitude \( T_V(\vp, \vp', E) \):
\begin{align}
\label{181207.3}
T_V(\vp, \vp', E) &= V(\vp, \vp') + \int \frac{d^3 q}{(2\pi)^3} \frac{V(\vp, \vq) T_V(\vq, \vp', E)}{q^2/2\mu - E - i\epsilon}~.
\end{align}

Next, we focus on contributions involving at least one bare state exchange, represented by a double line in the second and third rows of Fig.~\ref{fig.181207.1}. During the iteration process for computing the scattering amplitude, intermediate states consist of both bare and two-particle continuum states. After determining the bare-state self-energy, we obtain the standard Dyson resummation for the dressed propagator, shown in Fig.~\ref{fig.181207.1}$(b)$. Additionally, the bare coupling of the exchanged state to the continuum is dressed due to the final-state interactions (FSI) from direct scattering between the particles, as depicted in Fig.~\ref{fig.181207.1}$(c)$. Thus, the diagrams in rows $(b)$ and $(c)$ represent exchanges involving dressed propagators and couplings, which can be written as:
\begin{align}
\label{181207.4}
R(\vp,\vp',E) &= \frac{\Theta(\vp,E) \Theta(\vp',E)}{E - E_0 + \Sigma(E)}~.
\end{align}
Here, \( \Theta(\vp,E) \) is the dressed coupling, and \( \Sigma(E) \) is the self-energy. We now show that the scattering amplitude \( T(\vp,\vp',E) \) can be written as:
\begin{align}
\label{181207.5}
T(\vp,\vp',E) &= T_V(\vp,\vp',E) + \frac{\Theta(\vp,E)\Theta(\vp',E)}{E - E_0 + \Sigma(E)}~.
\end{align}
Substituting this into the LS equation in Eq.~\eqref{181207.2} leads to the equation for \( R(\vp,\vp',E) \) (using Eq.~\eqref{181207.3} for \( T_V(\vp,\vp',E) \)):
\begin{align}
\label{181207.6}
\frac{\Theta(\vp,E)\Theta(\vp',E)}{E - E_0 + \Sigma(E)} &= \frac{f(\vp) f(\vp')}{E - E_0} + \int \frac{d^3 q}{(2\pi)^3} \frac{1}{q^2/(2\mu) - E - i\epsilon} \Bigg[ V(\vp,\vq) \frac{\Theta(\vq,E) \Theta(\vp',E)}{E - E_0 + \Sigma(E)} \nn \\
&+ \frac{f(\vp) f(\vq)}{E - E_0} T_V(\vq,\vp',E) + \frac{f(\vp) f(\vq)}{E - E_0} \frac{\Theta(\vq,E) \Theta(\vp',E)}{E - E_0 + \Sigma(E)} \Bigg]~.
\end{align}
By taking \( E \to E_0 \) and \( E \to E_0 - \Sigma(E) \), we obtain the equations for \( \Sigma(E) \) and \( \Theta(\vp,E) \):
\begin{align}
\label{181207.7}
\Theta(\vp',E) \frac{-1}{\Sigma(E)} \int \frac{d^3 q}{(2\pi)^3} \frac{f(\vq) \Theta(\vq,E)}{q^2/(2\mu) - E - i\epsilon} &= f(\vp') + \int \frac{d^3 q}{(2\pi)^3} \frac{f(\vq) T_V(\vq,\vp',E)}{q^2/(2\mu) - E - i\epsilon}~. \\
\label{181207.7b}
\Theta(\vp,E) &= - \frac{f(\vp)}{\Sigma(E)} \int \frac{d^3 q}{(2\pi)^3} \frac{f(\vq) \Theta(\vq,E)}{q^2/(2\mu) - E - i\epsilon} + \int \frac{d^3 q}{(2\pi)^3} \frac{V(\vp,\vq) \Theta(\vq,E)}{q^2/(2\mu) - E - i\epsilon}~.
\end{align}
These equations are satisfied by identifying:
\begin{align}
\label{181207.8}
\Sigma(E) &= - \int \frac{d^3 q}{(2\pi)^3} \frac{f(\vq) \Theta(\vq,E)}{q^2/(2\mu) - E - i\epsilon}~,
\end{align}
and
\begin{align}
\label{181207.9}
\Theta(\vp',E) &= f(\vp') + \int \frac{d^3 q}{(2\pi)^3} \frac{f(\vq) T_V(\vq,\vp',E)}{q^2/(2\mu) - E - i\epsilon}~.
\end{align}
We discuss next a more formal derivation for representing \( T(\vp,\vp',E) \) as in Eq.~\eqref{181207.5} based on the theory of linear IEs \cite{tricomi.181021.1}. The Hamiltonian \( H \) is split into the free Hamiltonian \( H_0 \) and the potential \( V \), so that \( H = H_0 - V \), with \( |0\rangle \) being an eigenstate of \( H_0 \), i.e., \( H_0 |0\rangle = E_0 |0\rangle \). We define a modified \( T \)-matrix, \( T_1(E) \), obtained by excluding the state \( |0\rangle \) from the intermediate states in the LS equation as:
\begin{align}
\label{181207.11}
T_1(E) &= V + V (H_0 - E)^{-1} \theta T_1(E)~,
\end{align}
where \( \theta = I - |0\rangle \langle 0| \), with \( \theta |0\rangle = 0 \) and \( \langle 0|0\rangle = 1 \). Multiplying both sides of this equation on the right by \( (H_0 - E)^{-1} \theta \), we get:
\begin{align}
\label{181207.12}
T_1(E)(H_0 - E)^{-1} \theta &= V (H_0 - E)^{-1} \theta + V (H_0 - E)^{-1} \theta T_1(E) (H_0 - E)^{-1} \theta~.
\end{align}
This is the IE for the resolvent. Here the kernel  is \( V (H_0 - E)^{-1} \theta \), and the resolvent \( K_1(E) \) is:
\begin{align}
\label{181207.13}
K_1(E) &= T_1(E) (H_0 - E)^{-1} \theta~.
\end{align}
For the LS equation satisfied by the full \( T \)-matrix, \( T(E) \), we have:
\begin{align}
\label{181208.1}
T(E) &= V + V |0\rangle (E_0 - E)^{-1} \langle 0| T(E) + V (H_0 - E)^{-1} \theta T(E)~.
\end{align}
Here, the kernel from Eq.~\eqref{181207.11} appears, but with a different independent term, \( V + V |0\rangle (E_0 - E)^{-1} \langle 0| T(E) \). Since \( T_1(E) = V + K_1(E) V \), we can rewrite the solution for \( T(E) \) in Eq.~\eqref{181208.1} as:
\begin{align}
\label{181208.2}
T(E) &= T_1(E) + T_1(E) |0\rangle (E_0 - E)^{-1} \langle 0| T(E)~.
\end{align}

Multiplying the previous equation by \( \langle 0| \) gives an expression for \( \langle 0| T(E) \) in terms of known matrix elements:
\begin{align}
\label{181208.4}
\langle 0| T(E) &= \left[ 1 - \langle 0| T_1(E) |0\rangle (E_0 - E)^{-1} \right]^{-1} \langle 0| T_1(E)~.
\end{align}
Substituting this into Eq.~\eqref{181208.2} results in:
\begin{align}
\label{181208.5}
T(E) &= T_1(E) + \frac{T_1(E) |0\rangle \langle 0| T_1(E)}{E - E_0 - \langle 0| T_1(E) |0\rangle}~.
\end{align}
\( \langle 0| T_1(E) \) represents the coupling operator. Acting on continuum states,  
\begin{align}
\label{181209.1}
\Theta(\vp_n, E) &= \langle \vp_n | T_1(E) | 0 \rangle~.
\end{align}
Hence, the bare coupling is dressed due to FSI from the direct interaction between the continuum states. Equation~\eqref{181207.5} is a special case of Eq.~\eqref{181208.5}, which can be verified by taking matrix elements between continuum states. The expression in Eq.~\eqref{181208.5} was introduced in Ref.~\cite{Weinberg:1962hj}, although its derivation was not provided.

\section{Final (Initial)-State Interactions}
\def\theequation{\arabic{section}.\arabic{equation}}
\setcounter{equation}{0} 
\label{sec.241103.1}

Generally, relatively feeble interactions can lead to various final states that undergo strong mutual interactions, known as final-state interactions (FSI). By crossing symmetry, strong interactions in the initial state can also produce feeble probes, resulting in initial-state interactions. Some reactions involve both initial- and final-state interactions. Although the following focuses on FSI, the formalism applies equally to initial-state interactions.

\subsection{Unitarity Constraints in FSI}
\label{sec.241120.1}

The total \(T\) and \(S\) matrices account for both relatively-feeble and strong interactions. Retaining only terms linear in the former ones, represented by \(F_i\) (the form factor for producing channel \(i\)), unitarity yields:
\begin{align}
  \label{181118.1}
F_i - F_i^\dagger &= i \sum_j \int dQ_j \, \theta(s - s_{\text{th},j}) T^\dagger_{ij} F_j.
\end{align}
The sum includes only open channels and holds even if some final states \(|i\rangle\) are closed, provided the unitarity cut does not overlap with a crossed-channel cut (cf. Sec.~\ref{sec.1903291.1} when this does not hold).

In partial-wave analyses, the unitarity relation simplifies. Assuming time-reversal symmetry and restricting to two-body final states, the unitarity relation gives:
\begin{align}
\label{181118.3}
\Ima F_i(s) &= \sum_j F_j(s)\theta(s-s_{\text{th};j})\rho_j(s) T_{ij}(s)^* = \sum_j F_j(s)^*\theta(s-s_{\text{th};j})\rho_j(s) T_{ij}(s),
\end{align}
where \(i\) and \(j\) denote different partial-wave states. In the uncoupled case 
one has:
\begin{align}
\label{181119.1}
\Ima F_1(s) &= F_1(s)\rho_1(s)T_{11}(s)^*.
\end{align}
Since the left-hand side is real, the right-hand side must also be real. This implies that above the threshold and below the next higher threshold, the phase of \(F_1(s)\) matches that of \(T_{11}(s)\), modulo \(\pi\). This is Watson's theorem for FSI.

To generalize Watson's theorem to coupled channels, we rewrite the left-hand side of Eq.~\eqref{181118.3} as \((F_i(s) - F_i(s)^*) / 2i\) and grouping terms involving \(F_i(s)^*\). In matrix notation, one has ${F}(s) = \left[I + 2i {T}(s){\rho}(s)\right] {F}(s)^*\,.$ 
Substituting \({T}(s) = ({\cN}^{-1} + {G})^{-1}\), Eq.~\eqref{181109.6b}, and 
using \({G}(s) + 2i{\rho}(s) = {G}(s)^*\), it follows that:
\begin{align}
\label{181119.3}
\left({\cN}^{-1} + {G}\right){F} &= \left({\cN}^{-1} + {G}^*\right){F}^*.
\end{align}
Multiplying both sides by \({\cN}\), we get:
\begin{align}
\label{181119.4}
\left[I + {\cN}(s) {G}(s)\right]{F}(s) &= \left[I + {\cN}(s) {G}(s)^*\right]{F}(s)^*.
\end{align}
This generalization of Watson's theorem for FSI in coupled channels implies that \((I + \cN G)F\) is real along the unitarity cut. Thus it has no RHC, as it equals its complex conjugate for \(s > s_{\text{th},1}\).\footnote{Since \(\cN\) has no RHC, its complex conjugate could also be used in Eq.~\eqref{181119.4}.} Hence, \(F(s)\) can be written as:
\begin{align}
\label{181119.6}
F(s) &= \left[I + \cN(s) G(s)\right]^{-1} {L}(s),
\end{align}
where \({L}(s)\) is a vector containing only left-hand cuts (LHCs), if any.

These results can also be derived via the \(N/D\) method, where \({T} = {D}^{-1}{N}\) is substituted in ${F}(s) = \left[I + 2i {T}(s){\rho}(s)\right] {F}(s)^*\,$, and then analogous steps are followed taking into account that \(\Ima D(s) = -N(s)\rho(s)\). Therefore, \(F(s)\) can be also expressed as:
\begin{align}
\label{181120.2}
F(s) &= D(s)^{-1} {L}(s),
\end{align}
where \({L}(s)\) has at most LHC, if any. This form offers clearer separation between the RHC and LHC compared to Eq.~\eqref{181119.6}, as \(D(s)\) contains only RHC, while \(I + \cN(s) G(s)\) includes both RHC and LHC (with \(\cN(s)\) having LHC and \(G(s)\) RHC).


\subsection[The Omn\`es solution]{The Omn\`es solution}
\label{sec.181117.2}

Here, we focus on the uncoupled case, either because it is exact or serves as a good approximation. Assuming one-channel unitarity as in Eq.~\eqref{181119.1} and knowing the strong PWA, Watson's final-state theorem implies that the phase $\varphi(s)$ of the form factor \(F(s)\) is also determined. The solution for an analytic function in the cut complex \(s\)-plane with a RHC starting at the threshold \(s_{\text{th}}\) can be written using the Omn\`es function. We construct this by performing a DR on \(\log f(s) = F(s)Q(s)/P(s)\), where \(P(s)\) and \(Q(s)\) are polynomials accounting for the zeros and poles of \(F(s)\), so that  \(f(s)\) has no any of them. Then, the discontinuity of \(\omega(s)\equiv \log f(s)\) along the RHC is given by \(2i\varphi(s)\), and we can write the DR:
\begin{align}
\label{181120.8}
\omega(s) &= \sum_{i=0}^{n-1} a_i s^i + \frac{s^n}{\pi} \int_{s_{\text{th}}}^{+\infty} \frac{\varphi(s') \, ds'}{(s')^n(s'-s)}~,
\end{align}
where \(n\) subtractions are applied, assuming \(\varphi(s)\) does not diverge faster than \(s^{n-1-\nu}\) as \(s \to \infty\), with $\nu>0$. The Omn\`es function \(\Omega(s)\) is defined as:
\begin{align}
\label{181120.9}
\Omega(s) &= \exp{\omega(s)}~.
\end{align}
We normalize \(\Omega(s)\) such that \(\Omega(0) = 1\), fixing \(a_0 = 0\). Thus, the ratio:
\begin{align}
\label{181120.10}
R(s) &= \frac{F(s)}{\Omega(s)}~,
\end{align}
becomes a meromorphic function of \(s\) in the first Riemann sheet of the complex \(s\)-plane.

Hadronic form factors generally vanish as $s \to \infty$, reflecting the finite QCD scale $\Lambda_{QCD}$ and quark counting rules \cite{Brodsky:1973kr,Matveev:1973ra,Brodsky:2018snc}. The form factor phase is expected to approach a constant at infinity and, in such a case,  $\omega(s)$ in Eq.~\eqref{181120.8} requires only one subtraction, i.e., $|\varphi(s)/s| < s^{-\nu}$ for some $\nu > 0$ as $s \to \infty$. Thus, $F(s)$ is given by
\begin{align}
\label{181121.1}
F(s) &= \frac{P(s)}{Q(s)} \exp\left( \frac{s}{\pi} \int_{s_{\text{th}}}^\infty \frac{\varphi(s') \, ds'}{s'(s'-s)}\right)~.
\end{align}
 To analyze the asymptotic behavior of $\Omega(s)$ for $\varphi(\infty) < \infty$, we rewrite $\omega(s)$ as
\begin{align}
\label{181122.1}
\omega(s) &= \varphi(\infty) \frac{s}{\pi} \int_{s_{\text{th}}}^{+\infty} \frac{ds'}{s'(s'-s)} + \frac{s}{\pi} \int_{s_{\text{th}}}^{+\infty} \frac{\varphi(s') - \varphi(\infty)}{s'(s'-s)} \, ds'~.
\end{align}
For large $s$, this simplifies to
\begin{align}
\label{181122.2}
\omega(s+i\varepsilon) &\xrightarrow[s\to\infty]{} -\frac{\varphi(\infty)}{\pi} \log \frac{s}{s_{\text{th}}} + i\varphi(\infty) - \frac{1}{\pi} \int_{s_{\text{th}}}^{+\infty} \frac{\varphi(s') - \varphi(\infty)}{s'} \, ds'~.
\end{align}
The dominant logarithmic divergence gives
\begin{align}
\label{181122.3}
\Omega(s) &\xrightarrow[s\to \infty]{} {\cal C}_{\Omega}\, e^{i\varphi(\infty)} \left(\frac{s_{\text{th}}}{s}\right)^{\frac{\varphi(\infty)}{\pi}}~.
\end{align}
Substituting this into Eq.~\eqref{181121.1}, the asymptotic behavior of $F(s)$ is then
\begin{align}
\label{181122.4}
F(s) &\xrightarrow[s\to\infty]{} {\cal C}_{F}\, e^{i\varphi(\infty)} \, s^{p-q-\frac{\varphi(\infty)}{\pi}}~,
\end{align}
where ${\cal C}_{\Omega}$ and ${\cal C}_{F}$ are constants, and $p$, $q$ are the degrees of $P(s)$ and $Q(s)$, respectively. From Eq.~\eqref{181122.4}, we derive the following corollaries.

i) If the asymptotic behavior of $F(s)$ follows $s^\gamma$,  then
\begin{align}
\label{181122.5}
p-q-\frac{\varphi(\infty)}{\pi} = \gamma~,
\end{align}
which is a relativistic extension of the Levinson theorem \cite{levinson.181121.1,Weinberg:1963zza} for form factors.

ii) When modeling interactions, maintaining the constancy of Eq.~\eqref{181122.5} under continuous parameter variations may be crucial. If $\gamma$ is fixed, we have
\begin{align}
\label{181122.6}
p-q-\frac{\varphi(\infty)}{\pi} = {\rm fixed}~.
\end{align}
For example, if $\varphi(\infty)/\pi$ decreases by one and no bound states are present, an additional zero must be added to the form factor to maintain consistency with Eq.~\eqref{181122.6}. This has important implications even at low energies, as discussed in Refs.~\cite{Oller:2007xd} for the scalar pion form factor.

The pion scalar form factor illustrates the points above. Associated with the light-quark scalar source $\bar{u}u + \bar{d}d$, it is defined as
\begin{align}
\label{181122.7}
F(s) &= \hat{m} \int d^4 x \, e^{i(p+p')x} \langle 0 | \bar{u}(x)u(x) + \bar{d}(x)d(x) | 0 \rangle~,
\end{align}
where $u$ and $d$ are the up and down quarks, $\hat{m}$ is their average current mass, and $s = (p+p')^2$. Due to the quantum numbers of the scalar source, FSI stem from  the isoscalar scalar meson-meson scattering. At low energies, the $\pi\pi$ channel dominates, cf. Sec.~\ref{sec.241120.2}. 
At higher energies, the $K\bar{K}$ channel becomes relevant, with a threshold at 991.4~MeV \cite{ParticleDataGroup:2024cfk}, aligning with the $f_0(980)$ resonance. This resonance is relatively narrow \cite{ParticleDataGroup:2024cfk}, causing a rapid increase in the $\pi\pi$ isoscalar scalar phase shifts near the two-kaon threshold. Consequently, the elasticity parameter $\eta_{00}$ sharply decreases when the $K\bar{K}$ channel opens, as the $f_0(980)$ couples more strongly to $K\bar{K}$ than to $\pi\pi$ \cite{Guo:2012yt}, resulting in a significant conversion of pionic flux into kaonic flux.

The rise in the $\pi\pi$ phase shifts $\delta_{00}$ also increases the phase of the isoscalar scalar PWA, $\vh(s)$, which coincide below the $K\bar{K}$ threshold ($\sqrt{s}\leq \sqrt{s_K}=2m_K$). Above it, $\vh(s)$ decreases abruptly if $\delta_{00}(s_K)<\pi$; otherwise,  it continuous increasing if $\delta_{00}(s_K)>\pi$. These behaviors can flip with slight changes in the hadronic model, while keeping consistency with experimental phase shifts. This results in two different behaviors: If $\delta(s_K) > \pi$, $\Omega(s)$ grows large at $\delta(s) = \pi$; if $\delta(s_K) < \pi$, $\Omega(s)$ is zero  just below the $K\bar{K}$ threshold. The limit $\delta_{00}(s_K)\to \pi$ gives rise to a singularity in $\Omega(s)$. In this case, $\varphi(s)>\pi$ for $s>s_K$, and its discontinuity at $s_K$ when $\delta_{00}(s_K)\to\pi^+$ by an amount of $\pi/2$ results in a singularity in $\omega(s)$ at $s=s_K$, of the end-point type, due to the pole in the denominator of the integrand. This can be expressed as: 
\begin{align}
\label{181123.3}
&\frac{1}{\pi}\left[\int^{s_K-\Delta}\frac{\varphi(s_K-\ve)ds'}{s'-s_K}
+\int_{s_K+\Delta}\frac{\varphi(s_K+\ve)ds'}{s'-s_K}\right]\to \frac{1}{\pi}\left[\varphi(s_K-\ve)-\varphi(s_K+\ve)\right]\log\Delta
=\pm \frac{1}{2} \log\Delta~,
\end{align}
with $\Delta\to 0^+$ and $\delta(s_K-\ve)\to \pi^{\mp}$, respectively.
\noindent
Upon exponentiating $\omega(s)$ to obtain $\Omega(s)$, a factor of $(\sqrt{\Delta})^{\pm 1}$ appears, which leads to $\Omega(s_K)$ becoming infinite when $\delta(s_K)\to \pi^+$ and approaching zero when $\delta(s_K)\to \pi^-$. Notice that the transition $\delta(s_K-\ve)\to \pi^\pm$ implies a jump in Eq.~\eqref{181122.6}. To maintain this relation, a zero must be added when transitioning from $\delta(s_K)<\pi$ to $\delta_{00}(s_K)>\pi$ ($p$ increases by one). The zero of the pion scalar form factor for $\delta(s_K) > \pi$ occurs at $s_1 < s_K$ where $\delta(s_1) = \pi$.

Next, let us allow for LHC in  \( F(s) \), as it is the case of \( \gamma\gamma \to \pi\pi \). Following Eq.~\eqref{181120.2}, we write $F(s)=\Omega(s)L(s)$, with $\Ima L(s) = \Omega(s)^{-1} \Ima F(s)\,$,~$s < s_L$. Thus,  \( L(s) \) satisfies the DR
\begin{align}
\label{181125.6}
L(s) &= \sum_{i=1}^{n-1} a_i s^i + \frac{s^n}{\pi} \int_{-\infty}^{s_L} \frac{\Ima L(s') ds'}{(s')^n (s' - s)}~,
\end{align}
Once the Omn\`es function is implemented using the strong PWA along the RHC, the required input for this DR is \( \Ima F(s) \) along the LHC \cite{Morgan:1987gv, Morgan:1990kw, Pennington:2006dg, Danilkin:2018qfn}).  A variant of this method was used in Refs.~\cite{Pennington:2006dg, Oller:2007sh, Oller:2008kf} to study the $S$-wave of \( \gamma\gamma \to \pi^0\pi^0 \). A function \( {\cal F}_I(s) \), containing only the RHC, is formed by subtracting a function \( \widetilde{L}_I(s) \), which accounts for the LHC of \( F_I(s) \):
\begin{align}
\label{181125.7}
{\cal F}_I(s) &= \frac{F_I(s) - \widetilde{L}_I(s)}{\Omega_I(s)}~,
\end{align}
with $I=0,\, 2$. Next, a twice-subtracted DR is applied to this function:
\begin{align}
\label{181125.8}
F_I(s) &= \widetilde{L}_I(s) + a_I \Omega_I(s) + c_I s \Omega_I(s) + \Omega_I(s) \frac{s^2}{\pi} \int_{4m_\pi^2}^\infty \frac{\widetilde{L}_I(s') \sin \varphi_I(s') ds'}{(s')^2 (s' - s) |\Omega_I(s')|}~.
\end{align}
One subtraction constant in Eq.~\eqref{181125.8} is fixed using Low's theorem \cite{Low:1954kd}, $a_I=0$, ensuring that \( F_I(s) \) tends to its renormalized Born term contribution as \( s \to 0 \). The other constant is determined by matching with next-to-leading order (NLO) ChPT calculations \cite{Bijnens:1987dc, Donoghue:1988eea}.  A major improvement in Refs.~\cite{Oller:2007sh, Oller:2008kf} was the use of a stable \( \Omega_0(s) \) function by fulfilling Eq.~\eqref{181122.6}. This reduced the uncertainty in the \( \gamma\gamma \to \pi^0\pi^0 \) cross section by about a factor of 2 at \( \sqrt{s} \simeq M_\rho \) and 25\% at \( \sqrt{s} = 500 \) MeV.

\subsection[The Muskhelishvily-Omn\`es problem]{The Muskhelishvily-Omn\`es Problem in Coupled-Channel Form Factors}
\label{sec.181117.3}

Let \( \{ F_i(s), i=1,\ldots, n \} \) be a set of form factors  ordered by increasing thresholds \( s_{{\rm th};i} \). These form factors exhibit RHC behavior for \( s > s_{\text{th};1} \) and may have LHC  for \( s < s_L \). For \( s > s_{\text{th};1} \), the imaginary part of \( F_i(s) \) is determined by unitarity, as shown in Eq.~\eqref{181118.3}, and the $F_i(s)$ are assumed to satisfy the Schwarz reflection principle, $F_i(s^*)=F_i(s)^*$. 
 To further analyze the problem, we consider the matrix \( \mathcal{S}(s) \):
\begin{align}
\label{181125.13}
\mathcal{S}(s) &= I + 2i T(s) \rho(s)~.
\end{align}
Although \( T(s) \) is symmetric, \( \mathcal{S}(s) \) generally is not and, as as result, it is no unitary either for $n>1$. However, from the unitarity relation for \( T(s) \), $\Ima T=\rho T^* T\,,$ $s>s_{\text{th};1}$, we obtain $\mathcal{S}(s) \mathcal{S}(s)^* =\mathcal{S}(s)^* \mathcal{S}(s)= I\,.$ Using the matrix-$N/D$ method, \( T(s) = D(s)^{-1} N(s) \), we find:
\begin{align}
\label{181125.15}
\mathcal{S}(s) &= D(s)^{-1} D(s)^*~.
\end{align}
Since \( D(s)^* = D(s^*) \),  the complex-conjugation relation arises:
\begin{align}
\label{181126.1}
D(s)^{-1} &= \mathcal{S}(s) D(s^*)^{-1}~.
\end{align}
Multiplying both sides by \( L(s) \), it follows:
\begin{align}
\label{181126.1b}
F(s) &= \mathcal{S}(s) F(s^*)~.
\end{align}
Now, suppose we are given a matrix of PWAs, \( T(s) \). The task is to find an \( n \times n \) matrix \( D(s) \) with only  RHC such that Eq.~\eqref{181125.15} is fulfilled. This is known as the Hilbert problem. Equation~\eqref{181126.1} shows that each column of \( D(s)^{-1} \) corresponds to coupled form factors, satisfying Eq.~\eqref{181126.1b} along the RHC. The form factors \( F_i(s) \) are then obtained by a linear combination of the column vectors of \( D(s)^{-1} \), with \( L_i(s) \) as coefficients.

To start with, let us notice, as it is easy to prove, that the determinants of \( S(s) \) and \( \cS(s) \) are equal. 
 Once \( S \) is diagonalized, its determinant is 
 $\text{det} \, S = \exp \left( 2i \sum_{i=1}^{n} \varphi_i(s) \right)\,,$ where the $\varphi_i(s)$ are the eigen-phase shifts. Next, using Eq.~\eqref{181125.15}, we derive an Omnès representation for \( \text{det} D^{-1}(s) \), which, like \( D^{-1}(s) \) itself, only has RHC. The phase of \( \text{det} D^{-1}(s) \) along this cut is half the phase of \( \text{det} S \), with the latter denoted  as \( \Phi(s) \), as it follows from Eq.~\eqref{181125.15}. For the Omnès representation we introduce polynomials \( P(s) \) and \( Q(s) \), whose roots correspond to the zeros and poles of $\Delta(s)\equiv \text{det} D(s)$. 
Thus, we represent  \(  \Delta(s) \) with an Omn\'es construction as
\begin{align}
\label{181127.4}
\Delta(s) &= \frac{P(s)}{Q(s)} \exp\left( \frac{s - s_{\text{th};1}}{2\pi} \int_{s_{\text{th};1}}^\infty \frac{\Phi(s')}{(s' - s_{\text{th};1})(s' - s)} \, ds'\right)~.
\end{align}
The integral remains finite as \( s \to s_{\text{th};1} \) because \( \Phi(s_{\text{th};1}) = 0 \). Using Eq.~\eqref{181122.4}, we find the limiting behavior of \( \Delta(s) \) as \( s \to \infty \):
\begin{align}
\label{181127.5}
\Delta(s) &\xrightarrow[s \to \infty]{} s^{p - q - \frac{\Phi(\infty)}{2\pi}}~.
\end{align}
An interesting aspect of Eq.~\eqref{181127.5} is its connection to the asymptotic behavior of \( \Delta(s) \) and the leading power behavior in \( s \) of the columns of \( D(s)^{-1} \) \cite{musk.181126.1,warnock.181127.1,Babelon:1976kv}. Let \( \phi_i(s) \) denote the \( i \)-th column of \( D(s)^{-1} \), 
Assuming \( \cS(s) \to I \) as \( s \to \infty \), as in Refs.~\cite{warnock.181127.1,Babelon:1976kv}, we conclude that the leading behavior of \( \phi_i(s) \) is a power law in \( s \) for large \( s \) with leading exponent $\chi_i$. For $s\to \infty$ one has that  \( \Delta(s) \) behaves as \cite{Babelon:1976kv}:
\begin{align}
\label{181127.7}
\Delta(s) \xrightarrow[s \to \infty]{} s^{\chi_1 + \chi_2 + \ldots + \chi_n}~,
\end{align}
 with the different column vectors in \( D(s)^{-1} \) kept linearly independent. Comparing Eq.~\eqref{181127.7} with Eq.~\eqref{181127.5}, we derive:
\begin{align}
\label{181127.7b}
\chi_1 + \chi_2 + \ldots + \chi_n = p - q - \frac{\Phi(\infty)}{2\pi}~.
\end{align}
This formalism is used in Ref.~\cite{Jamin:2001zq} to describe the strangeness-changing scalar form factors for \( K\pi(1) \), \( K\eta(2) \), and \( K\eta'(3) \). It was also applied in the study of \( \pi\pi \) and \( K\bar{K} \) isoscalar scalar form factors \cite{Donoghue:1990xh}, and more recently in Ref.~\cite{Moussallam:1999aq}. The strangeness-changing scalar form factors are defined by:
\begin{align}
\label{181127.8}
\langle 0| \partial^\mu(\bar{s}\gamma_\mu u)(0)|K\phi_k\rangle &= -i\sqrt{\frac{3}{2}}\Delta_{K\pi}F_k(s)~, \\
\Delta_{K\pi} &= m_K^2 - m_\pi^2~,\nn
\end{align}
and have $I=1/2$, with \( |0\rangle \) the vacuum state. The scalar PWAs in coupled channels involving the $I=1/2$ states \( K\pi \), \( K\eta \), and \( K\eta' \) were studied in Ref.~\cite{Jamin:2000wn}, and the FSI were calculated in Ref.~\cite{Jamin:2001zq}. The results were shown to barely change with the inclusion or exclusion of the \( K\eta \) channel, so that we adopt a two-channel scenario here.
 Reference~\cite{Jamin:2001zq} assumes that the \( I = 1/2 \) scalar form factors vanish as \( s \to \infty \), consistent with QCD counting rules \cite{Brodsky:1973kr,Matveev:1973ra,Brodsky:2018snc}. Thus, unsubtracted DRs are used for \( F_1(s) \) and \( F_3(s) \), with the subscripts 1 and 3 referring to $K\pi$ and $K\eta'$, respectively:
\begin{align}
\label{181127.9}
F_1(s) &= \frac{1}{\pi} \int_{s_{{\rm th};1}}^\infty \frac{\rho_1(s') F_1(s') T_{11}(s')^*}{s' - s} + \frac{1}{\pi} \int_{s_{{\rm th};3}}^\infty \frac{\rho_3(s') F_3(s') T_{13}(s')^*}{s' - s}~, \\
F_3(s) &= \frac{1}{\pi} \int_{s_{{\rm th};1}}^\infty \frac{\rho_1(s') F_1(s') T_{13}(s')^*}{s' - s} + \frac{1}{\pi} \int_{s_{{\rm th};3}}^\infty \frac{\rho_3(s') F_3(s') T_{33}(s')^*}{s' - s}~.\nn
\end{align}
These coupled linear IEs are numerically solved in Ref.~\cite{Jamin:2001zq} by iteration. The PWAs used \cite{Jamin:2000wn} satisfy \(q = 0\) (no bound states) and \(p = 0\). In this case, Eq.~\eqref{181127.5} simplifies to
\[
\Delta(s) \xrightarrow[s\to\infty]{} s^{-\frac{\Phi(\infty)}{2\pi}}~.
\]
The first set of \(T\)-matrices in Ref.~\cite{Jamin:2000wn} gives \(\Phi(\infty) = 2\pi\) (\(\delta_1(\infty) = \pi\) and \(\delta_3(\infty) = 0\)), so that $\chi_1 + \chi_2 = -1$\,. 
Since \(\chi_1\) and \(\chi_2\) cannot both be negative integers, only one linearly independent solution can vanish at \(s \to \infty\), with \(\chi_1 = -1\). A second set of PWAs enable studying the transition from \(\Phi(\infty) = 2\pi\) to \(\Phi(\infty) = 4\pi\) by varying some \(K\)-matrix parameters, while matching experimental data up to \(\sqrt{s} = 2.5\)~GeV \cite{Aston:1987ir}. For \(\Phi(\infty) = 4\pi\), Eq.~\eqref{181127.7b} gives $\chi_1 + \chi_2 = -2\,.$  
In this case, \(\chi_1 = \chi_2 = -1\) allows two linearly independent solutions that vanish at infinity. To determine the linear combination, two constants are needed: \(F_1(0)\) from NLO ChPT and the value of the \(K\pi\) form factor at \(s = \Delta_{K\pi}\), the Callan-Treiman point \cite{Gasser:1984ux}.

Finally, returning to Eq.~\eqref{181120.2}, we express the form factors using the matrix of functions $D(s)^{-1}$ from the $N/D$ method. When \(N(s)\) is modeled without LHCs, as in Sec.~\ref{sec.181104.1}, explicit forms for \(\Omega(s)\) and \(D(s)\) are derived. In the uncoupled case, \(\Omega(s)\) is
\[
\Omega(s) = \frac{\prod_{i=1}^q (s - s_{P;i})}{\prod_{j=1}^p (s - s_{Z;j})} \frac{1}{1 + N(s)g(s)}~,
\]
where the zeros \(s_{Z;i}\) and poles \(s_{P;i}\) of \(1/[1 + N(s)g(s)]\) are removed. In the coupled-channel case, \(D(s)\) is
\[
D(s) = I + N(s)g(s)~.
\]
One can introduce an analogous function \(\cD^{-1}(s)\), which satisfies Eq.~\eqref{181126.1} and is holomorphic in the cut complex \(s\)-plane. From Eq.~\eqref{181126.1} we have that $
\cD(s+i\ve)D(s+i\ve)^{-1} - \cD(s-i\ve)D(s-i\ve)^{-1} = 0$\,. Therefore, the product \(\cD(s)D(s)^{-1}\) is a matrix of rational functions, \(R(s)\), and we can write that 
\[
D(s)^{-1} = \cD(s)^{-1}R(s)~.
\]

\subsection{The Khuri-Treiman approach}
\label{sec.1903291.1}

This section presents the Khuri-Treiman (KT) framework to study FSI. Originally developed for $K\to 3\pi$ decays \cite{Khuri:1960zz}, the KT formalism has been widely applied to $\eta\to 3\pi$ decays. Relying on elastic $\pi\pi$ unitarity and crossing symmetry, it has been refined by many studies \cite{Gribov:1962fu, Bronzan:1963mby, Kacser:1963zz, Aitchison:1965zz, Pasquier:1969dt, Neveu:1970tn,Kambor:1995yc, Anisovich:1996tx, Guo:2016wsi, Albaladejo:2017hhj}, and extended to coupled channels in Refs.~\cite{Albaladejo:2017hhj,Moussallam:2016evb}. Here, we focus on the uncoupled case and refer to \cite{Oller:2019opk} for our study of the coupled-channel case. 

In QCD, $\eta \to 3\pi$ decays violate isospin symmetry due to the $G$-parity difference between $\eta$ and $\pi$. The amplitude is proportional to $m_u - m_d$,
 and electromagnetic corrections are small but non-negligible \cite{Urech:1994hd, Ditsche:2008cq}, contributing at the percent level. Early ChPT underestimated the decay width, while NLO calculations raised it to $(160 \pm 50)$ eV \cite{Gasser:1984gg}, still below the observed $(300 \pm 12)$ eV \cite{ParticleDataGroup:2024cfk}. Similar discrepancies appear in the Dalitz plot parameter $\alpha$ \cite{Gasser:1984gg}, prompting refinements using unitarized ChPT and KT formalism \cite{Roiesnel:1980gd, Beisert:2003zs, Borasoy:2005du}.

For $\eta(p_0)\to \pi(p_1)\pi(p_2)\pi(p_3)$ the Mandelstam variables $s$, $t$, and $u$ are defined as $  s=(p_0-p_3)^2=(p_1+p_2)^2\,,$ $t=(p_0-p_1)^2=(p_2+p_3)^2\,$, $u=(p_0-p_2)^2=(p_1+p_3)^2\,.$ For $\eta \to \pi^+ \pi^- \pi^0$, the amplitude $A(s,t,u)$ is used, while $B(s,t,u)$ applies to $\eta \to 3\pi^0$, being symmetric under variable permutations because of Bose-Einstein symmetry. Scattering amplitudes for different channels can be obtained by applying crossing symmetry \cite{Oller:2019opk}. 
The relations between $s$, $t$, and $u$ in the $s$-channel depend on the scattering angle $\theta$:
\begin{align}
\label{190330.2}
t,\,u(s,\cos\theta) &= \frac{1}{2}\left(m_\eta^2 + 3m_\pi^2 - s \pm \cos\theta \sqrt{\lambda(s,m_\pi^2,m_\eta^2)\sigma(s)}\right)~.
\end{align}
The quantity $\kappa(s) = \sqrt{\sigma(s)\lambda(s,m_\pi^2,m_\eta^2)}$ is also defined, with $t - u=\kappa(s) \cos\theta$. Details for the different physical regions in $s$, $t$- and $u$-channels are provided in Ref.~\cite{Oller:2019opk}.

We now consider the isospin decomposition of decay amplitudes. The isospin-breaking QCD operator $- \frac{1}{2}(m_u - m_d)(\bar{u}u - \bar{d}d)$ is an isovector with $t_3=0$. The ${\cal O}(e^2)$ operator from photon exchange is also isovector with $t_3=0$ as it stems from the squared charge matrix ${\rm diag}(4/9, 2/9) = 5I/18 + {\rm diag}(1, -1)/6$.  In this way, the isospin amplitudes $M^I(s,t,u)$ arise by combining the initial state with the isovector isospin-breaking operator, defining $I$ attached to the final state:
\begin{align}
  \label{190403.1}
A(\eta\pi^0\to \pi^+\pi^-) = A(s,t,u) &= -\frac{1}{3} M^2(s,t,u) + \frac{1}{3} M^0(s,t,u)~,\\
A(\eta\pi^+\to \pi^+\pi^0) = A(u,t,s) &= +\frac{1}{2} M^2(s,t,u) + \frac{1}{2} M^1(s,t,u)~,\nn\\
A(\eta\pi^-\to \pi^0\pi^-) = A(t,s,u) &= +\frac{1}{2} M^2(s,t,u) - \frac{1}{2} M^1(s,t,u)~,\nn\\
A(\eta\to\pi^0\pi^0\pi^0) = B(s,t,u) &= \frac{2}{3} M^2(s,t,u) + \frac{1}{3} M^0(s,t,u)= A(s,t,u) + A(t,s,u) + A(u,t,s)~.
\end{align}
The isospin amplitudes are then expressed as:
\begin{align}
  \label{190403.4}
  M^0(s,t,u) &= 3 A(s,t,u) + A(u,t,s) + A(t,s,u)~,\\
  M^1(s,t,u) &= A(u,t,s) - A(t,s,u)~,\nn\\
  M^2(s,t,u) &= A(u,t,s) + A(t,s,u)~,\nn
 \end{align}
Next, we consider the PWAs of the  \( M^I(s,t,u) \). Since all particles involved are spin-0, the angular projection formula is
\begin{align}
\label{190403.8}
M^{(IJ)}(s) &= \frac{1}{2} \int_{-1}^{+1} d\cos\theta \, P_J(\cos\theta) M^I(s,t,u)~,\\
M^I(s,t,u)
&= \sum_{J=0}^\infty (2J+1) P_J(\cos\theta) M^{(IJ)}(s)~.\nn 
\end{align}
For \( m_\eta < 3m_\pi \), the PWAs \( M^{(IJ)}(s) \) satisfy the standard unitarity relation for PWAs of form factors, shown in Eq.~\eqref{181118.3}, considering the process \( \eta\pi \to \pi\pi \) to first-order isospin violation. Hence, for \( m_\eta < 3m_\pi \) and \( s \geq 4m_\pi^2 \),
\begin{align}
\label{190403.10}
\Delta M^{(IJ)}(s)&\equiv M^{(IJ)}(s+i\vep)-M^{(IJ)}(s-i\vep)
=i\frac{\sigma(s)^{1/2}}{8\pi} M^{(IJ)}(s+i\vep) T^{(IJ)}(s+i\vep)^*\\
&=2i e^{-i\delta^{(IJ)}(s)}\sin\delta^{(IJ)}(s) M^{(IJ)}(s+i\vep)~.\nn
\end{align}
Here, \( T^{(IJ)}(s) \) represents the \( \pi\pi \) PWA for isospin \( I \) and angular momentum \( J \), and \( \delta^{(IJ)}(s) \) is the corresponding phase shift. For \( m_\eta > 3m_\pi \), a small positive imaginary part is added to \( m_\eta^2 \) in the external legs, \( m_\eta^2 \to m_\eta^2 + i\epsilon \) \cite{Mandelstam:1960zz}. This ensures proper analytic continuation of \( t \), \( u \), and \( s \) relations in Eq.~\eqref{190330.2}, and the PWAs in Eq.~\eqref{190403.8}. At \( \epsilon = 0 \), when \( s = (m_\eta^2 - m_\pi^2)/2 \) and \( \cos\theta = \mp 1 \), both \( t \) and \( u \) approach \( 4m_\pi^2 \), creating a threshold branch-point singularity in crossed channels.  For \( s \in [4m_\pi^2, (m_\eta - m_\pi)^2] \), the pions in all three channels are physical. Using the complex \( \eta \)-mass prescription (\( m_\eta^2 + i\epsilon \)) avoids overlapping branch points at \( t, u = 4m_\pi^2 \) with the \( s \)-channel unitarity cut. Equation~\eqref{190403.10} describes the discontinuity of \( M^{IJ}(s) \) across the real \( s \)-axis for \( s \geq 4m_\pi^2 \), while its imaginary part arises from physical particles across the RHC and from contributions due to the  crossed cuts \cite{Anisovich:1996tx}.

A critical step in the KT method represents \( A(s,t,u) \) as a sum of functions of only one argument \( M_0(s) \), \( M_1(s) \), and \( M_2(s) \) \cite{Albaladejo:2017hhj,Colangelo:2018jxw}:
\begin{align}
\label{190407.1}
A(s,t,u)&=M_0(s)-\frac{2}{3}M_2(s)+(s-u)M_1(t)+(s-t)M_1(u)+M_2(t)+M_2(u)~.
\end{align}
This structure can be deduced just from crossing symmetry and isospin decomposition, summarized in Eq.~\eqref{190403.1},  by keeping at most up to $J=1$ in the partial-wave expansions for the $s-$, $t-$ and $u$-channels, Eq.~\eqref{190403.8}.  Equation~\eqref{190407.1} is symmetric  under \( t \leftrightarrow u \)  due to charge conjugation symmetry, consistent with the exchange \( \pi^+ \leftrightarrow \pi^- \). The amplitude for \( \eta \to \pi^0 \pi^0 \pi^0 \)  is then, Eq.~\eqref{190403.1}, 
\begin{align}
\label{190408.1}
B(s,t,u)&=\frac{1}{3}\left[M_n(s)+M_n(t)+M_n(u)\right]~,\\
M_n(s)&=M_0(s)+4M_2(s)~.\nn
\end{align}
The isospin amplitudes \( M^{I}(s,t,u) \) in Eq.~\eqref{190403.4} can also be written straightforwardly in terms of the $M_i$ functions, cf. Eq.~\eqref{190403.4}. 

Next, we perform an $S$-wave projection for the $I=0$ and $I=2$ amplitudes, and a $P$-wave projection for the isovector amplitude, using Eq.~\eqref{190403.8}.
As commonly done \cite{Anisovich:1996tx}, the angular projections of the functions involving the variables $t$ and $u$ are denoted by the symbol \( \langle z^n M_I \rangle(s) \), which is defined as:
\begin{align}
\label{190408.3}
\langle z^n M_I \rangle(s) &= \frac{1}{2} \int_{-1}^{+1} dz\, z^n\, M^I(s, t(s,z), u(s,z))~.
\end{align}
Here, \( t(s,z) \) and \( u(s,z) \) are computed using Eq.~\eqref{190330.2} with \( m_\eta^2 + i\epsilon \). From these expressions, the following relations hold:
\begin{align}
\label{190408.4}
M^{(00)}(s) &\equiv 3\big[M_0(s) + \hat{M}_0(s)\big]~, \\
M^{(11)}(s) &\equiv -\frac{2}{3} \kappa \big[M_1(s) + \hat{M}_1(s)\big]~,\nn \\
M^{(20)}(s) &\equiv 2\big[M_2(s) + \hat{M}_2(s)\big]~.\nn
\end{align}
The functions \( \hat{M}_0(s) \), \( \hat{M}_1(s) \) and \( \hat{M}_2(s) \) stem from the angular projection of $t$ and $u$ functions and do not have RHC. For explicit expressions see  \cite{Oller:2019opk}.  
Using Eqs.~\eqref{190408.4} and \eqref{190403.10}, the discontinuities of \( M_I(s) \) along the RHC can be easily derived \cite{Oller:2019opk}. 
These discontinuities allow us to derive DRs for the \( M_I(s) \) functions under specific high-energy assumptions. Ref.~\cite{Albaladejo:2017hhj} posits that \( M_0(s) \) and \( M_2(s) \) grow at most linearly with \( s \) as \( s \to \infty \), while \( M_1(s) \) remains bounded. Notably, \( M_1(s) \) is multiplied by \( \kappa \), as in Eq.~\eqref{190408.4}, ensuring that \( A(s,t,u) \) grows linearly in all directions as \( s, t, u \to \infty \), as required by Ref.~\cite{Anisovich:1996tx}.

Within these constraints, \( M_I(s) \) can be redefined without altering \( A(s,t,u) \):
\begin{align}
\label{190409.1}
M_1(s) &\to M_1(s) + a_1~, \\
M_2(s) &\to M_2(s) + a_2 + b_2 s~,\nn
\end{align}
requiring the following shift in \( M_0(s) \):
\begin{align}
\label{190409.2}
M_0(s) &\to M_0(s) + a_0 + b_0 s~,
\end{align}
where $a_0 = -\frac{4}{3} a_2 + 3s_0(a_1 - b_1)\,$, $b_0 = -3a_1 + \frac{5}{3} b_2\,$, with $3s_0 = m_\eta^2 + 3m_\pi^2$. To eliminate such ambiguity, we impose:
\begin{align}
\label{190409.4}
M_1(0) &= 0~, \\
M_2(0) &= 0~,\nn \\
M'_2(0) &= 0~.\nn
\end{align}
Combining the RHC discontinuities of the $M_I(s)$ with the above conditions and asymptotic behavior leads to DRs:
\begin{align}
\label{190409.5}
M_0(s) &= \widetilde{\alpha}_0 + \widetilde{\beta}_0 s + \frac{s^2}{\pi} \int_{4m_\pi^2}^\infty ds' \frac{\Delta M_0(s')}{(s')^2(s' - s)}~, \\
M_1(s) &= \frac{s}{\pi} \int_{4m_\pi^2}^\infty ds' \frac{\Delta M_1(s')}{(s')(s' - s)}~,\nn \\
M_2(s) &= \frac{s^2}{\pi} \int_{4m_\pi^2}^\infty ds' \frac{\Delta M_2(s')}{(s')^2(s' - s)}~.\nn
\end{align}
Since \( \Delta M_{I}(s)/s^{2-J} \) decays as \( s^{-\nu} \) (\( \nu>0 \)) for large \( s \), logarithmic divergences are avoided, ensuring a power-law behavior for \( M_I(s) \) in agreement with expectations and previous studies \cite{Anisovich:1996tx, Albaladejo:2018gif, Colangelo:2018jxw}.

Equations  \eqref{190409.5} form a coupled linear IE system for \( M_0(s) \), \( M_1(s) \), and \( M_2(s) \), valid for \( s \geq 4m_\pi^2 \), taking into account that the $\Delta M_I$ are also expressed in terms of the very same functions. Solving this system addresses the KT problem for given subtraction constants \( \widetilde{\alpha}_0 \), \( \widetilde{\beta}_0 \), and input \( \pi\pi \) phase shifts \( \delta^{(IJ)}(s) \). Reference~\cite{Albaladejo:2017hhj} established an iterative numerical procedure for its solutions to which we refer, also discussed in Ref.~\cite{Oller:2019opk}. 


\subsection[Non-relativistic case set-up: elastic, inelastic and bare states]{Framework for Parameterizing Near-threshold States Considering Open Inelastic Channels and Bare States}
\label{sec.190224.1}

\begin{figure}
 	\begin{center}
\includegraphics[width=.4\textwidth]{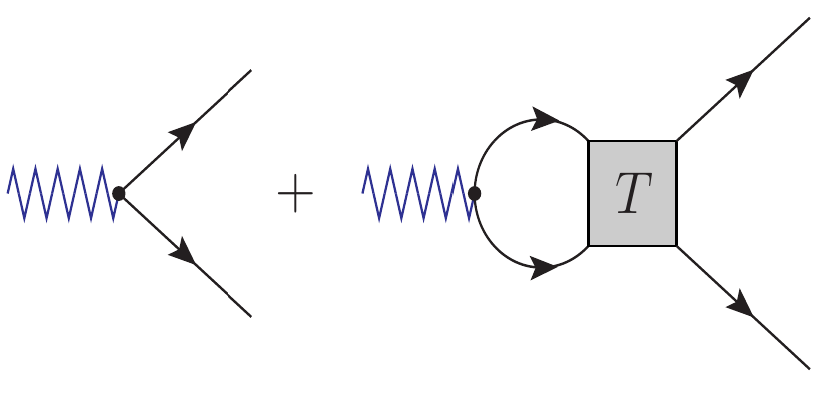}
\caption{{\small Production process including FSI (gray box). Left vertices represent Born terms ${\cal F}$, and blue wiggly lines indicate the probe.}
\label{fig.190225.1}}
        \end{center}
        \end{figure}

We elaborate on the formalism from Ref.~\cite{Hanhart:2015cua}, focused on near-threshold state, particularly for charmonium and bottomonium. Our exposition follows the more compact line of presentation in  Ref.~\cite{Oller:2019opk}. 
 Coupled channels are divided into elastic ones  ($L$), whose thresholds are those for interest, and lighter inelastic channels  ($I$), alongside the bare state $|0\rangle$. The Hamiltonian is split as $H = H_0 - V$, with $V$ coupling $L$, $I$, and $|0\rangle$ ($H_0 |0\rangle = E_0 |0\rangle$). The potential $V$ is represented as:
\begin{align}
\label{190225.1}
V &= \left(
\begin{matrix}
V^{00} & V^{0L} & V^{0I} \\
V^{L0} & V^{LL} & V^{LI} \\
V^{I0} & V^{IL} & V^{II}
\end{matrix}
\right)~,
\end{align}
where superscripts denote coupled subsets. As in Ref.~\cite{Hanhart:2015cua} we take a symmetric potential. This is the case for partial-wave projected potentials due to time-reversal symmetry. 
The projector $\theta_1$ excludes intermediate states $|0\rangle$ and $I$, and $T_1$ is the $T$ matrix solving the LS equation without them, $T_1 = V + V (H_0 - E)^{-1} \theta_1 T_1$, whose formal solution is:
\begin{align}
\label{190225.3}
T_1 &= \left[ I - V J^L \right]^{-1} V~,
\end{align}
with $J^L = (H_0 - E)^{-1} \theta_1$. 
Incorporating intermediate states from $I$, the $T$ matrix $T_2(E)$ satisfies $T_2 = V + V (H_0 - E)^{-1} \theta T_2$~, with $\theta$ the projector excluding the bare state $|0\ra$, cf. Sec.~\ref{sec.190210.1}. This equation can be rewritten as:
\begin{align}
\label{180225.6}
T_2 &= V + V J^I T_2 + V (H_0 - E)^{-1} \theta_1 T_2~,
\end{align}
where $J^I = (H_0 - E)^{-1} (\theta - \theta_1)$. From the LS satisfied by $T_1$,  $T_2 = T_1 + T_1 J^I T_2$\,, whose solution is
\begin{align}
\label{180225.9}
T_2 &= \left(I - T_1 J^I\right)^{-1} T_1~.
\end{align}
The full $T$ matrix, with the hole set of intermediate states, satisfies:
\begin{align}
 \label{190225.12}
T &= V + V |0\rangle (E_0 - E)^{-1} \langle 0| T + V (H_0 - E)^{-1} \theta T~.
\end{align}
Following steps analogous to deriving Eq.~\eqref{181208.5}, the solution is:
\begin{align}
\label{190225.13}
T &= T_2 + \frac{T_2 |0\rangle \langle 0| T_2}{E - E_0 - \langle 0| T_2 |0\rangle}~.
\end{align}
For production amplitudes, the column vector $F$, accounting for the FSI as depicted in Fig.~\ref{fig.190225.1}, is:
\begin{align}
  \label{190226.1}
F&={\cal F}+{\cal F}(H_0-E)^{-1} T={\cal F}(I-(H_0-E)^{-1}V)^{-1}~.
\end{align}
Reference~\cite{Hanhart:2015cua} simplifies the FSI by considering only $L$ states in intermediate sums, yielding:
\begin{align}
\label{190226.3}
F&={\cal F} + {\cal F}^L J^L T~.
\end{align}
Equation~\eqref{190225.13} can be generalized to include $n_0$ bare states that span  the subspace $\Pi_0$, annihilated by the projectors $\theta_1$ and $\theta$. In this case, Eq.~\eqref{190225.12} becomes $T=V+\sum_a V|a\rangle (E_a-E)^{-1}\langle a|T+V(H_0-E)^{-1}\theta T$\,.  Note that $\langle a|(H_0-E)^{-1}|b\rangle=(E_a-E)^{-1}\delta_{ab}$.  Following an analogous procedure as after Eq.~\eqref{190225.12}, the solution is  
$T=T_2+\sum_a T_2|a\rangle(E_a-E)^{-1}\langle a|T$\,.
 Multiplying to the left  by $\langle b|\in \Pi_0$, we obtain $T=T_2+T_2 J^0(I-T_2^{00}J^0)^{-1}T_2\,,$
where $J^0$ and $T_2^{00}$ are the restrictions of $(H_0-E)^{-1}$ and $T_2$ in $\Pi_0$, respectively.

\section{Event Distributions in the Resonant Region, Branching Ratios, and Resonance-Pole Properties}
\def\theequation{\arabic{section}.\arabic{equation}}
\setcounter{equation}{0}
\label{sec.241110.1}

The relationship between event distributions,  branching ratios measured and  resonance pole properties, its pole position $s_R$ and couplings $\widetilde{g}_a$, $a=1,\ldots,n_c$, is further studied. A two-potential model is introduced as a phenomenological framework for analyzing event distributions, in particular, for resonant line shapes.  We also define branching ratios intrinsic to the resonance, not attached to any particular reaction. Understanding the implications of the RS in which the poles lies is not a minor point,  as it impacts the resonance signal.

\subsection{The Two-Potential Model}
\label{sec.241101.1}

The interaction between resonant and non-resonant (background) contributions in scattering and production amplitudes affected by FSI  was explored in Ref.~\cite{Heuser:2024biq} using a two-potential model \cite{Nakano:1982bc,Ray:1980ck}. We summarize the approach from Ref.~\cite{Heuser:2024biq}, provide additional steps in the derivation of some its results, and introduce some new developments. 

The background amplitude $M_B(s)$ is parameterized as in Eq.~\eqref{181109.6}, with $\lim_{s \to \infty} M_B(s) = 0$. The subtraction constant in each unitarity-loop function, cf. Eq.~\eqref{240815.4}, is chosen so that it vanishes at the threshold for channel $a$, $s_{\text{th};a}$. Similarly as in Ref.~\cite{Heuser:2024biq}, $M_B(s)$ is taken to be diagonal. As discussed in Sec.~\ref{sec.241102.1}, the matrix ${\cal N}(s)$ has no RHC, following Ref.~\cite{Heuser:2024biq}. 
An analytic function in the complex $s$-plane, cut along the LHC for $s<s_L$, can be written as a Taylor expansion around $\omega(s) = 0$, with $\omega(s)$ defined as
\begin{align}
  \label{241101.2}
\omega(s) &= \frac{\sqrt{s - s_L} - \sqrt{s_E - s_L}}{\sqrt{s - s_L} + \sqrt{s_E - s_L}}\,.
\end{align}
This approach is commonly used in hadron physics (see Ref.~\cite{Yndurain:2002ud}). Here we take  $\pi\pi$ scattering which has $s_L = 0$. The Taylor series in powers of $\omega(s)$ is applied to ${\cal N}_{aa}(s)^{-1}$, with $s_E = \Rea\, s_R$, to maximize the sensitivity to the resonance region. For overlapping resonances, a compromised choice for $s_E$ between resonance masses would be taken. The complex $s$-plane, cut to exclude the LHC, is mapped to the region $|\omega(s)| < 1$, with $|\omega(s)| = 1$ corresponding to the LHC for $s<s_L$. The expressions for $\omega(s)$ and $|\omega(s)|$, with $s=|s|e^{i\phi}$ and $\phi\in(-\pi,\pi]$, are 
\begin{align}
  \label{241102.3}
  \omega(s) &= \frac{|s| - s_E + i 2\sqrt{|s|s_E} \sin \frac{\phi}{2}}{|s| + s_E + 2\sqrt{|s|s_E} \cos \frac{\phi}{2}}\,, ~
  |\omega(s)|^2 = \frac{2}{1 + 2 \cos\frac{\phi}{2}\sqrt{|s|s_E}/(|s| + s_E)} - 1\,.
\end{align}
Thus, $|\omega(s)| < 1$ for $\phi \neq \pi$ and $|\omega(s)| = 1$ for $\phi = \pi$. The RHC, with $\phi = 0$, maps to the real axis of $\omega(s)$, with $\omega(s) < 0$ for $0 \leq |s| < s_E$ and $\omega(s) \geq 0$ for $|s| \geq s_E$. In Ref.~\cite{Heuser:2024biq}, ${\cal N}_{aa}(s)^{-1}$ is approximated by
\begin{align}
  \label{241101.4}
  {\cal N}_{aa}(s)^{-1} &= \frac{1}{f_{0;a}} + \sum_{k=1}^{k_{\rm max}} \frac{f_{k;a}}{f_{0;a}} \left( 2\omega(s) + \omega(s)^2 \right)^k + f_{x;a} s\,,
\end{align}
where $k_{\rm max} = 0, 1,$ or 2. The last term added ensures that $M_B(s)$ vanishes as $1/s$ at large $s$. The expansion is in terms of $2\omega(s) + \omega(s)^2$ rather than just $\omega(s)$, to account for the $s^{3/2}$ singularity arising from two-pion exchange in the $t$- and $u$-channels in $\pi\pi$ PWA \cite{Heuser:2024biq}. Indeed, by expanding around $s = 0$, we get
\begin{align}
  \label{241101.5}
  \alpha \omega(s) + \omega(s)^2 &= (1 - \alpha) - 2(\alpha - 2)\left( \frac{s}{s_E} \right)^{1/2} + (8 - 2\alpha) \frac{s}{s_E} + 2(\alpha - 6) \left( \frac{s}{s_E} \right)^{3/2} + {\cal O}(s^2)\,,
\end{align}
showing that only for $\alpha = 2$ can the $\sqrt{s}$ branch-point singularity be removed, leaving the $s^{3/2}$-type singularity  for the onset of the LHC as $s \to 0$. Using Eqs.~\eqref{181109.6} and \eqref{241101.4}, we calculate the background phase shift $\delta_B^a(s)$, and then compute the Omn\`es function $\gamma_a(s)$:
\begin{align}
  \label{241101.1}
  \gamma_a(s) &= \exp\left(\frac{s}{\pi}\int_{s_{\text{th};a}}^{\infty} ds' \frac{\delta_B^a(s')}{s'(s' - s)}\right)\,,
\end{align}
with $s_{\text{th};a}$ and $\delta_B^a(s)$ being the threshold and background phase shifts for the $a$th channel, respectively. Only one subtraction is needed for convergence since $\delta^a_B(s)$ tends to a constant at large $s$. The dressed resonance couplings are written as $g_a\gamma_a(s)\xi_a(s)$, where $g_a$ is the bare coupling, $\xi_a(s)$ is a centrifugal factor $\propto p_a^{\ell_a}$, cut  for $\ell_a>0$ by a phenomenological barrier factor. Here, $\ell_a$ is the orbital angular momentum and $p_a(s)$ is the momentum of channel $a$. Using these couplings the resonance self-energy reads:
\begin{align}
  \label{241101.6a}
  \Sigma(s) &= \sum_a g_a^2 \Sigma_a(s)\,, \\
  \label{241101.6}
  \Sigma_a(s) &= \frac{s}{\pi} \int_{s_{\text{th};a}}^{\infty} ds' \frac{|\gamma_a(s')|^2 \xi_a(s')^2 \rho_a(s')}{s'(s' - s)}\,, 
\end{align}
The self-energy is subtracted such that $\Sigma(0) = 0$, and  the resonance propagator can be written as
\begin{align}
  \label{241101.7a}
  G_R(s) &= \left( s - m^2 + \sum_{a=1}^{n_c} g_a^2 \Sigma_a(s) \right)^{-1}\,.
\end{align}
Thus, the resonance-exchange contribution to the PWA $a \to b$ is
\begin{align}
  \label{241101.7}
  M_{R;ab}(s) &= -g_a \gamma_a(s) \xi_a(s) G_R(s) g_b \gamma_b(s) \xi_b(s)\,,
\end{align}
which is added to the background amplitude to give the total PWA:
\begin{align}
  \label{241101.8}
  M_{ab}(s) &= M_{B;ab}(s) + M_{R;ab}(s)\,.
\end{align}
The separation of the interaction into a background term and a resonance contribution is analogous to Eq.~\eqref{181207.5}, derived from non-relativistic dynamics, cf. Fig.~\ref{fig.181207.1}. Regarding the production amplitudes $F_a(s)$, the formula used is \cite{Heuser:2024biq,Hanhart:2012wi}: 
\begin{align}
  \label{241101.9}
F_{R;a}(s)&=-g_a\gamma_a(s)\xi_a(s) G_R(s) \alpha \,, 
\end{align}
where $\alpha$ is the bare coupling of the production source to the resonance.

To apply this formalism, we need to determine the bare parameters $m$, $\alpha$, and $g_a$, for $a=1,\ldots,n_c$. Ref.~\cite{Heuser:2024biq} uses known resonance-pole parameters, such as the pole position $s_R$, residues $\widetilde{g}_a$ for each channel, and production-related residues $\widetilde{\alpha}_a$. The scattering amplitudes at the resonance pole position are obtained by analytically extrapolating $\gamma_a(s)$ and $\Sigma_a(s)$ to the unphysical RS where the pole lies. Note that there are $2^{n_c}$ possible RSs, arising from the two choices for each channel: physical or second RS. For clarity, we discuss the transition to the second RS for channel $a$. The following derivations use the Schwarz reflection principle for all involved functions—$M_{B;aa}(s)$, $\gamma_a(s)$, and $\Sigma_a(s)$—and assume $M_B(s)$ is diagonal. The background $T$-matrix $M_B(s)$, as given in Eqs.~\eqref{181109.6} and \eqref{241101.4}, changes across RSs as described in Sec.~\ref{sec.241102.1}, cf. Eq.~\eqref{181110.1}.

As mentioned, $\gamma_a(s)$ arises from FSI due to background interactions $M_B(s)$. Thus, its discontinuity along the RHC is,
\begin{align}
  \label{241109.1}
 \Delta \gamma_a(s+i\ep)&=2i \gamma_{a}(s+i\ep)\rho_a(s+i\ep)M_{B;aa}(s+i\ep)^*~,~s>s_{\text{th};a}\,.
\end{align}
For the self-energy $\Sigma_a(s)$ one has a unitarity relation analogous to that for PWAs:
\begin{align}
  \label{241109.2}
\Delta \Sigma_a(s+i\ep)&=2i\rho_a(s+i\ep)|\gamma_a(s+i\ep)\xi_a(s)|^2~,~s>s_{\text{th};a}\,.
\end{align}
We denote by  $\Sigma^{(-)}_a(s)$ and $\gamma_a^{(-)}(s)$ the corresponding functions in the second RS.
We consider  the left-hand side of Eq.~\eqref{241109.1} and rewrite it as $\gamma_a(s+i\ep)-\gamma_a(s-i\ep)=\gamma_a(s+i\ep)-\gamma_a^{(-)}(s+i\ep).$ Its right-hand side becomes $2i \gamma_{a}(s+i\ep)\rho_a(s+i\ep)M_{B;aa}^{(-)}(s+i\ep)$, so that
\begin{align}
  \label{241109.5}
\gamma_a^{(-)}(s)&=\gamma_a(s)\left(1-2i\rho_a(s)M_{B;aa}^{(-)}(s)\right)\,.
\end{align}
For $\Sigma_a(s)$, we follow analogous steps and, therefore,
\begin{align}
  \label{241109.7}
\Sigma_a^{(-)}(s)&=\Sigma_a(s)-2i\rho(s)|\xi_a(s)|^2\gamma_a(s)\gamma_a^{(-)}(s)\,,~s\in \mathbb{C}\,.
\end{align}
Once $\gamma_a^{(-)}(s)$ is known from Eq.~\eqref{241109.5}, $\Sigma_a^{(-)}(s)$ can be calculated from Eq.~\eqref{241109.7}. This allows us to obtain $M_R(s)$ in its second RS from Eq.~\eqref{241101.7}. Around the resonance pole in the appropriate RS, $M_R(s)$ has the Laurent series
\begin{align}
  \label{241109.8}
  M^{(-)}_{R;ab}(s)&=-\frac{R_{ab}}{s-s_R}+\ldots~,
\end{align}
where $R_{ab}$ is the residue matrix, factorized as $R_{ab}=\widetilde{g}_a\widetilde{g}_b$ (see Eq.~\eqref{190225.13})), and the ellipsis indicate regular terms. The physical resonance couplings $\widetilde{g}_a$ can be expressed in terms of the bare ones $g_a$ by taking the limit
\begin{align}
  \label{241109.9}
  \widetilde{g}_a\widetilde{g}_b&=-
\lim_{s\to s_R}(s-s_R)M^{(-)}_{R;ab}(s)=Z\,g_ag_b\gamma_a^{(-)}(s_R)\gamma_b^{(-)}(s_R)\xi_a(s_R)\xi_b(s_R) \,
\end{align}
where $Z$ is the field renormalization constant, $Z=\lim_{s\to s_R}(s-s_R)G_R(s)$\,. Reference~\cite{Heuser:2024biq} defines the branching ratios for a resonance following Ref.~\cite{Burkert:2022bqo}, and the total width $\Gamma^{R\text{(tot)}}$ is identified with the sum over all partial-decay widths. 
The branching ratio $Br_a$ is then computed as $Br_a=N_a/N_{\text{tot}}$. A key issue is calculating the partial-decay widths by using a normalized spectral function for the resonance, expressed  as $-\Ima G_R(s)/\pi$. The normalization condition is
\begin{align}
  \label{241109.11}
-\frac{1}{\pi}\int_{s_{\text{th};1}}^{+\infty}ds\, \Ima G_R(s)&=-\frac{1}{\pi}\sum_{a=1}^{n_c}\int_{s_{\text{th};1}}^{+\infty}ds\, \Ima G_{R;a}(s)\rho_a(s)\theta(s-s_{\text{th};a})=1\,.
\end{align}
Here,
\begin{align}
\label{241109.14}
\Ima G_{R;a}(s)\rho_a(s)\theta(s-s^a_{\rm{th}})&=-g_a^2|G_R(s)|^2\Ima \Sigma_a(s)=g_a^2|G_R(s)|^2|\gamma_a(s)\xi_a(s)|^2\rho_a(s)\theta(s-s^a_{{\rm thr}})\,,
\end{align}
with  $G_R(s)$  given in Eq.~\eqref{241101.7a}. Therefore, the branching ratio for channel $a$ is defined as
\begin{align}
\label{241109.12}
Br_a&=-\frac{1}{\pi}\int_{s_{\text{th};a}}^{+\infty}ds \,\Ima G_{R;a}(s)\rho_a(s)\,,
\end{align}
and due to the normalization condition in Eq.~\eqref{241109.11}, we have that $\sum_{a=1}^{n_c}Br_a=1\,.$

What remains to be seen is that Eq.~\eqref{241109.11} is satisfied. The point to notice is that $\Sigma_a(s)$  from Eq.~\eqref{241101.6} tends to constant for $s\to\infty$,  so that $G_R(s)$, Eq.~\eqref{241101.7a},  vanishes as $1/s$ for $s\to\infty$. Actually, for the steps that follow it is only required that $\Sigma_a(s)/s\to 0$ for $s\to \infty$. We use this fact to apply the Cauchy integration theorem to a closed line integral in the complex $s$ plane along the circuit ${\cal C}$ consisting of a circle of radius $R\to \infty$ engulfing the RHC, 
\begin{align}
  \label{241109.15}
  \oint_{{\cal C}} G_R(s')ds'&=0=i\lim_{R\to \infty}R\int_0^{2\pi}d\phi e^{i\phi}G_R(Re^{i\phi})
  +2i\int_{s_{\text{th};1}}^{+\infty}ds'\Ima G_R(s'+i\ep)\,.
\end{align}
Now, as $G_R(s)\to s$ for $s\to\infty$, it follows that the first integral after the last equal sign in Eq.~\eqref{241109.15} is $i 2\pi$, so that  Eq.~\eqref{241109.11} is fulfilled. 

Other less elaborated approaches take only the imaginary part of $\Sigma_a$ and write
\begin{align}
  \label{241109.16}
  G_R(s)&=\left(s-m^2+i\sum_a h_a^2\rho_a(s)\xi_a(s)^2\right)^{-1}\,,\\
  -\Ima G_R(s)&=\frac{\sum_a h_a^2\rho_a(s)\xi_a(s)^2}{\left|s-m^2+i\sum_a h_a^2\rho_a(s)\xi_a(s)^2\right|^2}\,,\nn
\end{align}
Here, $h_a$ is an effective coupling, not necessarily equal to $|\widetilde{g}_a|$ or $g_a$. E.g. in Ref.~\cite{Burkert:2022bqo} they differ from $|\widetilde{g}_a|$ in a common constant $f$, which is adjusted together with $m^2$ to get $s_R$ at a given value.  
The problem with the calculation of $Br_a$ as in Eq.~\eqref{241109.12} is that the sum over all of them is not equal to 1 because $-\Ima G_R(s)/\pi$ is not properly normalized, Eq.~\eqref{241109.11}. It is worth explaining why the demonstration provided above, based on Eq.~\eqref{241109.15}, concerning the normalization of $-\Ima G_R(s)/\pi$, with $G_R(s)$ given in Eq.~\eqref{241101.7a}, does not work in this case with $G_R(s)$ as in Eq.~\eqref{241109.16}. 
The novelty concerns the analytical properties of $\rho_a(s)$. 
 Generally, $\rho_a(s)$, as defined by Eq.~\eqref{241205.1}, has a RHC for $s>(m_{1;a}+m_{2;a})^2$ and a LHC for $s<(m_{1;a}-m_{2;a})^2$, with the first RS taken for the square root, so that $\sqrt{z}$ is calculated with $\text{arg}z\in [0,2\pi)$. Because of the latter LHC the closed line integration over $G_R(s)$ of Eq.~\eqref{241109.15} picks up an extra contribution which invalidates the normalization condition, Eq.~\eqref{241109.11}.

\subsection{Poles in contiguous and non-contiguous Riemann sheets}
\label{sec.241115.1}

We review here on the findings of Refs.~\cite{Wang:2022vga,Burkert:2022bqo}, and  exemplify the discussions that follow with the example of the $f_0(980)$ \cite{ParticleDataGroup:2024cfk}, though the results would apply to any resonance whose pole position lies close to an inelastic threshold. 
 The pole parameters of the $f_0(980)$ resonance were obtained from a thorough dispersive analysis of experimental data \cite{Garcia-Martin:2011iqs}: 
\begin{align}
  \label{241115.1}
  E_R=(996 \pm 7 - i\,25^{+10}_{-6})\,\text{MeV}~, ~\vert \widetilde{g}_{\pi\pi} \vert = 2.3 \pm 0.2\,\text{GeV}\,.
\end{align}
Following conventional reasoning, one might interpret the total width as negative twice the imaginary part of the pole position, and calculate the partial width to two pions simply as
\begin{equation}
  \label{241205.2}
\Gamma_{\pi \pi} = \frac{g_{\pi \pi}^2}{M} \rho_{\pi \pi}(M^2) = 100^{+20}_{-17}\, \text{MeV},
\end{equation}
 This narrow-resonance equation is expected to be accurate because the two pion state is much lighter than the resonance. Then,  finite-width corrections of the $f_0(980)$, with a pole width $\Gamma_{\rm pole}=-2\Ima E_R$ centered around 50~MeV, should be suppressed for this decay. 
However, Eq.~\eqref{241205.2}  gives $\Gamma_{\pi\pi}$  twice as large as the pole width, $\Gamma_{\rm pole} = 50^{+20}_{-12}\,$MeV, Eq.~\eqref{241115.1}. Of course, this result is at odds with the usual identification of the total width of a resonance with its pole width (minus twice the imaginary part of $E_R$).

This discrepancy was initially explained in~\cite{Wang:2022vga}, which derived the relationship between pole width and residues (Eq.~\eqref{eq:GIInottot} below). In practice, there are no observable four-pion state effects in $\pi\pi$ scattering below 1 GeV, so the first inelastic threshold for the $\pi\pi$ $S$-wave is the $K \bar K$ threshold. The issue with the $f_0(980)$ resonance is that its pole, despite having a mass of $996 \pm 7~$MeV, above the $K \bar K$ threshold, lies in the second RS, not the expected fourth one. This changes the interpretation of the total width, as the kaon phase-space contribution changes sign in the second RS compared to the fourth, a scenario emphasized in \cite{Wang:2022vga} and exemplified through a toy model here.


\bigskip
              {\bf  A Toy Model}
              \bigskip

We construct a toy model for resonant $S$-wave scattering with two channels, $a=1(\pi\pi)$ and 2\,($K\bar{K}$). Using an energy-dependent relativistic Breit-Wigner parameterization \cite{Burkert:2022bqo}, we write the $t_{11}$ matrix element on the physical or first RS sheet ($Im \, s \geq 0$) as:
\begin{equation}\label{RS-I}
t_{11}^{I}(s) = \frac{g_1^2 \, \rho_1(s)}{M^2 - s - i \,(g_1^2 \, \rho_1(s) + g_2^2 \, \rho_2(s))}\,,
\end{equation}
The analytical continuation of this function to the adjacent RS is obtained by crossing  the physical $s$ axis into the lower half-plane ($\Ima \, s < 0$). To simplify the discussion, we take here the cut of $\sqrt{s - s_{\text{th};a}}$ in $\rho_a$ to run along $s \le s_{\text{th};a}$, as it is usually the case in  standard programming languages. In particular, the contiguous sheet for $s_{\text{th};1} \leq s \leq s_{\text{th};2}$ is  the second RS, and for $\Ima s <0$  the amplitude becomes:
\begin{equation}\label{RS-II}
t_{11}^{II}(s) = \frac{g_1^2 \, \rho_1(s)}{M^2 - s - i \,(g_1^2 \, \rho_1(s) - g_2^2 \, \rho_2(s))}.
\end{equation}
Note the sign change in front of $\rho_2(s)$ upon crossing into the second sheet, while $\rho_1(s)$ retains the same sign. Above the $K \bar K$ threshold ($s > s_2$), the adjacent sheet is  the fourth RS, cf. Sec.~\ref{sec.241102.1}. Therefore, for $\Ima s<0$, 
\begin{equation}\label{RS-III}
t_{11}^{IV}(s) = \frac{g_1^2 \, \rho_1(s)}{M^2 - s - i \,(g_1^2 \, \rho_1(s) + g_2^2 \, \rho_2(s))}\,.
\end{equation}
Next, we consider a scenario where the PWA near \( s = M^2 \) is dominated by an isolated pole in the fourth RS, located at \( s_{R,IV} \). The partial-decay widths are defined as \( \Gamma_a \equiv \frac{g_a^2 \rho_a(M^2)}{M} \), with the approximation \( \rho_a(s) \simeq \rho_a(M^2) \) and assuming that \( \frac{\Gamma_{\rm pole}^2}{M^2} \ll 1 \). From Eq.~\eqref{RS-III}, along the resonance-energy region, we obtain:
\begin{align}
t^{IV}_{11}(s) &\simeq \frac{g_1^2 \, \rho_1(s)}{s_{R,IV} - s} \simeq \frac{g_1^2 \, \rho_1(s)}{(M - i (\Gamma_1 + \Gamma_2)/2)^2 - s} = \frac{g_1^2 \, \rho_1(s)}{(M - i \Gamma_{\rm tot}/2)^2 - s}. \nonumber
\label{eq:twithGammas}
\end{align}
Thus, the usual pole-resonance parameter relation applies, $\sqrt{s_{R,IV}} \simeq M - i \frac{\Gamma_{\rm tot}}{2} = M - i \frac{\Gamma_1 + \Gamma_2}{2}$. It also follows that the branching ratios fulfill  \( Br_a \equiv {\Gamma_a}/{\Gamma_{\rm tot}} \leq  1 \) and \( \sum_a Br_a = 1 \).

However, if there were a pole in the second sheet at \( s_{R,II} \), close to but just above \( s_2 \) (as is the case for the \( f_0(980) \) pole from Ref.~\cite{Garcia-Martin:2011nna}), it could still significantly influence real values of \( s \) below and even slightly above this threshold. This is explicitly reflected in the right panel of Fig.~\ref{fig.240816.1}, where the far-right pole corresponds precisely to the \( f_0(980) \), as discussed there. For such a pole, applying Eq.~\eqref{RS-II} yields:
\begin{align}
t^{II}_{11}(s)
&\simeq \frac{g_1^2\,\rho_1(s)}{s_{R,II}-s} \simeq 
\frac{g_1^2\,\rho_1(s)}{(M - i (\Gamma_1 - \Gamma_2)/2)^2 - s} = \frac{g_1^2\,\rho_1(s)}{(M - i \Gamma_{\rm{II}}/2)^2 - s}.
\end{align}
Thus, for poles in the second sheet under the described conditions, the usual relationship between the pole position and resonance parameters  no longer applies, \( \Gamma_{\rm{II}} \neq \Gamma_{\rm tot} = \Gamma_1 + \Gamma_2 \). Instead, one has:
\begin{align}
\label{eq:GIInottot}
\sqrt{s_{R,II}} &\simeq M - i \frac{\Gamma_{\rm{II}}}{2}, \\
\Gamma_{\rm{II}} &= \Gamma_1 - \Gamma_2.
\end{align}
 Therefore, the branching ratios cannot be written as \( {\Gamma_a}/{\Gamma_{\rm{II}}} \), as this would imply \( \Gamma_1 > \Gamma_{\rm{II}} \) and \( BR_1 > 1 \), which is not physically meaningful.  This result was also illustrated within a two-coupled channel Flatt\'e parameterization in Ref.~\cite{Wang:2022vga}. Nonetheless, when calculating cross sections influenced by the exchange of such a resonance, it is still reasonable to define the branching ratio as \( BR_a = \Gamma_a/\Gamma_{\rm tot} \) \cite{Burkert:2022bqo}.

\bigskip
\textbf{Branching Ratios from the Pole Parameters}
\bigskip

The $f_0(980)$ pole in Eq.~\eqref{241115.1} from Ref.~\cite{Garcia-Martin:2011nna} lies in the second RS. Therefore, and in light of the previous model, it is clear that \( \Gamma_{\text{pole}} = \Gamma_{\rm{II}} = 50^{+20}_{-12}\, \text{MeV} \) does not correspond to the total width \( \Gamma_{\text{tot}} \). As explained earlier, the partial width to \( K\bar{K} \) contributes negatively to $\Gamma_{\text{pole}}$, cf. Eq.~\eqref{eq:GIInottot}. Including the correlations from the Madrid-Krakow dispersive calculation, Ref.~\cite{Burkert:2022bqo} finds:
\begin{eqnarray}
  \label{241205.3}
&& \Gamma_{K \bar K} \simeq \Gamma_{\pi \pi} - \Gamma_{\rm{II}} = 50^{+26}_{-21}\, \text{MeV}, \\
&& \Gamma_{\text{tot}} \simeq \Gamma_{\pi \pi} + \Gamma_{K \bar K} = 151^{+44}_{-37}\, \text{MeV}, \nonumber \\
&& {\rm BR}_{\pi \pi} \simeq 0.67 \pm 0.07, \nonumber \\
&& {\rm BR}_{K \bar K} \simeq 0.33 \pm 0.07, \nonumber \\
&& r_{K \bar K/\pi \pi} = \frac{{\rm BR}_{K \bar K}}{{\rm BR}_{\pi \pi}} \simeq 0.49 \pm 0.11, \nonumber
\end{eqnarray}
where the approximate symbol arises because we have neglected channels beyond \( \pi \pi \) and \( K \bar K \). These results align well with the values listed in the PDG \cite{ParticleDataGroup:2024cfk}, where the central values for \( \frac{\Gamma_{\pi \pi}}{\Gamma_{\pi \pi} + \Gamma_{K \bar K}} \) range from 0.52 to 0.84. They are also compatible  with the value of \( 0.75 \pm 0.20 \) obtained from the chiral unitary approach in \cite{Wang:2022vga,Guo:2012yt,Guo:2012ym}.

Next, following Ref.~\cite{Burkert:2022bqo}, we use the integration method in Eq.~(\ref{241109.12}) to determine the branching ratios of the \( f_0(980) \) to \( K \bar K \) and \( \pi \pi \), using the simpler spectral function \( - \text{Im} G_R / \pi \) corresponding to Eq.~\eqref{241109.16}. The scalar-isoscalar amplitude, derived from a fit to data on \( J/\psi \) radiative decays and other data, is employed. 
For the ratio, we obtain:
\begin{equation}
r_{K \bar K/\pi \pi} = 0.40 \pm 0.07,
\end{equation}
which is in good agreement with the previous value in Eq.~\eqref{241205.3}.

\section{Conclusions}

This article provides a pedagogical introduction to coupled-channel scattering, an indispensable tool for understanding the complex interactions that occur in hadronic physics. By expanding upon single-channel models, the article highlights how coupled channels, including both elastic and inelastic scattering processes, as well as resonance formation, can be effectively described. The discussion covered key theoretical approaches, such as the $N/D$ method, the Castillejo-Dalitz-Dyson poles, general parameterizations for partial-wave amplitudes, the Lippmann-Schwinger equation with potentials given, which together ensure unitarity and are compatible with analyticity in the description of scattering amplitudes. The article also explored the role of subtraction constants, effective-range expansions, and Flatté parameterizations. Several methods related to the study of final-state interactions in a process stemming from an external probe were also covered. After discussing general constraints from unitarity and analyticity, we have discussed the Omn\`es solution and Muskhelishvily-Omn\`es problem, the Khuri-Treiman approach, and the use in this context of the Lippmann-Schwinger equations with general potentials.    The importance of these methods was further demonstrated in the study of resonant event distributions, and in the physical implications of resonance poles not lying in contiguous Riemann sheets. In this connection, we have also explained  in detail  how to reach the resonance poles by accessing to the different unphysical Riemann sheets.  Overall, this work offers an accessible introduction to some essentials of coupled-channel scattering and reflects into its theoretical framework, highlighting its significance in phenomenological applications.

\begin{ack}[Acknowledgments]%
  I would like to thank Christoph Hanhart for reading the manuscript and engaging in discussions, particularly with respect to Sec.~6.1. I also acknowledge partial financial support to the Grant PID2022-136510NB-C32 funded
by \\MCIN/AEI/10.13039/501100011033/ and FEDER, UE, and to the EU Horizon 2020 research and innovation program, STRONG-2020 project, under grant agreement no. 824093.
\end{ack}

\bibliographystyle{Numbered-Style} 
\bibliography{references2}

\begin{thebibliography*}{100}
\providecommand{\bibtype}[1]{}
\providecommand{\url}[1]{{\tt #1}}
\providecommand{\urlprefix}{URL }
\expandafter\ifx\csname urlstyle\endcsname\relax
  \providecommand{\doi}[1]{doi:\discretionary{}{}{}#1}\else
  \providecommand{\doi}{doi:\discretionary{}{}{}\begingroup
  \urlstyle{rm}\Url}\fi
\providecommand{\bibinfo}[2]{#2}
\providecommand{\eprint}[2][]{\url{#2}}
\makeatletter\def\@biblabel#1{\bibinfo{label}{[#1]}}\makeatother

\bibtype{Book}%
\bibitem{martin.200705.1}
\bibinfo{author}{A.~D. Martin}, \bibinfo{author}{T.~D. Spearman},
  \bibinfo{title}{Elementary Particle Theory},
  \bibinfo{publisher}{North-Holland Publishing Company, Amsterdam}
  \bibinfo{year}{1970}.

\bibtype{Book}%
\bibitem{Barton:1965dr}
\bibinfo{author}{G. Barton}, \bibinfo{title}{{Introduction to dispersion
  techniques in field theory}}, \bibinfo{publisher}{W.A. Benjamin}
  \bibinfo{year}{1965}.

\bibtype{Book}%
\bibitem{Weinberg:1995mt}
\bibinfo{author}{Steven Weinberg}, \bibinfo{title}{{The Quantum theory of
  fields. Vol. 1: Foundations}}, \bibinfo{publisher}{Cambridge University
  Press} \bibinfo{year}{2005}, ISBN \bibinfo{isbn}{978-0-521-67053-1,
  978-0-511-25204-4}, \bibinfo{doi}{\doi{10.1017/CBO9781139644167}}.

\bibtype{Book}%
\bibitem{Schweber:1961zz}
\bibinfo{author}{Silvan~S. Schweber}, \bibinfo{title}{An Introduction to
  Relativistic Quantum Field Theory}, \bibinfo{publisher}{Dover Publications,
  Inc., N.Y.} \bibinfo{year}{1961}.

\bibtype{Book}%
\bibitem{Weinberg:2013qm}
\bibinfo{author}{Steven Weinberg}, \bibinfo{title}{{Lectures on Quantum
  Mechanics}}, \bibinfo{publisher}{Cambridge University Press}
  \bibinfo{year}{2013}.

\bibtype{Book}%
\bibitem{olive.181102.1}
\bibinfo{author}{R.~J. Eden}, \bibinfo{author}{P.~V. Landshoff},
  \bibinfo{author}{D.~I. Olive}, \bibinfo{author}{J.~C. Polkinghorne},
  \bibinfo{title}{The Analytic $S$-matrix}, \bibinfo{publisher}{Cambridge
  University Press, Cambridge} \bibinfo{year}{1966}.

\bibtype{Book}%
\bibitem{Oller:2019rej}
\bibinfo{author}{Jos\'e~Antonio Oller}, \bibinfo{title}{{A Brief Introduction
  to Dispersion Relations}}, SpringerBriefs in Physics,
  \bibinfo{publisher}{Springer} \bibinfo{year}{2019}, ISBN
  \bibinfo{isbn}{978-3-030-13581-2, 978-3-030-13582-9},
  \bibinfo{doi}{\doi{10.1007/978-3-030-13582-9}}.

\bibtype{Article}%
\bibitem{Oller:2020guq}
\bibinfo{author}{Jos\'e~Antonio Oller}, \bibinfo{title}{{Unitarization Technics
  in Hadron Physics with Historical Remarks}}, \bibinfo{journal}{Symmetry}
  \bibinfo{volume}{12} (\bibinfo{number}{7}) (\bibinfo{year}{2020})
  \bibinfo{pages}{1114}, \bibinfo{doi}{\doi{10.3390/sym12071114}},
  \eprint{2005.14417}.

\bibtype{Article}%
\bibitem{Oller:2019opk}
\bibinfo{author}{J.~A. Oller}, \bibinfo{title}{{Coupled-channel approach in
  hadron\textendash{}hadron scattering}}, \bibinfo{journal}{Prog. Part. Nucl.
  Phys.} \bibinfo{volume}{110} (\bibinfo{year}{2020}) \bibinfo{pages}{103728},
  \bibinfo{doi}{\doi{10.1016/j.ppnp.2019.103728}}, \eprint{1909.00370}.

\bibtype{Article}%
\bibitem{Oller:2024lrk}
\bibinfo{author}{J.~A. Oller}, \bibinfo{title}{{Lectures on scattering theory
  in partial-wave amplitudes}}  (\bibinfo{year}{2024}), \eprint{2409.16790}.

\bibtype{Article}%
\bibitem{Chew:1960iv}
\bibinfo{author}{Geoffrey~F. Chew}, \bibinfo{author}{Stanley Mandelstam},
  \bibinfo{title}{{Theory of low-energy pion pion interactions}},
  \bibinfo{journal}{Phys. Rev.} \bibinfo{volume}{119} (\bibinfo{year}{1960})
  \bibinfo{pages}{467--477}, \bibinfo{doi}{\doi{10.1103/PhysRev.119.467}}.

\bibtype{Article}%
\bibitem{Castillejo:1955ed}
\bibinfo{author}{L. Castillejo}, \bibinfo{author}{R.~H. Dalitz},
  \bibinfo{author}{F.~J. Dyson}, \bibinfo{title}{{Low's scattering equation for
  the charged and neutral scalar theories}}, \bibinfo{journal}{Phys. Rev.}
  \bibinfo{volume}{101} (\bibinfo{year}{1956}) \bibinfo{pages}{453--458},
  \bibinfo{doi}{\doi{10.1103/PhysRev.101.453}}.

\bibtype{Article}%
\bibitem{Oller:1998zr}
\bibinfo{author}{J.~A. Oller}, \bibinfo{author}{E. Oset}, \bibinfo{title}{{N/D
  description of two meson amplitudes and chiral symmetry}},
  \bibinfo{journal}{Phys. Rev. D} \bibinfo{volume}{60} (\bibinfo{year}{1999})
  \bibinfo{pages}{074023}, \bibinfo{doi}{\doi{10.1103/PhysRevD.60.074023}},
  \eprint{hep-ph/9809337}.

\bibtype{Article}%
\bibitem{Entem:2016ipb}
\bibinfo{author}{D.~R. Entem}, \bibinfo{author}{J.~A. Oller},
  \bibinfo{title}{{The N/D method with non-perturbative left-hand-cut
  discontinuity and the $^1S_0$ $NN$ partial wave}}, \bibinfo{journal}{Phys.
  Lett. B} \bibinfo{volume}{773} (\bibinfo{year}{2017})
  \bibinfo{pages}{498--504},
  \bibinfo{doi}{\doi{10.1016/j.physletb.2017.09.012}}, \eprint{1610.01040}.

\bibtype{Article}%
\bibitem{Oller:2018zts}
\bibinfo{author}{J.~A. Oller}, \bibinfo{author}{D.~R. Entem},
  \bibinfo{title}{{The exact discontinuity of a partial wave along the
  left-hand cut and the exact $N/D$ method in non-relativistic scattering}},
  \bibinfo{journal}{Annals Phys.} \bibinfo{volume}{411} (\bibinfo{year}{2019})
  \bibinfo{pages}{167965}, \bibinfo{doi}{\doi{10.1016/j.aop.2019.167965}},
  \eprint{1810.12242}.

\bibtype{Article}%
\bibitem{Sanchez:2024xzl}
\bibinfo{author}{M.~S. S\'anchez}, \bibinfo{author}{J.~A. Oller},
  \bibinfo{author}{D.~R. Entem}, \bibinfo{title}{{Confronting the
  Lippmann-Schwinger equation and the $N/D$ method for coupled-wave separable
  potentials}}  (\bibinfo{year}{2024}), \eprint{2403.05486}.

\bibtype{Article}%
\bibitem{Blas:2020dyg}
\bibinfo{author}{Diego Blas}, \bibinfo{author}{Jorge Martin~Camalich},
  \bibinfo{author}{Jose~Antonio Oller}, \bibinfo{title}{{Unitarization of
  infinite-range forces: graviton-graviton scattering}},
  \bibinfo{journal}{JHEP} \bibinfo{volume}{08} (\bibinfo{year}{2022})
  \bibinfo{pages}{266}, \bibinfo{doi}{\doi{10.1007/JHEP08(2022)266}},
  \eprint{2010.12459}.

\bibtype{Article}%
\bibitem{Blas:2020och}
\bibinfo{author}{D. Blas}, \bibinfo{author}{J. Martin~Camalich},
  \bibinfo{author}{J.~A. Oller}, \bibinfo{title}{{Scalar resonance in
  graviton-graviton scattering at high-energies: The graviball}},
  \bibinfo{journal}{Phys. Lett. B} \bibinfo{volume}{827} (\bibinfo{year}{2022})
  \bibinfo{pages}{136991}, \bibinfo{doi}{\doi{10.1016/j.physletb.2022.136991}},
  \eprint{2009.07817}.

\bibtype{Article}%
\bibitem{Bjorken:1960zz}
\bibinfo{author}{J.~D. Bjorken}, \bibinfo{title}{{Construction of Coupled
  Scattering and Production Amplitudes Satisfying Analyticity and Unitarity}},
  \bibinfo{journal}{Phys. Rev. Lett.} \bibinfo{volume}{4}
  (\bibinfo{year}{1960}) \bibinfo{pages}{473--474},
  \bibinfo{doi}{\doi{10.1103/PhysRevLett.4.473}}.

\bibtype{Article}%
\bibitem{Albaladejo:2012sa}
\bibinfo{author}{M. Albaladejo}, \bibinfo{author}{J.~A. Oller},
  \bibinfo{title}{{Nucleon-Nucleon Interactions from Dispersion Relations:
  Coupled Partial Waves}}, \bibinfo{journal}{Phys. Rev. C} \bibinfo{volume}{86}
  (\bibinfo{year}{2012}) \bibinfo{pages}{034005},
  \bibinfo{doi}{\doi{10.1103/PhysRevC.86.034005}}, \eprint{1201.0443}.

\bibtype{Article}%
\bibitem{Gulmez:2016scm}
\bibinfo{author}{D. G\"ulmez}, \bibinfo{author}{U.~G. Mei\ss{}ner},
  \bibinfo{author}{J.~A. Oller}, \bibinfo{title}{{A chiral covariant approach
  to $\rho\rho$ scattering}}, \bibinfo{journal}{Eur. Phys. J. C}
  \bibinfo{volume}{77} (\bibinfo{number}{7}) (\bibinfo{year}{2017})
  \bibinfo{pages}{460}, \bibinfo{doi}{\doi{10.1140/epjc/s10052-017-5018-z}},
  \eprint{1611.00168}.

\bibtype{Article}%
\bibitem{Du:2018gyn}
\bibinfo{author}{Meng-Lin Du}, \bibinfo{author}{Dilege G\"ulmez},
  \bibinfo{author}{Feng-Kun Guo}, \bibinfo{author}{Ulf-G. Mei\ss{}ner},
  \bibinfo{author}{Qian Wang}, \bibinfo{title}{{Interactions between vector
  mesons and dynamically generated resonances}}, \bibinfo{journal}{Eur. Phys.
  J. C} \bibinfo{volume}{78} (\bibinfo{number}{12}) (\bibinfo{year}{2018})
  \bibinfo{pages}{988}, \bibinfo{doi}{\doi{10.1140/epjc/s10052-018-6475-8}},
  \eprint{1808.09664}.

\bibtype{Article}%
\bibitem{ParticleDataGroup:2024cfk}
\bibinfo{author}{S. Navas}, et al. (\bibinfo{collaboration}{Particle Data
  Group}), \bibinfo{title}{{Review of particle physics}},
  \bibinfo{journal}{Phys. Rev. D} \bibinfo{volume}{110} (\bibinfo{number}{3})
  (\bibinfo{year}{2024}) \bibinfo{pages}{030001},
  \bibinfo{doi}{\doi{10.1103/PhysRevD.110.030001}}.

\bibtype{Article}%
\bibitem{Gasser:1984gg}
\bibinfo{author}{J. Gasser}, \bibinfo{author}{H. Leutwyler},
  \bibinfo{title}{{Chiral Perturbation Theory: Expansions in the Mass of the
  Strange Quark}}, \bibinfo{journal}{Nucl. Phys. B} \bibinfo{volume}{250}
  (\bibinfo{year}{1985}) \bibinfo{pages}{465--516},
  \bibinfo{doi}{\doi{10.1016/0550-3213(85)90492-4}}.

\bibtype{Article}%
\bibitem{Oller:1997ti}
\bibinfo{author}{J.~A. Oller}, \bibinfo{author}{E. Oset},
  \bibinfo{title}{{Chiral symmetry amplitudes in the S wave isoscalar and
  isovector channels and the $\sigma$, f$_0$(980), a$_0$(980) scalar mesons}},
  \bibinfo{journal}{Nucl. Phys. A} \bibinfo{volume}{620} (\bibinfo{year}{1997})
  \bibinfo{pages}{438--456},
  \bibinfo{doi}{\doi{10.1016/S0375-9474(97)00160-7}}, \bibinfo{note}{[Erratum:
  Nucl.Phys.A 652, 407--409 (1999)]}, \eprint{hep-ph/9702314}.

\bibtype{Article}%
\bibitem{Oller:2000wa}
\bibinfo{author}{J.~A. Oller}, \bibinfo{title}{{Scalar mesons and chiral
  symmetry}}, \bibinfo{journal}{Soryushiron Kenkyu} \bibinfo{volume}{102}
  (\bibinfo{number}{5}) (\bibinfo{year}{2000}) \bibinfo{pages}{33--40},
  \eprint{hep-ph/0007349}.

\bibtype{Article}%
\bibitem{Oller:2000fj}
\bibinfo{author}{J.~A. Oller}, \bibinfo{author}{Ulf~G. Mei{\ss}ner},
  \bibinfo{title}{{Chiral dynamics in the presence of bound states: Kaon
  nucleon interactions revisited}}, \bibinfo{journal}{Phys. Lett. B}
  \bibinfo{volume}{500} (\bibinfo{year}{2001}) \bibinfo{pages}{263--272},
  \bibinfo{doi}{\doi{10.1016/S0370-2693(01)00078-8}}, \eprint{hep-ph/0011146}.

\bibtype{Article}%
\bibitem{Bernard:1991zc}
\bibinfo{author}{Veronique Bernard}, \bibinfo{author}{Norbert Kaiser},
  \bibinfo{author}{Ulf~G. Mei{\ss}ner}, \bibinfo{title}{{Chiral perturbation
  theory in the presence of resonances: Application to pi pi and pi K
  scattering}}, \bibinfo{journal}{Nucl. Phys. B} \bibinfo{volume}{364}
  (\bibinfo{year}{1991}) \bibinfo{pages}{283--320},
  \bibinfo{doi}{\doi{10.1016/0550-3213(91)90586-M}}.

\bibtype{Article}%
\bibitem{Kawarabayashi:1966kd}
\bibinfo{author}{Ken Kawarabayashi}, \bibinfo{author}{Mahiko Suzuki},
  \bibinfo{title}{{Partially conserved axial vector current and the decays of
  vector mesons}}, \bibinfo{journal}{Phys. Rev. Lett.} \bibinfo{volume}{16}
  (\bibinfo{year}{1966}) \bibinfo{pages}{255},
  \bibinfo{doi}{\doi{10.1103/PhysRevLett.16.255}}.

\bibtype{Article}%
\bibitem{Riazuddin:1966sw}
\bibinfo{author}{Riazuddin}, \bibinfo{author}{Fayyazuddin},
  \bibinfo{title}{{Algebra of current components and decay widths of $\rho$ and
  $K^*$ mesons}}, \bibinfo{journal}{Phys. Rev.} \bibinfo{volume}{147}
  (\bibinfo{year}{1966}) \bibinfo{pages}{1071--1073},
  \bibinfo{doi}{\doi{10.1103/PhysRev.147.1071}}.

\bibtype{Article}%
\bibitem{Kang:2016jxw}
\bibinfo{author}{Xian-Wei Kang}, \bibinfo{author}{J.~A. Oller},
  \bibinfo{title}{{Different pole structures in line shapes of the $X(3872)$}},
  \bibinfo{journal}{Eur. Phys. J. C} \bibinfo{volume}{77} (\bibinfo{number}{6})
  (\bibinfo{year}{2017}) \bibinfo{pages}{399},
  \bibinfo{doi}{\doi{10.1140/epjc/s10052-017-4961-z}}, \eprint{1612.08420}.

\bibtype{Article}%
\bibitem{Flatte:1976xu}
\bibinfo{author}{Stanley~M. Flatte}, \bibinfo{title}{{Coupled - Channel
  Analysis of the pi eta and K anti-K Systems Near K anti-K Threshold}},
  \bibinfo{journal}{Phys. Lett. B} \bibinfo{volume}{63} (\bibinfo{year}{1976})
  \bibinfo{pages}{224--227}, \bibinfo{doi}{\doi{10.1016/0370-2693(76)90654-7}}.

\bibtype{Article}%
\bibitem{Guo:2015daa}
\bibinfo{author}{Zhi-Hui Guo}, \bibinfo{author}{J.~A. Oller},
  \bibinfo{title}{{Probabilistic interpretation of compositeness relation for
  resonances}}, \bibinfo{journal}{Phys. Rev. D} \bibinfo{volume}{93}
  (\bibinfo{number}{9}) (\bibinfo{year}{2016}) \bibinfo{pages}{096001},
  \bibinfo{doi}{\doi{10.1103/PhysRevD.93.096001}}, \eprint{1508.06400}.

\bibtype{Article}%
\bibitem{Kang:2016ezb}
\bibinfo{author}{Xian-Wei Kang}, \bibinfo{author}{Zhi-Hui Guo},
  \bibinfo{author}{J.~A. Oller}, \bibinfo{title}{{General considerations on the
  nature of $Z_b(10610)$ and $Z_b(10650)$ from their pole positions}},
  \bibinfo{journal}{Phys. Rev. D} \bibinfo{volume}{94} (\bibinfo{number}{1})
  (\bibinfo{year}{2016}) \bibinfo{pages}{014012},
  \bibinfo{doi}{\doi{10.1103/PhysRevD.94.014012}}, \eprint{1603.05546}.

\bibtype{Article}%
\bibitem{Morgan:1992ge}
\bibinfo{author}{D. Morgan}, \bibinfo{title}{{Pole counting and resonance
  classification}}, \bibinfo{journal}{Nucl. Phys. A} \bibinfo{volume}{543}
  (\bibinfo{year}{1992}) \bibinfo{pages}{632--644},
  \bibinfo{doi}{\doi{10.1016/0375-9474(92)90550-4}}.

\bibtype{Article}%
\bibitem{Oller:2017alp}
\bibinfo{author}{J.~A. Oller}, \bibinfo{title}{{New results from a number
  operator interpretation of the compositeness of bound and resonant states}},
  \bibinfo{journal}{Annals Phys.} \bibinfo{volume}{396} (\bibinfo{year}{2018})
  \bibinfo{pages}{429--458}, \bibinfo{doi}{\doi{10.1016/j.aop.2018.07.023}},
  \eprint{1710.00991}.

\bibtype{Article}%
\bibitem{Weinberg:1962hj}
\bibinfo{author}{Steven Weinberg}, \bibinfo{title}{{Elementary particle theory
  of composite particles}}, \bibinfo{journal}{Phys. Rev.} \bibinfo{volume}{130}
  (\bibinfo{year}{1963}) \bibinfo{pages}{776--783},
  \bibinfo{doi}{\doi{10.1103/PhysRev.130.776}}.

\bibtype{Article}%
\bibitem{Weinberg:1965zz}
\bibinfo{author}{Steven Weinberg}, \bibinfo{title}{{Evidence That the Deuteron
  Is Not an Elementary Particle}}, \bibinfo{journal}{Phys. Rev.}
  \bibinfo{volume}{137} (\bibinfo{year}{1965}) \bibinfo{pages}{B672--B678},
  \bibinfo{doi}{\doi{10.1103/PhysRev.137.B672}}.

\bibtype{Article}%
\bibitem{Baru:2003qq}
\bibinfo{author}{V. Baru}, \bibinfo{author}{J. Haidenbauer},
  \bibinfo{author}{C. Hanhart}, \bibinfo{author}{Yu. Kalashnikova},
  \bibinfo{author}{Alexander~Evgenyevich Kudryavtsev},
  \bibinfo{title}{{Evidence that the $a_0(980)$ and $f_0(980)$ are not
  elementary particles}}, \bibinfo{journal}{Phys. Lett. B}
  \bibinfo{volume}{586} (\bibinfo{year}{2004}) \bibinfo{pages}{53--61},
  \bibinfo{doi}{\doi{10.1016/j.physletb.2004.01.088}}, \eprint{hep-ph/0308129}.

\bibtype{Article}%
\bibitem{Hyodo:2011qc}
\bibinfo{author}{Tetsuo Hyodo}, \bibinfo{author}{Daisuke Jido},
  \bibinfo{author}{Atsushi Hosaka}, \bibinfo{title}{{Compositeness of
  dynamically generated states in a chiral unitary approach}},
  \bibinfo{journal}{Phys. Rev. C} \bibinfo{volume}{85} (\bibinfo{year}{2012})
  \bibinfo{pages}{015201}, \bibinfo{doi}{\doi{10.1103/PhysRevC.85.015201}},
  \eprint{1108.5524}.

\bibtype{Article}%
\bibitem{Aceti:2012dd}
\bibinfo{author}{F. Aceti}, \bibinfo{author}{E. Oset}, \bibinfo{title}{{Wave
  functions of composite hadron states and relationship to couplings of
  scattering amplitudes for general partial waves}}, \bibinfo{journal}{Phys.
  Rev. D} \bibinfo{volume}{86} (\bibinfo{year}{2012}) \bibinfo{pages}{014012},
  \bibinfo{doi}{\doi{10.1103/PhysRevD.86.014012}}, \eprint{1202.4607}.

\bibtype{Article}%
\bibitem{Sekihara:2016xnq}
\bibinfo{author}{Takayasu Sekihara}, \bibinfo{title}{{Two-body wave functions
  and compositeness from scattering amplitudes. I. General properties with
  schematic models}}, \bibinfo{journal}{Phys. Rev. C} \bibinfo{volume}{95}
  (\bibinfo{number}{2}) (\bibinfo{year}{2017}) \bibinfo{pages}{025206},
  \bibinfo{doi}{\doi{10.1103/PhysRevC.95.025206}}, \eprint{1609.09496}.

\bibtype{Article}%
\bibitem{Hernandez:1984zzb}
\bibinfo{author}{E. Hernandez}, \bibinfo{author}{A. Mondragon},
  \bibinfo{title}{{Resonant states in momentum representation}},
  \bibinfo{journal}{Phys. Rev. C} \bibinfo{volume}{29} (\bibinfo{year}{1984})
  \bibinfo{pages}{722--738}, \bibinfo{doi}{\doi{10.1103/PhysRevC.29.722}}.

\bibtype{Article}%
\bibitem{Kamiya:2015aea}
\bibinfo{author}{Yuki Kamiya}, \bibinfo{author}{Tetsuo Hyodo},
  \bibinfo{title}{{Structure of near-threshold quasibound states}},
  \bibinfo{journal}{Phys. Rev. C} \bibinfo{volume}{93} (\bibinfo{number}{3})
  (\bibinfo{year}{2016}) \bibinfo{pages}{035203},
  \bibinfo{doi}{\doi{10.1103/PhysRevC.93.035203}}, \eprint{1509.00146}.

\bibtype{Article}%
\bibitem{Sekihara:2015gvw}
\bibinfo{author}{Takayasu Sekihara}, \bibinfo{author}{Takashi Arai},
  \bibinfo{author}{Junko Yamagata-Sekihara}, \bibinfo{author}{Shigehiro Yasui},
  \bibinfo{title}{{Compositeness of baryonic resonances: Application to the
  $\Delta$(1232),$N$(1535), and $N$(1650) resonances}}, \bibinfo{journal}{Phys.
  Rev. C} \bibinfo{volume}{93} (\bibinfo{number}{3}) (\bibinfo{year}{2016})
  \bibinfo{pages}{035204}, \bibinfo{doi}{\doi{10.1103/PhysRevC.93.035204}},
  \eprint{1511.01200}.

\bibtype{Article}%
\bibitem{Gao:2018jhk}
\bibinfo{author}{Rui Gao}, \bibinfo{author}{Zhi-Hui Guo},
  \bibinfo{author}{Xian-Wei Kang}, \bibinfo{author}{J.~A. Oller},
  \bibinfo{title}{{Effective-range-expansion study of near threshold
  heavy-flavor resonances}}, \bibinfo{journal}{Adv. High Energy Phys.}
  \bibinfo{volume}{2019} (\bibinfo{year}{2019}) \bibinfo{pages}{4651908},
  \bibinfo{doi}{\doi{10.1155/2019/4651908}}, \eprint{1812.07323}.

\bibtype{Article}%
\bibitem{Wang:2022vga}
\bibinfo{author}{Ze-Qiang Wang}, \bibinfo{author}{Xian-Wei Kang},
  \bibinfo{author}{J.~A. Oller}, \bibinfo{author}{Lu Zhang},
  \bibinfo{title}{{Analysis on the composite nature of the light scalar mesons
  f0(980) and a0(980)}}, \bibinfo{journal}{Phys. Rev. D} \bibinfo{volume}{105}
  (\bibinfo{number}{7}) (\bibinfo{year}{2022}) \bibinfo{pages}{074016},
  \bibinfo{doi}{\doi{10.1103/PhysRevD.105.074016}}, \eprint{2201.00492}.

\bibtype{Article}%
\bibitem{Phillips:1997xu}
\bibinfo{author}{Daniel~R. Phillips}, \bibinfo{author}{Silas~R. Beane},
  \bibinfo{author}{Thomas~D. Cohen}, \bibinfo{title}{{Nonperturbative
  regularization and renormalization: Simple examples from nonrelativistic
  quantum mechanics}}, \bibinfo{journal}{Annals Phys.} \bibinfo{volume}{263}
  (\bibinfo{year}{1998}) \bibinfo{pages}{255--275},
  \bibinfo{doi}{\doi{10.1006/aphy.1997.5771}}, \eprint{hep-th/9706070}.

\bibtype{Article}%
\bibitem{vanKolck:1998bw}
\bibinfo{author}{U. van Kolck}, \bibinfo{title}{{Effective field theory of
  short range forces}}, \bibinfo{journal}{Nucl. Phys. A} \bibinfo{volume}{645}
  (\bibinfo{year}{1999}) \bibinfo{pages}{273--302},
  \bibinfo{doi}{\doi{10.1016/S0375-9474(98)00612-5}}, \eprint{nucl-th/9808007}.

\bibtype{Article}%
\bibitem{Hanhart:2007yq}
\bibinfo{author}{C Hanhart}, \bibinfo{author}{Yu.~S Kalashnikova},
  \bibinfo{author}{Alexander~Evgenyevich Kudryavtsev}, \bibinfo{author}{A.~V
  Nefediev}, \bibinfo{title}{{Reconciling the X(3872) with the near-threshold
  enhancement in the D0 anti-D*0 final state}}, \bibinfo{journal}{Phys. Rev. D}
  \bibinfo{volume}{76} (\bibinfo{year}{2007}) \bibinfo{pages}{034007},
  \bibinfo{doi}{\doi{10.1103/PhysRevD.76.034007}}, \eprint{0704.0605}.

\bibtype{Book}%
\bibitem{tricomi.181021.1}
\bibinfo{author}{F.~G. Tricomi}, \bibinfo{title}{Integral Equations},
  \bibinfo{publisher}{Dover Publications, Inc., N.Y.} \bibinfo{year}{1985}.

\bibtype{Article}%
\bibitem{Brodsky:1973kr}
\bibinfo{author}{Stanley~J. Brodsky}, \bibinfo{author}{Glennys~R. Farrar},
  \bibinfo{title}{{Scaling Laws at Large Transverse Momentum}},
  \bibinfo{journal}{Phys. Rev. Lett.} \bibinfo{volume}{31}
  (\bibinfo{year}{1973}) \bibinfo{pages}{1153--1156},
  \bibinfo{doi}{\doi{10.1103/PhysRevLett.31.1153}}.

\bibtype{Article}%
\bibitem{Matveev:1973ra}
\bibinfo{author}{V.~A. Matveev}, \bibinfo{author}{R.~M. Muradian},
  \bibinfo{author}{A.~N. Tavkhelidze}, \bibinfo{title}{{Automodellism in the
  large - angle elastic scattering and structure of hadrons}},
  \bibinfo{journal}{Lett. Nuovo Cim.} \bibinfo{volume}{7}
  (\bibinfo{year}{1973}) \bibinfo{pages}{719--723},
  \bibinfo{doi}{\doi{10.1007/BF02728133}}.

\bibtype{Article}%
\bibitem{Brodsky:2018snc}
\bibinfo{author}{Stanley~J. Brodsky}, \bibinfo{author}{Felipe~J.
  Llanes-Estrada}, \bibinfo{title}{{Using QCD Counting rules to Identify the
  Production of Gluonium}}, \bibinfo{journal}{Phys. Lett. B}
  \bibinfo{volume}{793} (\bibinfo{year}{2019}) \bibinfo{pages}{405--410},
  \bibinfo{doi}{\doi{10.1016/j.physletb.2019.05.011}}, \eprint{1810.08772}.

\bibtype{Article}%
\bibitem{levinson.181121.1}
\bibinfo{author}{N. Levinson}, \bibinfo{title}{{}}, \bibinfo{journal}{Kgl.
  Danske Videnskab. Selskab, Mat. Fys. Medd.} \bibinfo{volume}{25}
  (\bibinfo{year}{1949}) \bibinfo{pages}{N$^{\text{o}}$ 9}.

\bibtype{Article}%
\bibitem{Weinberg:1963zza}
\bibinfo{author}{Steven Weinberg}, \bibinfo{title}{{Quasiparticles and the Born
  Series}}, \bibinfo{journal}{Phys. Rev.} \bibinfo{volume}{131}
  (\bibinfo{year}{1963}) \bibinfo{pages}{440--460},
  \bibinfo{doi}{\doi{10.1103/PhysRev.131.440}}.

\bibtype{Article}%
\bibitem{Oller:2007xd}
\bibinfo{author}{Jose~A. Oller}, \bibinfo{author}{Luis Roca},
  \bibinfo{title}{{Scalar radius of the pion and zeros in the form factor}},
  \bibinfo{journal}{Phys. Lett. B} \bibinfo{volume}{651} (\bibinfo{year}{2007})
  \bibinfo{pages}{139--146},
  \bibinfo{doi}{\doi{10.1016/j.physletb.2007.06.023}}, \eprint{0704.0039}.

\bibtype{Article}%
\bibitem{Guo:2012yt}
\bibinfo{author}{Zhi-Hui Guo}, \bibinfo{author}{J.~A. Oller},
  \bibinfo{author}{J. Ruiz~de Elvira}, \bibinfo{title}{{Chiral dynamics in form
  factors, spectral-function sum rules, meson-meson scattering and semi-local
  duality}}, \bibinfo{journal}{Phys. Rev. D} \bibinfo{volume}{86}
  (\bibinfo{year}{2012}) \bibinfo{pages}{054006},
  \bibinfo{doi}{\doi{10.1103/PhysRevD.86.054006}}, \eprint{1206.4163}.

\bibtype{Article}%
\bibitem{Morgan:1987gv}
\bibinfo{author}{D. Morgan}, \bibinfo{author}{M.~R. Pennington},
  \bibinfo{title}{{What Can We Learn From $\gamma \gamma \to \pi \pi$, $K
  \bar{K}$ in the Resonance Region}}, \bibinfo{journal}{Z. Phys. C}
  \bibinfo{volume}{37} (\bibinfo{year}{1988}) \bibinfo{pages}{431},
  \bibinfo{doi}{\doi{10.1007/BF01578139}}, \bibinfo{note}{[Erratum: Z.Phys.C
  39, 590 (1988)]}.

\bibtype{Article}%
\bibitem{Morgan:1990kw}
\bibinfo{author}{D. Morgan}, \bibinfo{author}{M.~R. Pennington},
  \bibinfo{title}{{Amplitude Analysis of $\gamma \gamma \to \pi \pi$ From
  Threshold to 1.4-{GeV}}}, \bibinfo{journal}{Z. Phys. C} \bibinfo{volume}{48}
  (\bibinfo{year}{1990}) \bibinfo{pages}{623--632},
  \bibinfo{doi}{\doi{10.1007/BF01614697}}.

\bibtype{Article}%
\bibitem{Pennington:2006dg}
\bibinfo{author}{M.~R. Pennington}, \bibinfo{title}{{Sigma coupling to photons:
  Hidden scalar in gamma gamma ---\ensuremath{>} pi0 pi0}},
  \bibinfo{journal}{Phys. Rev. Lett.} \bibinfo{volume}{97}
  (\bibinfo{year}{2006}) \bibinfo{pages}{011601},
  \bibinfo{doi}{\doi{10.1103/PhysRevLett.97.011601}}.

\bibtype{Article}%
\bibitem{Danilkin:2018qfn}
\bibinfo{author}{Igor Danilkin}, \bibinfo{author}{Marc Vanderhaeghen},
  \bibinfo{title}{{Dispersive analysis of the $\gamma\gamma^{*} \to \pi \pi$
  process}}, \bibinfo{journal}{Phys. Lett. B} \bibinfo{volume}{789}
  (\bibinfo{year}{2019}) \bibinfo{pages}{366--372},
  \bibinfo{doi}{\doi{10.1016/j.physletb.2018.12.047}}, \eprint{1810.03669}.

\bibtype{Article}%
\bibitem{Oller:2007sh}
\bibinfo{author}{Jose~A. Oller}, \bibinfo{author}{Luis Roca},
  \bibinfo{author}{Carlos Schat}, \bibinfo{title}{{Improved dispersion
  relations for $\gamma \gamma \to \pi^0 \pi^0$}}, \bibinfo{journal}{Phys.
  Lett. B} \bibinfo{volume}{659} (\bibinfo{year}{2008})
  \bibinfo{pages}{201--208},
  \bibinfo{doi}{\doi{10.1016/j.physletb.2007.11.030}}, \eprint{0708.1659}.

\bibtype{Article}%
\bibitem{Oller:2008kf}
\bibinfo{author}{J.~A. Oller}, \bibinfo{author}{L. Roca}, \bibinfo{title}{{Two
  photons into pi0 pi0}}, \bibinfo{journal}{Eur. Phys. J. A}
  \bibinfo{volume}{37} (\bibinfo{year}{2008}) \bibinfo{pages}{15--32},
  \bibinfo{doi}{\doi{10.1140/epja/i2008-10600-0}}, \eprint{0804.0309}.

\bibtype{Article}%
\bibitem{Low:1954kd}
\bibinfo{author}{F.~E. Low}, \bibinfo{title}{{Scattering of light of very low
  frequency by systems of spin 1/2}}, \bibinfo{journal}{Phys. Rev.}
  \bibinfo{volume}{96} (\bibinfo{year}{1954}) \bibinfo{pages}{1428--1432},
  \bibinfo{doi}{\doi{10.1103/PhysRev.96.1428}}.

\bibtype{Article}%
\bibitem{Bijnens:1987dc}
\bibinfo{author}{Johan Bijnens}, \bibinfo{author}{Fernando Cornet},
  \bibinfo{title}{{Two Pion Production in Photon-Photon Collisions}},
  \bibinfo{journal}{Nucl. Phys. B} \bibinfo{volume}{296} (\bibinfo{year}{1988})
  \bibinfo{pages}{557--568}, \bibinfo{doi}{\doi{10.1016/0550-3213(88)90032-6}}.

\bibtype{Article}%
\bibitem{Donoghue:1988eea}
\bibinfo{author}{John~F. Donoghue}, \bibinfo{author}{Barry~R. Holstein},
  \bibinfo{author}{Y.~C. Lin}, \bibinfo{title}{{The Reaction gamma Gamma
  ---\ensuremath{>} pi0 pi0 and Chiral Loops}}, \bibinfo{journal}{Phys. Rev. D}
  \bibinfo{volume}{37} (\bibinfo{year}{1988}) \bibinfo{pages}{2423},
  \bibinfo{doi}{\doi{10.1103/PhysRevD.37.2423}}.

\bibtype{Book}%
\bibitem{musk.181126.1}
\bibinfo{author}{W.~I. Muskhelishvili}, \bibinfo{title}{Singular Integral
  Equations}, \bibinfo{publisher}{North-Holland, Amsterdam, 1958}
  \bibinfo{year}{1958}.

\bibtype{Article}%
\bibitem{warnock.181127.1}
\bibinfo{author}{R.~L. Warnock}, \bibinfo{title}{{Matrix $N/D$ method with
  absorption and the unitarity problem in coupled-channel Regge theory}},
  \bibinfo{journal}{Phys. Rev. D} \bibinfo{volume}{22} (\bibinfo{year}{1980})
  \bibinfo{pages}{2077},
  \bibinfo{doi}{\doi{https://doi.org/10.1103/PhysRevD.23.1832}}.

\bibtype{Article}%
\bibitem{Babelon:1976kv}
\bibinfo{author}{O. Babelon}, \bibinfo{author}{J.~L. Basdevant},
  \bibinfo{author}{D. Caillerie}, \bibinfo{author}{G. Mennessier},
  \bibinfo{title}{{Unitarity and Inelastic Final State Interactions}},
  \bibinfo{journal}{Nucl. Phys. B} \bibinfo{volume}{113} (\bibinfo{year}{1976})
  \bibinfo{pages}{445--476}, \bibinfo{doi}{\doi{10.1016/0550-3213(76)90137-1}}.

\bibtype{Article}%
\bibitem{Jamin:2001zq}
\bibinfo{author}{Matthias Jamin}, \bibinfo{author}{Jose~Antonio Oller},
  \bibinfo{author}{Antonio Pich}, \bibinfo{title}{{Strangeness changing scalar
  form-factors}}, \bibinfo{journal}{Nucl. Phys. B} \bibinfo{volume}{622}
  (\bibinfo{year}{2002}) \bibinfo{pages}{279--308},
  \bibinfo{doi}{\doi{10.1016/S0550-3213(01)00605-8}}, \eprint{hep-ph/0110193}.

\bibtype{Article}%
\bibitem{Donoghue:1990xh}
\bibinfo{author}{John~F. Donoghue}, \bibinfo{author}{J. Gasser},
  \bibinfo{author}{H. Leutwyler}, \bibinfo{title}{{The Decay of a Light Higgs
  Boson}}, \bibinfo{journal}{Nucl. Phys. B} \bibinfo{volume}{343}
  (\bibinfo{year}{1990}) \bibinfo{pages}{341--368},
  \bibinfo{doi}{\doi{10.1016/0550-3213(90)90474-R}}.

\bibtype{Article}%
\bibitem{Moussallam:1999aq}
\bibinfo{author}{Bachir Moussallam}, \bibinfo{title}{{N(f) dependence of the
  quark condensate from a chiral sum rule}}, \bibinfo{journal}{Eur. Phys. J. C}
  \bibinfo{volume}{14} (\bibinfo{year}{2000}) \bibinfo{pages}{111--122},
  \bibinfo{doi}{\doi{10.1007/s100520050738}}, \eprint{hep-ph/9909292}.

\bibtype{Article}%
\bibitem{Jamin:2000wn}
\bibinfo{author}{Matthias Jamin}, \bibinfo{author}{Jose~Antonio Oller},
  \bibinfo{author}{Antonio Pich}, \bibinfo{title}{{S wave K pi scattering in
  chiral perturbation theory with resonances}}, \bibinfo{journal}{Nucl. Phys.
  B} \bibinfo{volume}{587} (\bibinfo{year}{2000}) \bibinfo{pages}{331--362},
  \bibinfo{doi}{\doi{10.1016/S0550-3213(00)00479-X}}, \eprint{hep-ph/0006045}.

\bibtype{Article}%
\bibitem{Aston:1987ir}
\bibinfo{author}{D. Aston}, et al., \bibinfo{title}{{A Study of K- pi+
  Scattering in the Reaction K- p ---\ensuremath{>} K- pi+ n at 11-GeV/c}},
  \bibinfo{journal}{Nucl. Phys. B} \bibinfo{volume}{296} (\bibinfo{year}{1988})
  \bibinfo{pages}{493--526}, \bibinfo{doi}{\doi{10.1016/0550-3213(88)90028-4}}.

\bibtype{Article}%
\bibitem{Gasser:1984ux}
\bibinfo{author}{J. Gasser}, \bibinfo{author}{H. Leutwyler},
  \bibinfo{title}{{Low-Energy Expansion of Meson Form-Factors}},
  \bibinfo{journal}{Nucl. Phys. B} \bibinfo{volume}{250} (\bibinfo{year}{1985})
  \bibinfo{pages}{517--538}, \bibinfo{doi}{\doi{10.1016/0550-3213(85)90493-6}}.

\bibtype{Article}%
\bibitem{Khuri:1960zz}
\bibinfo{author}{N.~N. Khuri}, \bibinfo{author}{S.~B. Treiman},
  \bibinfo{title}{{Pion-Pion Scattering and K + /- --\ensuremath{>} 3pi
  Decay}}, \bibinfo{journal}{Phys. Rev.} \bibinfo{volume}{119}
  (\bibinfo{year}{1960}) \bibinfo{pages}{1115--1121},
  \bibinfo{doi}{\doi{10.1103/PhysRev.119.1115}}.

\bibtype{Article}%
\bibitem{Gribov:1962fu}
\bibinfo{author}{V.~N. Gribov}, \bibinfo{author}{V.~V. Anisovich},
  \bibinfo{author}{A.~A. Anselm}, \bibinfo{title}{{Contribution to the theory
  of the PI + N --\ensuremath{>} N + PI + PI and GAMMA + N --\ensuremath{>} N +
  PI + PI reactions near threshold}}, \bibinfo{journal}{Sov. Phys. JETP}
  \bibinfo{volume}{15} (\bibinfo{year}{1962}) \bibinfo{pages}{159}.

\bibtype{Article}%
\bibitem{Bronzan:1963mby}
\bibinfo{author}{John~B. Bronzan}, \bibinfo{author}{Claude Kacser},
  \bibinfo{title}{{Khuri-Treiman Representation and Perturbation Theory}},
  \bibinfo{journal}{Phys. Rev.} \bibinfo{volume}{132} (\bibinfo{number}{6})
  (\bibinfo{year}{1963}) \bibinfo{pages}{2703},
  \bibinfo{doi}{\doi{10.1103/PhysRev.132.2703}}.

\bibtype{Article}%
\bibitem{Kacser:1963zz}
\bibinfo{author}{C. Kacser}, \bibinfo{title}{{Analytic Structure of
  Partial-Wave Amplitudes for Production and Decay Processes}},
  \bibinfo{journal}{Phys. Rev.} \bibinfo{volume}{132} (\bibinfo{number}{6})
  (\bibinfo{year}{1963}) \bibinfo{pages}{2712},
  \bibinfo{doi}{\doi{10.1103/PhysRev.132.2712}}.

\bibtype{Article}%
\bibitem{Aitchison:1965zz}
\bibinfo{author}{I.~J.~R. Aitchison}, \bibinfo{title}{{Dispersion Theory Model
  of Three-Body Production and Decay Processes}}, \bibinfo{journal}{Phys. Rev.}
  \bibinfo{volume}{137} (\bibinfo{year}{1965}) \bibinfo{pages}{B1070--B1084},
  \bibinfo{doi}{\doi{10.1103/PhysRev.137.B1070}}.

\bibtype{Article}%
\bibitem{Pasquier:1969dt}
\bibinfo{author}{R. Pasquier}, \bibinfo{author}{J.~Y. Pasquier},
  \bibinfo{title}{{Khuri-treiman-type equations for three-body decay and
  production processes. 2.}}, \bibinfo{journal}{Phys. Rev.}
  \bibinfo{volume}{177} (\bibinfo{year}{1969}) \bibinfo{pages}{2482--2493},
  \bibinfo{doi}{\doi{10.1103/PhysRev.177.2482}}.

\bibtype{Article}%
\bibitem{Neveu:1970tn}
\bibinfo{author}{A. Neveu}, \bibinfo{author}{J. Scherk},
  \bibinfo{title}{{Final-state interaction and current algebra in k-3pi and
  eta-3pi decays}}, \bibinfo{journal}{Annals Phys.} \bibinfo{volume}{57}
  (\bibinfo{year}{1970}) \bibinfo{pages}{39--64},
  \bibinfo{doi}{\doi{10.1016/0003-4916(70)90268-X}}.

\bibtype{Article}%
\bibitem{Kambor:1995yc}
\bibinfo{author}{J. Kambor}, \bibinfo{author}{C. Wiesendanger},
  \bibinfo{author}{D. Wyler}, \bibinfo{title}{{Final state interactions and
  Khuri-Treiman equations in eta --\ensuremath{>} 3 pi decays}},
  \bibinfo{journal}{Nucl. Phys. B} \bibinfo{volume}{465} (\bibinfo{year}{1996})
  \bibinfo{pages}{215--266}, \bibinfo{doi}{\doi{10.1016/0550-3213(95)00676-1}},
  \eprint{hep-ph/9509374}.

\bibtype{Article}%
\bibitem{Anisovich:1996tx}
\bibinfo{author}{A.~V. Anisovich}, \bibinfo{author}{H. Leutwyler},
  \bibinfo{title}{{Dispersive analysis of the decay eta ---\ensuremath{>} 3
  pi}}, \bibinfo{journal}{Phys. Lett. B} \bibinfo{volume}{375}
  (\bibinfo{year}{1996}) \bibinfo{pages}{335--342},
  \bibinfo{doi}{\doi{10.1016/0370-2693(96)00192-X}}, \eprint{hep-ph/9601237}.

\bibtype{Article}%
\bibitem{Guo:2016wsi}
\bibinfo{author}{P. Guo}, \bibinfo{author}{I.~V. Danilkin}, \bibinfo{author}{C.
  Fern\'andez-Ram\'\i{}rez}, \bibinfo{author}{V. Mathieu},
  \bibinfo{author}{A.~P. Szczepaniak}, \bibinfo{title}{{Three-body final state
  interaction in \ensuremath{\eta} \textrightarrow{} 3 \ensuremath{\pi}
  updated}}, \bibinfo{journal}{Phys. Lett. B} \bibinfo{volume}{771}
  (\bibinfo{year}{2017}) \bibinfo{pages}{497--502},
  \bibinfo{doi}{\doi{10.1016/j.physletb.2017.05.092}}, \eprint{1608.01447}.

\bibtype{Article}%
\bibitem{Albaladejo:2017hhj}
\bibinfo{author}{M. Albaladejo}, \bibinfo{author}{B. Moussallam},
  \bibinfo{title}{{Extended chiral Khuri-Treiman formalism for $\eta\to 3\pi$
  and the role of the $a_0(980)$, $f_0(980)$ resonances}},
  \bibinfo{journal}{Eur. Phys. J. C} \bibinfo{volume}{77} (\bibinfo{number}{8})
  (\bibinfo{year}{2017}) \bibinfo{pages}{508},
  \bibinfo{doi}{\doi{10.1140/epjc/s10052-017-5052-x}}, \eprint{1702.04931}.

\bibtype{Article}%
\bibitem{Moussallam:2016evb}
\bibinfo{author}{B. Moussallam}, \bibinfo{author}{M. Albaladejo},
  \bibinfo{title}{{Role of the $a_{0}(980), f_{0}(980)$ resonances in $\eta \to
  3\pi$ from the Khuri-Treiman formalism}}, \bibinfo{journal}{EPJ Web Conf.}
  \bibinfo{volume}{130} (\bibinfo{year}{2016}) \bibinfo{pages}{03007},
  \bibinfo{doi}{\doi{10.1051/epjconf/201613003007}}.

\bibtype{Article}%
\bibitem{Urech:1994hd}
\bibinfo{author}{Res Urech}, \bibinfo{title}{{Virtual photons in chiral
  perturbation theory}}, \bibinfo{journal}{Nucl. Phys. B} \bibinfo{volume}{433}
  (\bibinfo{year}{1995}) \bibinfo{pages}{234--254},
  \bibinfo{doi}{\doi{10.1016/0550-3213(95)90707-N}}, \eprint{hep-ph/9405341}.

\bibtype{Article}%
\bibitem{Ditsche:2008cq}
\bibinfo{author}{Christoph Ditsche}, \bibinfo{author}{Bastian Kubis},
  \bibinfo{author}{Ulf-G. Mei{\ss}ner}, \bibinfo{title}{{Electromagnetic
  corrections in $\eta \to 3 \pi$ decays}}, \bibinfo{journal}{Eur. Phys. J. C}
  \bibinfo{volume}{60} (\bibinfo{year}{2009}) \bibinfo{pages}{83--105},
  \bibinfo{doi}{\doi{10.1140/epjc/s10052-009-0863-z}}, \eprint{0812.0344}.

\bibtype{Article}%
\bibitem{Roiesnel:1980gd}
\bibinfo{author}{C. Roiesnel}, \bibinfo{author}{Tran~N. Truong},
  \bibinfo{title}{{Resolution of the $\eta \to 3 \pi$ Problem}},
  \bibinfo{journal}{Nucl. Phys. B} \bibinfo{volume}{187} (\bibinfo{year}{1981})
  \bibinfo{pages}{293--300}, \bibinfo{doi}{\doi{10.1016/0550-3213(81)90275-3}}.

\bibtype{Article}%
\bibitem{Beisert:2003zs}
\bibinfo{author}{N. Beisert}, \bibinfo{author}{B. Borasoy},
  \bibinfo{title}{{Hadronic decays of eta and eta-prime with coupled
  channels}}, \bibinfo{journal}{Nucl. Phys. A} \bibinfo{volume}{716}
  (\bibinfo{year}{2003}) \bibinfo{pages}{186--208},
  \bibinfo{doi}{\doi{10.1016/S0375-9474(02)01585-3}}, \eprint{hep-ph/0301058}.

\bibtype{Article}%
\bibitem{Borasoy:2005du}
\bibinfo{author}{B. Borasoy}, \bibinfo{author}{R. Nissler},
  \bibinfo{title}{{Hadronic eta and eta-prime decays}}, \bibinfo{journal}{Eur.
  Phys. J. A} \bibinfo{volume}{26} (\bibinfo{year}{2005})
  \bibinfo{pages}{383--398}, \bibinfo{doi}{\doi{10.1140/epja/i2005-10188-9}},
  \eprint{hep-ph/0510384}.

\bibtype{Article}%
\bibitem{Mandelstam:1960zz}
\bibinfo{author}{S. Mandelstam}, \bibinfo{title}{{Unitarity Condition Below
  Physical Thresholds in the Normal and Anomalous Cases}},
  \bibinfo{journal}{Phys. Rev. Lett.} \bibinfo{volume}{4}
  (\bibinfo{year}{1960}) \bibinfo{pages}{84--87},
  \bibinfo{doi}{\doi{10.1103/PhysRevLett.4.84}}.

\bibtype{Article}%
\bibitem{Colangelo:2018jxw}
\bibinfo{author}{Gilberto Colangelo}, \bibinfo{author}{Stefan Lanz},
  \bibinfo{author}{Heinrich Leutwyler}, \bibinfo{author}{Emilie Passemar},
  \bibinfo{title}{{Dispersive analysis of $\eta \rightarrow 3 \pi $}},
  \bibinfo{journal}{Eur. Phys. J. C} \bibinfo{volume}{78}
  (\bibinfo{number}{11}) (\bibinfo{year}{2018}) \bibinfo{pages}{947},
  \bibinfo{doi}{\doi{10.1140/epjc/s10052-018-6377-9}}, \eprint{1807.11937}.

\bibtype{Article}%
\bibitem{Albaladejo:2018gif}
\bibinfo{author}{M. Albaladejo}, \bibinfo{author}{N. Sherrill},
  \bibinfo{author}{C. Fern\'andez-Ram\'\i{}rez}, \bibinfo{author}{A. Jackura},
  \bibinfo{author}{V. Mathieu}, \bibinfo{author}{M. Mikhasenko},
  \bibinfo{author}{J. Nys}, \bibinfo{author}{A. Pilloni},
  \bibinfo{author}{A.~P. Szczepaniak} (\bibinfo{collaboration}{JPAC}),
  \bibinfo{title}{{Khuri\textendash{}Treiman equations for $\pi \pi $
  scattering}}, \bibinfo{journal}{Eur. Phys. J. C} \bibinfo{volume}{78}
  (\bibinfo{number}{7}) (\bibinfo{year}{2018}) \bibinfo{pages}{574},
  \bibinfo{doi}{\doi{10.1140/epjc/s10052-018-6045-0}}, \eprint{1803.06027}.

\bibtype{Article}%
\bibitem{Hanhart:2015cua}
\bibinfo{author}{C. Hanhart}, \bibinfo{author}{Yu.~S. Kalashnikova},
  \bibinfo{author}{P. Matuschek}, \bibinfo{author}{R.~V. Mizuk},
  \bibinfo{author}{A.~V. Nefediev}, \bibinfo{author}{Q. Wang},
  \bibinfo{title}{{Practical Parametrization for Line Shapes of Near-Threshold
  States}}, \bibinfo{journal}{Phys. Rev. Lett.} \bibinfo{volume}{115}
  (\bibinfo{number}{20}) (\bibinfo{year}{2015}) \bibinfo{pages}{202001},
  \bibinfo{doi}{\doi{10.1103/PhysRevLett.115.202001}}, \eprint{1507.00382}.

\bibtype{Article}%
\bibitem{Heuser:2024biq}
\bibinfo{author}{L.~A. Heuser}, \bibinfo{author}{G. Chanturia},
  \bibinfo{author}{F.~K. Guo}, \bibinfo{author}{C. Hanhart},
  \bibinfo{author}{M. Hoferichter}, \bibinfo{author}{B. Kubis},
  \bibinfo{title}{{From pole parameters to line shapes and branching ratios}},
  \bibinfo{journal}{Eur. Phys. J. C} \bibinfo{volume}{84} (\bibinfo{number}{6})
  (\bibinfo{year}{2024}) \bibinfo{pages}{599},
  \bibinfo{doi}{\doi{10.1140/epjc/s10052-024-12884-6}}, \eprint{2403.15539}.

\bibtype{Article}%
\bibitem{Nakano:1982bc}
\bibinfo{author}{K. Nakano}, \bibinfo{title}{{Two potential formalisms and the
  Coulomb-nuclear interference}}, \bibinfo{journal}{Phys. Rev. C}
  \bibinfo{volume}{26} (\bibinfo{year}{1982}) \bibinfo{pages}{1123--1131},
  \bibinfo{doi}{\doi{10.1103/PhysRevC.26.1123}}.

\bibtype{Article}%
\bibitem{Ray:1980ck}
\bibinfo{author}{L. Ray}, \bibinfo{author}{G.~W. Hoffmann},
  \bibinfo{author}{R.~M. Thaler}, \bibinfo{title}{{Coulomb interaction in
  multiple scattering theory}}, \bibinfo{journal}{Phys. Rev. C}
  \bibinfo{volume}{22} (\bibinfo{year}{1980}) \bibinfo{pages}{1454--1467},
  \bibinfo{doi}{\doi{10.1103/PhysRevC.22.1454}}.

\bibtype{Article}%
\bibitem{Yndurain:2002ud}
\bibinfo{author}{F.~J. Yndurain}, \bibinfo{title}{{Low-energy pion physics}}
  (\bibinfo{year}{2002}), \eprint{hep-ph/0212282}.

\bibtype{Article}%
\bibitem{Hanhart:2012wi}
\bibinfo{author}{C. Hanhart}, \bibinfo{title}{{A New Parameterization for the
  Pion Vector Form Factor}}, \bibinfo{journal}{Phys. Lett. B}
  \bibinfo{volume}{715} (\bibinfo{year}{2012}) \bibinfo{pages}{170--177},
  \bibinfo{doi}{\doi{10.1016/j.physletb.2012.07.038}}, \eprint{1203.6839}.

\bibtype{Article}%
\bibitem{Burkert:2022bqo}
\bibinfo{author}{V. Burkert~{\it et al.}}, \bibinfo{title}{{Note on the
  definitions of branching ratios of overlapping resonances}},
  \bibinfo{journal}{Phys. Lett. B} \bibinfo{volume}{844} (\bibinfo{year}{2023})
  \bibinfo{pages}{138070}, \bibinfo{doi}{\doi{10.1016/j.physletb.2023.138070}},
  \eprint{2207.08472}.

\bibtype{Article}%
\bibitem{Garcia-Martin:2011iqs}
\bibinfo{author}{R. Garcia-Martin}, \bibinfo{author}{R. Kaminski},
  \bibinfo{author}{J.~R. Pelaez}, \bibinfo{author}{J. Ruiz~de Elvira},
  \bibinfo{author}{F.~J. Yndurain}, \bibinfo{title}{{The Pion-pion scattering
  amplitude. IV: Improved analysis with once subtracted Roy-like equations up
  to 1100 MeV}}, \bibinfo{journal}{Phys. Rev. D} \bibinfo{volume}{83}
  (\bibinfo{year}{2011}) \bibinfo{pages}{074004},
  \bibinfo{doi}{\doi{10.1103/PhysRevD.83.074004}}, \eprint{1102.2183}.

\bibtype{Article}%
\bibitem{Garcia-Martin:2011nna}
\bibinfo{author}{R. Garcia-Martin}, \bibinfo{author}{R. Kaminski},
  \bibinfo{author}{J.~R. Pelaez}, \bibinfo{author}{J. Ruiz~de Elvira},
  \bibinfo{title}{{Precise determination of the f0(600) and f0(980) pole
  parameters from a dispersive data analysis}}, \bibinfo{journal}{Phys. Rev.
  Lett.} \bibinfo{volume}{107} (\bibinfo{year}{2011}) \bibinfo{pages}{072001},
  \bibinfo{doi}{\doi{10.1103/PhysRevLett.107.072001}}, \eprint{1107.1635}.

\bibtype{Article}%
\bibitem{Guo:2012ym}
\bibinfo{author}{Zhi-Hui Guo}, \bibinfo{author}{J.~A. Oller},
  \bibinfo{author}{J. Ruiz~de Elvira}, \bibinfo{title}{{Chiral dynamics in U(3)
  unitary chiral perturbation theory}}, \bibinfo{journal}{Phys. Lett. B}
  \bibinfo{volume}{712} (\bibinfo{year}{2012}) \bibinfo{pages}{407--412},
  \bibinfo{doi}{\doi{10.1016/j.physletb.2012.05.021}}, \eprint{1203.4381}.

\end{thebibliography*}

\end{document}